\documentclass[a4paper, amsfonts, amssymb, amsmath, onecolumn, showkeys, nofootinbib, twoside, superscriptaddress]{revtex4-2}
\usepackage[english]{babel}
\usepackage[utf8]{inputenc}
\usepackage[colorinlistoftodos, color=green!40, prependcaption]{todonotes}
\usepackage{amsthm}
\usepackage{mathtools}
\usepackage{physics}
\usepackage{xcolor}
\usepackage{graphicx}
\usepackage{gensymb}
\usepackage{multirow}
\usepackage[left=23mm,right=13mm,top=35mm,columnsep=15pt]{geometry} 
\usepackage{adjustbox}
\usepackage{placeins}
\usepackage[T1]{fontenc}
\usepackage{csquotes}
\usepackage[pdftex, pdftitle={Article}, pdfauthor={Author}]{hyperref} 
\usepackage[final]{changes}

\begin{document}
\title{Intramolecular Structural Heterogeneity altered by Long-range Contacts in an Intrinsically Disordered Protein}

\author{Gil Koren}
    \affiliation{Raymond \& Beverly Sackler School of Physics \& Astronomy, Tel Aviv University, Tel Aviv 6997801, Israel}
    \affiliation{The Center for Physics \& Chemistry of Living Systems, Tel Aviv University, Tel Aviv 6997801, Israel}
    \affiliation{The Center for NanoTechnology \& NanoScience, Tel Aviv University, Tel Aviv 6997801, Israel}
\author{Sagi Meir}
    \affiliation{Raymond \& Beverly Sackler School of Physics \& Astronomy, Tel Aviv University, Tel Aviv 6997801, Israel}
    \affiliation{The Center for Physics \& Chemistry of Living Systems, Tel Aviv University, Tel Aviv 6997801, Israel}
    \affiliation{The Center for NanoTechnology \& NanoScience, Tel Aviv University, Tel Aviv 6997801, Israel}
\author{Lennard Holschuh}
    \affiliation{Applied Theoretical Physics-Computational Physics, Physikalisches Institut, Albert-Ludwigs-Universit Freiburg, D-79104 Freiburg, Germany}
\author{Haydyn D.T. Mertens}
    \affiliation{European Molecular Biology Laboratory, Hamburg Unit, 22607 Hamburg, Germany}
\author{Tamara Ehm}
    \affiliation{Raymond \& Beverly Sackler School of Physics \& Astronomy, Tel Aviv University, Tel Aviv 6997801, Israel}
    \affiliation{The Center for Physics \& Chemistry of Living Systems, Tel Aviv University, Tel Aviv 6997801, Israel}
    \affiliation{The Center for NanoTechnology \& NanoScience, Tel Aviv University, Tel Aviv 6997801, Israel}
    \affiliation{Faculty of Physics and Center for NanoScience, Ludwig-Maximilians-Universit{\"a}t, M{\"u}nchen D-80539, Germany}
\author{Nadav Yahalom}
    \affiliation{The Center for Physics \& Chemistry of Living Systems, Tel Aviv University, Tel Aviv 6997801, Israel}
    \affiliation{The Center for NanoTechnology \& NanoScience, Tel Aviv University, Tel Aviv 6997801, Israel}
    \affiliation{School of Chemistry, Raymond and Beverly Sackler Faculty of Exact Sciences and Tel Aviv University Center for Light-Matter Interaction, Tel Aviv University, Tel Aviv 6997801, Israel}
\author{Adina Golombek}
    \affiliation{The Center for Nanoscience and Nanotechnology, Tel Aviv University, Tel Aviv 69978, Israel}
    \affiliation{School of Chemistry, Raymond and Beverly Sackler Faculty of Exact Sciences and Tel Aviv University Center for Light-Matter Interaction, Tel Aviv University, Tel Aviv 6997801, Israel}
\author{Tal Schwartz}
    \affiliation{The Center for Nanoscience and Nanotechnology, Tel Aviv University, Tel Aviv 69978, Israel}
    \affiliation{School of Chemistry, Raymond and Beverly Sackler Faculty of Exact Sciences and Tel Aviv University Center for Light-Matter Interaction, Tel Aviv University, Tel Aviv 6997801, Israel}
\author{Dmitri I. Svergun}
    \affiliation{European Molecular Biology Laboratory, Hamburg Unit, 22607 Hamburg, Germany}
\author{Omar A. Saleh}
    \affiliation{BMSE Program, University of California, Santa Barbara, CA 93110, United States of America}
    \affiliation{Materials Department, University of California, Santa Barbara, CA 93110, United States of America}
\author{Joachim Dzubiella}
    \affiliation{Applied Theoretical Physics-Computational Physics, Physikalisches Institut, Albert-Ludwigs-Universit Freiburg, D-79104 Freiburg, Germany}
    \affiliation{Cluster of Excellence livMatS @ FIT–Freiburg Center for Interactive Materials and Bioinspired Technologies, Albert-Ludwigs-Universit Freiburg, D-79104 Freiburg, Germany}
\author{Roy Beck}
    \email[Correspondence email address: ]{roy@tauex.tau.ac.il}
    \affiliation{Raymond \& Beverly Sackler School of Physics \& Astronomy, Tel Aviv University, Tel Aviv 6997801, Israel}
    \affiliation{The Center for Physics \& Chemistry of Living Systems, Tel Aviv University, Tel Aviv 6997801, Israel}
    \affiliation{The Center for NanoTechnology \& NanoScience, Tel Aviv University, Tel Aviv 6997801, Israel}

\date{\today} 

\begin{abstract}
Short-range interactions and long-range contacts drive the 3D folding of structured proteins. The proteins' structure has a direct impact on their biological function. However, nearly \replaced{40\% of the eukaryotes proteome}{half of the proteome} is composed of intrinsically disordered proteins (IDPs) and protein regions that fluctuate between ensembles of numerous conformations. Therefore, to understand their biological function, it is critical to depict how the structural ensemble statistics correlate to the IDPs' amino acid sequence. Here, using small-angle x-ray scattering (SAXS) and time-resolved F{\"o}rster resonance energy transfer (trFRET), we study the intra-molecular structural heterogeneity of the neurofilament low intrinsically disordered tail domain (NFLt). Using theoretical \replaced{results}{analogs} of polymer physics, we find that the Flory scaling exponent of NFLt sub-segments correlates linearly with their net charge, ranging from statistics of ideal to self-avoiding chains. \deleted{Further analysis reveals specific structural elements arising from di-proline bending and transient loop formation.} Surprisingly, measuring the same segments in the context of the whole NFLt protein, we find that regardless of the peptide sequence, the segments' structural statistics are more expanded than when measured independently. Our findings show that while polymer physics can, to some level, relate the IDP's sequence to its ensemble conformations, long-range contacts between distant amino acids play a crucial role in determining intra-molecular structures. This emphasizes the necessity of advanced polymer theories to fully describe IDPs ensembles with the hope it will allow us to model their biological function.
\end{abstract}


\maketitle

It has been more than two decades since the structure-to-function paradigm was challenged by the discovery that intrinsically disordered proteins (IDPs), that lack fixed 3D structures, have significant biological functions \cite{kriwacki1996structural, wright1999intrinsically}. Yet, many questions relating the primary sequence to biological function remain open \cite{xue2012orderly,uversky2019intrinsically,ehm2021intrinsically}. IDPs are known to participate in many cellular functions \cite{dyson2005intrinsically, forman2013sequence} and to remain unstructured even in the crowded environment of the cell \cite{theillet2016structural}. IDPs can become structured or remain unstructured while participating in protein-protein interaction \cite{mollica2016binding}. Either way, the interaction can be very strong while fully retaining the IDP's structural disorder \cite{borgia2018extreme}. 

For IDPs, a dynamic ensemble of conformations characterizes its `structure' \cite{metskas2020single}. But, a fundamental yet unresolved question is \replaced{to what extent}{how much} IDPs can be treated as conventional polymers, \replaced{versus requiring consideration of}{or should we consider} intramolecular structural heterogeneity \deleted{(ISH)} derived from amino acid sequence\deleted{to describe IDPs better}? Some IDPs have minimal sequence heterogeneity (e.g., PAS repeats \cite{breibeck2018polypeptide,gebauer2018prospects},  GRGDSPYS repeats \cite{dzuricky2020novo}, SR and RRRRR repeats \cite{mao2010net}, FG repeats \cite{milles2011single}), and are thus expected  to behave as conventional long homopolymers. However, other more heterogeneous IDPs show discrepancies \replaced{with}{to} these polymeric models \cite{baul2019sequence,riback2017innovative}. Therefore, evaluating \replaced{intramolecular structural heterogeneity}{ISH} may highlight fundamental properties and their relation to the IDPs' sequence, \replaced{allowing unraveling of}{leading to unravel} their function. 

IDPs can be divided into three major archetype sequences \cite{van2014classification}: polar tracts, polyelectrolytes, and polyampholytes. Each type could exhibit \replaced{intramolecular structural heterogeneity}{ISH} depending on its sequence \cite{baul2019sequence}. For example, it was shown that polyelectrolyte regions are locally stretched due to electrostatic repulsion as opposed to electrostatic bridges in the polyampholyte regime \cite{muller2010charge}. Moreover, it was found that IDPs can include collapsed domains due to the hydrophobic interaction \cite{milles2011single}. 

The conformational ensemble (i.e., sizes and shapes) of many polyelectrolytic IDPs vary, as expected, as a function of the net charge per residue (NCPR) \cite{muller2010charge,mao2010net,bianchi2022distribution,marsh2010sequence}. The NCPR is a simple average metric, which by definition cannot capture heterogeneity. Therefore, in the case of IDPs that are more polyampholitic or richer in polar residues, NCPR might not fully capture its heterogeneity. \deleted{Following, }The next level of description would be the variances of charge and hydrophobicity patterning parameters, which can capture additional heterogeneity \added{and long-range contacts} \cite{das2013conformations,sawle2015theoretical,zheng2020hydropathy, devarajan2022effect, wessen2022analytical}. 

\added {Unlike the coarse-grained metrics above, the impact of long-range contacts on an IDP's ensemble of structures is difficult to evaluate.}
\deleted{a more elusive property that contributes to the IDPs' ensemble of structures are long-range contacts between distant amino acids.} These contacts contribute on case-by-case bases, similarly to structural folded proteins. For example, long-range contacts were found to contribute to the intramolecular heterogeneity of the IDP prion monomer, which \replaced{adopts}{has adopted} collapsed and extended regions \cite{mukhopadhyay2007natively}. \added{Recently, long-range contacts in the cold denatured state made the ensemble contract, deviating from the homopolymer model \cite{stenzoski2020cold}. Similarly, long-range stickers of aromatic residues tend to contract IDPs \cite{martin2020valence}. Another work showed that long-range interactions can contract an IDP after swapping $<2\%$ of the residues \cite{bowman2020properties}. These results show that  IDPs' ensembles are highly sequenced dependent.}\deleted{
Similarly, the IDP $\alpha$-synuclein showed correlated long-range dynamics atypical for a Markovian random-walk polymer \cite{parigi2014long}. This result is in agreement with weak long-ranged contacts present in $\alpha$-synuclein \cite{schwalbe2014predictive}.}

\added{IDPs are often measured in isolation. This raises the question: to what extent does their context influence their behavior? Here, we will show IDP segments measured in the context of flanking sequences have long-range contacts that can expand the ensemble of conformations. We will} \deleted{Here, we will show that long-range contacts are a crucial component for understanding IDPs, and} demonstrate the conditions that such contacts are rather common \added{phenomena} and influence the ensemble in a consistent manner.  

The structural flexibility of IDPs encourages us to approximate and describe them with mean-field polymer physics theories. Treating IDPs as polymers suggests that, for large intermonomer spacing, they would display fractal-like behavior with an approximate scaling relation $\langle(\vec r_i - \vec r_j)^2\rangle \propto |i - j|^{2\nu}$. Here, $<..>$ denotes the ensemble average, $\vec r_x$ are the coordinates of monomers $x\in \{i,j\}$, and $\nu$ is the Flory scaling exponent. For rigid rod, self-avoiding, and ideal polymer, $\nu$ is expected to be $1, 3/5,$ and $1/2$, respectively.\cite{rubinstein2003polymer} 

The polymer radius of gyration ($R_g$) or end-to-end distance ($R_{ee}$) are common measurable properties of the polymer's ensemble average structure. However, the Flory exponent is a more convenient parameter to compare \replaced{the}{between} structural compactness of polymers with a different number of residues ($N$) as $R_g \propto R_{ee} \propto N^{\nu}$ \cite{brant1965configuration}. For example, this was used to show that denatured unfolded protein\added{s} behave as excluded volume random coils with $\nu=0.598 \pm 0.028$ \cite{kohn2004random}. On average across different sequences, IDPs behave as ideal chain (or chain in $\theta$-solvent) with $\nu=0.522 \pm 0.01$ \cite{bernado2009self}. The latter indicated that IDPs are more compact than chemically denatured proteins and that individual IDPs can show large deviations from the average value. Hofmann {\it et al}. have shown that the unfolded state of a globular protein can be modeled using polymer-like statistics with $\nu=0.46 \pm 0.05$ and that IDPs are more expanded \cite{hofmann2012polymer}. Recently, using molecular dynamics (MD) simulation, it was shown that IDPs scale as $\nu=0.53 \pm 0.03$ \cite{zerze2019evolution} opposed to another MD simulation that showed scaling behaviour  $\nu \approx0.588$ \cite{baul2019sequence}.  \added{We emphasize that the Flory exponent needs to be taken with caution as it represents the projection of a finite-sized heteropolymer within a theoretical framework developed for long homopolymers. Consequently, the Flory exponent of complex heteropolymers might not describe their scaling, yet, is a sufficient parameter for structural compactness.}

Experimentally, IDP ensemble structures have been investigated using combined biophysical methods such as single molecule F\"orster resonance energy transfer (smFRET), Nuclear magnetic resonance (NMR), and small angle x-ray scattering (SAXS) \cite{naudi2021synergies,gomes2017insights}. For example, Borgia {\it et al}. \cite{borgia2016consistent} combined smFRET and SAXS on the IDP ACTR to solve the discrepancy of IDP expansion in denatured chemical condition. Gomes {\it et al}. combined SAXS, smFRET, and NMR, and showed overall compactness and large end-to-end distance fluctuations of the disordered N-terminal region of the Sic1 protein \cite{gomes2020conformational}. Recently, using smFRET and NMR, it was shown that global structural characteristics of the IDP Tau could be tuned with a small mutation \cite{stelzl2022global}. Here, we combine SAXS and ensemble time-resolved FRET (trFRET) techniques to evaluate the structural heterogeneity within an IDP model system.

SAXS intensity over the wave transferred momentum, $I (q)$, measures the ensemble average Fourier transform \replaced{of}{to} the auto-correlation of the electron density. As such, it is a powerful technique suited for polymers and IDP structural characterization \cite{bernado2012structural, ehm2021intrinsically}. Kratky analysis ($q^2I$ vs. $q$) gives a qualitative characterization of the protein's shape (folded-unfolded, globular-IDP, number of domains, etc.) while the Guinier approximation, via the intensity at low angles, immediately reports on $R_g$. Recently, Zheng {\it et al.}\cite{zheng2018extended} showed that extended Guinier approximation can also report on the scaling exponent $\nu$.

Additionally, techniques such as the ensemble optimization method (EOM) and the molecular form-factor (MFF) consider the protein's sequence or the number of residues to evaluate the ensemble structural distribution a priori from the scattering profile  \cite{bernado2007structural, tria2015advanced, riback2017innovative}. Unfortunately, SAXS has some limitations. Protein solutions need to be measured in relatively high concentrations (compared to FRET), which may result in interactions (aggregation, repulsion) affecting the proteins' measured structure. 

By dipole-dipole interaction, FRET measures the distance between fluorescence molecules. Recently, smFRET has become a well-established method for studying biomolecular conformations with a robust distance quantification \cite{hellenkamp2018precision}. We chose to use the ensemble trFRET method for several reasons. While smFRET has a limitation of detecting distances larger than 30 \AA \cite{spiegel2016failure}, using trFRET with smaller dyes can decrease the distance resolution down to 5 \AA. This enables the study of relatively small segments in the polypeptide chains. Furthermore, while it has been shown that smFRET dyes \added{have modest influence \cite{zerze2014modest} but in some cases} can promote collapse \cite{riback2019commonly,reinartz2020fret},  trFRET benefits from the lesser biased structural characterization by using small dyes such as natural amino acid (e.g., Tryptophan) \cite{orevi2016sequential}. Moreover, while smFRET triumphs with direct observation to the ensemble sub-population it is less sensitive to the distance distributions' width as evaluated by trFRET \cite{schuler2018perspective,haas2005study}.

The carboxy tail domain of mouse Neurofilament Light (NFLt) protein is an interesting IDP model system for its sequence heterogeneity properties. The entire NFL protein is a critical component of the neuronal-specific intermediate filaments network \cite{laser2015neurofilament,kornreich2015order}. Specifically, the carboxy tail domain has been shown to regulate the structure and interaction between the self-assembled neurofilaments \cite{beck2010gel,kornreich2015composite,kornreich2016neurofilaments,malka2017phosphorylation,heins1993rod} and nanoparticles grafted with such domain \cite{pregent2015probing}. 

Mean field calculations \cite{leermakers2010projection} and molecular dynamics simulations \cite{stevens2011interactions} showed that the NFLt plays a critical role in regulating the full filament's structural properties, compressibility and frequency of cross-bridges between the three different neurofilament proteins. Moreover, Monte-Carlo simulations demonstrated a tendency of the NFLt to fold back and form loops or locally coiled structurees upon the increase of the salt concentration \cite{jeong2016monte,lee2013effects}. Recently, NFLt was shown to have glassy dynamics with the response to tension \cite{morgan2020glassy}. Such dynamics were associated with multiple weakly interacting domains and structural heterogeneity. 

Here, by combining SAXS and trFRET, we experimentally study structural heterogeneity of NFLt at equilibrium. We find that each sub-segment has \replaced{intramolecular structural heterogeneity}{ISH} that needs to be considered when translating primary sequence into the ensemble of structures. Furthermore, we show that the structural properties of each segment significantly alter when measured in the context of the \added{full length} protein. Together, these results demonstrate that polymeric scaling arguments should be taken with care of the primary sequence properties and suggest that long-range contacts between the distant amino acids must play a significant role in such IDPs.


\section*{Results}

\subsection*{Experimental Design}
\subsubsection*{NFLt sequence description}

\begin{figure}
\centering
\includegraphics[width=1\linewidth]{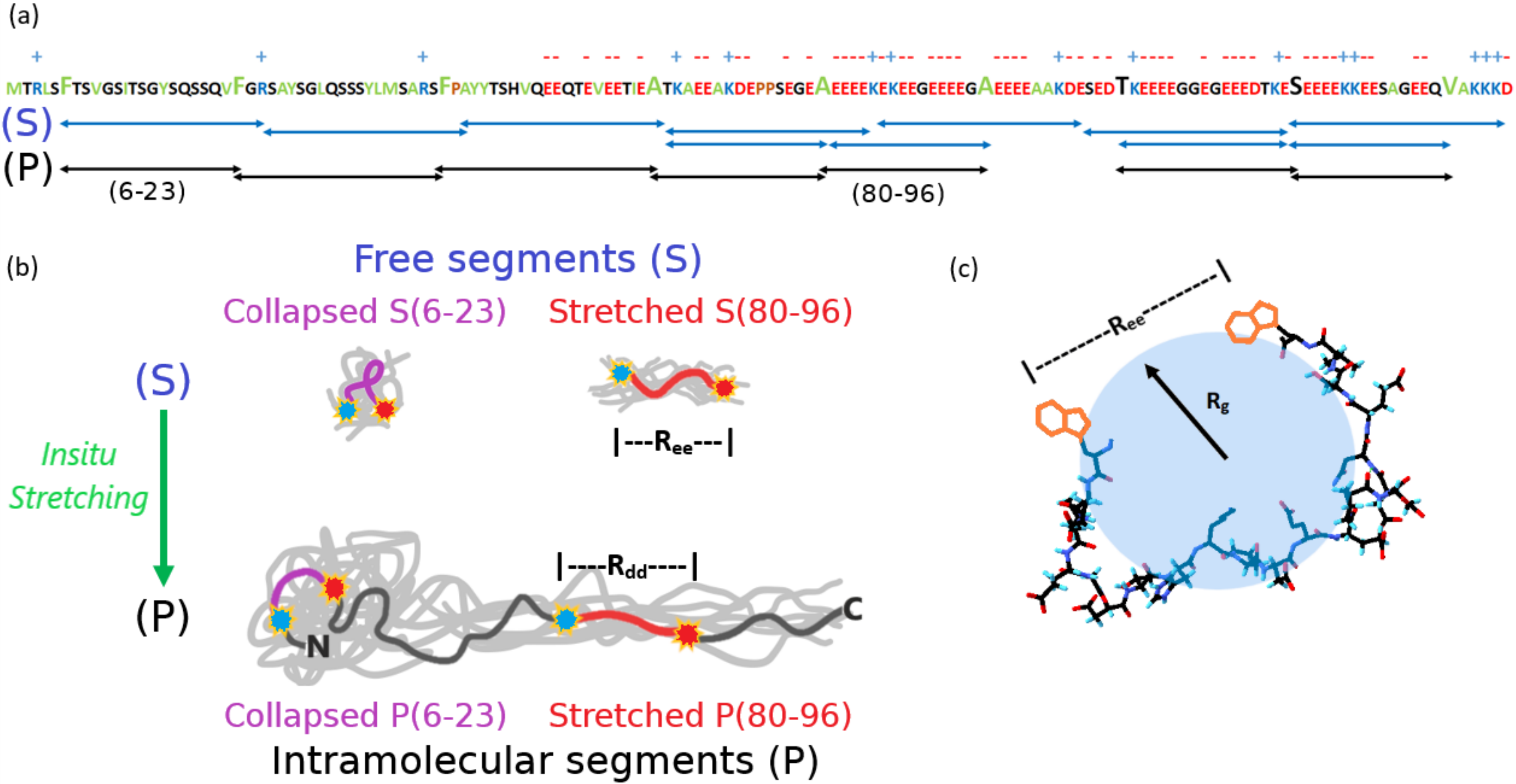}
\caption{
Experimental design. (A) Amino acid composition of the NFLt model protein. Negatively charged, positively charged, or hydrophobic amino acids are colored red, blue and green, respectively. The charge states of the amino acids at pH 8.0 are denoted above. S and P segments are shown below the sequence. S segments were measured as free segments without the protein context, and P segments were measured as a part of the full protein. Large letters denote the residues that were replaced with Trp or labeled Cys for trFRET measurements. (B)  Schematic illustration of the experimental result. For example, S segments (6-23) and (80-96) were found to be collapsed and stretched, respectively. Both were additionally stretched while measured as segments in the protein context, still showing structural heterogeneity. (C) The combined SAXS/trFRET approach. Labeled S segments were measured via trFRET to extract $R_{ee}$, and unlabeled S segments were measured via SAXS to extract the $R_g$. P segments were measured only by trFRET to extract $R_{dd}$.}
\label{fig:1}
\end{figure}

The NFLt sequence can be divided into three diverse regions (Fig. \ref{fig:1}a). An uncharged region (residue 1-50), a mildly negative charged region (residue 50-80), and a highly negative charged region (residue 80-146). An additional domain with a distinguished sequence is the positively charged tip composed of three lysines at the C-terminal. This tip is known to influence hydrogel contraction of the neurofilament hydrogel network \cite{kornreich2016neurofilaments}.

Nonetheless, by calculating NCPR for different segments of different lengths, we find that a segment length of $\sim 20$ residues results in a large NCPR diversity (Fig. \ref{fig:NCPR_VS_sequence}). Furthermore, with a segment length of 20 amino acids, the NCPR value changes from 0 to $\sim 0.3$ at position 50, and by another increase at position 70 to a NCPR of $\sim 0.5$. Similarly, \added{the sequence charge decoration (SCD)\cite{sawle2015theoretical} and} the patterning parameter \cite{das2013conformations}, $\kappa$, shows large variability along the sequence and correlates with NCPR (Fig. \ref{fig:Kappa_VS_NCPR}). 

\subsubsection*{Segments design}
\deleted{Accordingly,}We design a set of 11 different segments derived from the NFLt sequence to evaluate the structural heterogeneity \added{(Table \ref{tab:table1})}\deleted{(Supplementary Table \ref{tab:S_Sequence})}. A set of seven segments, named S(n-m), are chosen consecutively, with 20 residues for SAXS and 23 residues for FRET, covering the entire NFLt sequence. Here, we mark (n-m) as the first and last amino acid positions that are counted from the beginning N-terminus of the tail domain. In addition, we purify from \textit{E. Coli} the entire NFLt with seven mutants, named P(n-m) \added{(Table \ref{tab:table1})}\deleted{(Supplementary Table \ref{tab:P_Sequence})}. Each mutant is designed to incorporate FRET pairs while avoiding replacing charged amino acids. Following, additional four S segments are designed to probe equal length sequences to match the P segments' lengths and to evaluate special cases. The segments' properties, including their NCPR, fraction of charged residues (FCR) and $\kappa$ are listed in the Supplementary Tables \ref{tab:S_NCPR_KAPPA} and \ref{tab:P_NCPR_KAPPA}. 

\subsubsection*{FRET labeling design}
To ensure a similar spectral environment, each S segment for FRET includes additional Ala residues at the N- and two Ala residues at the C- terminals. The acceptor dye (5-(Dimethylamino)naphthalene-1-sulfonyl, Dansyl) is attached to the N-terminal while the donor dye (Naphthyl) is attached to the C-terminal. For P segments, we use Trp as a FRET donor and  7-acetamido-4-coumarincarboxylic acid (Coumarin), coupled to cysteine residue, as a FRET acceptor. The dye pairs of both S and P segments have a F{\"o}rster distance, $R_0$, of 24 \AA \cite{woodard2018intramolecular, rahamim2015resolution}. As a control, one labeled S segment is also measured using SAXS and compared with unlabeled (but with equal sequence) S segments (Fig. \ref{fig:SI-un_and_labeled-S5p}). The labels increase the $R_g$ by about $\sim$ 1 \AA.

All measurements are conducted in the presence of a 20 mM Tris, pH 8.0, to fully deprotonate the Histidine residue. All S segments and the full NFLt are validated to be fully disordered by circular dichroism (Fig. \ref{fig:CD}). Bioinformatic analysis confirms the high disorder probability for residues at positions 60-146 (Fig. \ref{fig:PrDOS}).

\begin{figure}
\centering
\includegraphics[width=0.6\linewidth]{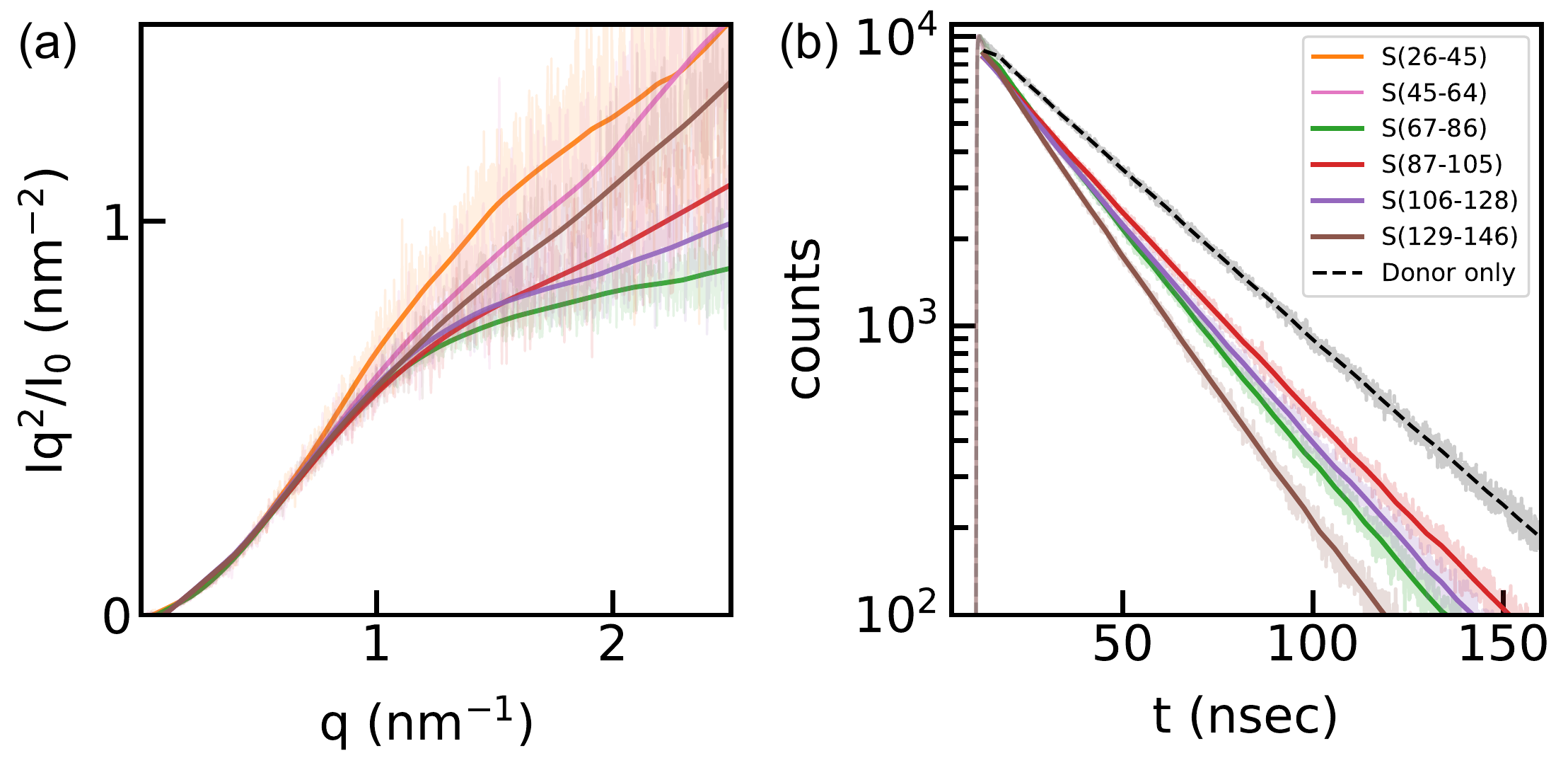}
\caption{Intrinsic structural heterogeneity demonstrated in the raw data. All bold solid lines are moving averages of the raw data behind them. (a) Kratky plot of the SAXS measurements. Each line color is a different S segment (which is 20 amino acids long). Kratky plot indicates that all segments are unfolded as $Iq^2/I_0$ intensity diverge at large $q$. Moreover, segments with smaller $R_g$ show overall larger Kratky intensities. \added{Additional Kratky plot at Supplementary Fig. \ref{fig:SI-EOM-Kratky-Tris},\ref{fig:SI-EOM-Kratky-50mM},\ref{fig:SI-EOM-Kratky-150mM} and \ref{fig:SI-EOM-Kratky-500mM}.} (b) Time-resolved fluorescent decays of segments with only a donor (DO) (black dashed line) and donor in the presence of an acceptor (DA) (continuous lines). While all DO decays for the presented segments are equal, the DA decays show significant heterogeneity.}
\label{fig:2}
\end{figure}

\subsection*{Divide and conquer - structural heterogeneity of NFLt segments}

\subsubsection*{Raw data heterogeneity}
Fluorescence decay of labeled peptides are measured by trFRET and the unlabeled segments by SAXS (Fig. \ref{fig:1}c). Structural heterogeneity can already be demonstrated in the raw data, of equal length segments, shown by the Kratky plot (Fig. \ref{fig:2}a) and trFRET fluorescence decay (Fig. \ref{fig:2}b). While fluorescence decay of segments with donor only (DO) is identical for all charged S segments, the donor in the presence of an acceptor (DA) fluorescence decays (for S segments with equal length) shows a significant difference.

\subsubsection*{FRET Scaling parameter with SAW}
To evaluate and distinguish structural heterogeneity between the segments, we follow a polymer physics-like analysis to extract the Flory scaling parameter, $\nu$, from our trFRET data set. Using the Zheng {\it et al}. procedure \cite{zheng2018inferring} the self-avoiding walk (SAW) model is used with dye-to-dye distance distribution:

\begin{equation}
    P(r)=A\frac{4 \pi}{R}\left(\frac{r}{R}\right)^{2+g} \exp \left(-\alpha\left(\frac{r}{R}\right)^{\delta}\right).
    \label{eqSAW}
\end{equation}

Here, the mean dye-to-dye distance is $R=bN^{\nu}$, where $N$ is the number of peptide bonds between the dyes, $b=0.55$ nm, $g=0.1615/\nu$, $\delta=1/(1-\nu)$ and the constants $A$ and $\alpha$ are determined for a given value of $\nu$ and $R$, from the constrains $\int_{0}^{\infty} P(r) d r=1$ and  $\int_{0}^{\infty} P(r) r^{2} d r=R^{2}$. Zheng's model was strictly derived for a real chain in good solvent, but the previous analysis indicated that it provides a useful approximation even outside the good solvent regime\cite{zheng2018inferring}. Importantly, Zheng's model includes a single free parameter, $\nu$. Effectively $\nu$ quantifies the chain's compactness level where $\nu= 1/3,\; 1/2,\; 3/5$ are for real polymer in poor, theta, and good solvent, respectively.  The average lifetime is extracted from the fluorescence decays and used to calculate the mean energy transfer $\langle E\rangle=1-\frac{\tau_{DA}}{\tau_{DO}}$ where $\tau_{DA}$ and $\tau_{DO}$ are the donor lifetime with and without the presence of an acceptor, respectively. The energy transfer is thus related to Eq. \ref{eqSAW} by:
\begin{equation}
 \langle E\rangle = \int_{0}^{l_c} \frac{R_0^6}{R_0^6+r^6}P(r) dr.   
 \label{eqEt}
\end{equation}

Here, $R_0$ is the F{\"o}rster radius \added{and $l_c$ is the contour length}. 

\subsubsection*{SAXS Scaling parameter with SAW}
Based on Zheng's SAW model, we can also extract the radius of gyration ($R_g$) and $\nu$ from the SAXS data using an extended Guinier analysis \cite{zheng2018extended}:
\begin{equation}
    \begin{split}
        I(q)=I_{0} \exp \left\{-\frac{1}{3} R_{g}^{2} q^{2}+0.0479(\nu-0.212) R_{g}^{4} q^{4}\right\},\\
        R_{g}=\sqrt{\frac{\gamma(\gamma+1)}{2(\gamma+2 \nu)(\gamma+2 \nu+1)}} b N^{\nu},\\
        \gamma = 1.1615 \text{ .}
    \end{split}
    \label{Eq-extendedGunier}
\end{equation}
The extended Guinier analysis fits well for the segments until $qR_g = 2$, as expected \cite{zheng2018extended}. As a control we compare $R_g$ to the traditional Guinier approximation ($ I(q) \approx I_0\exp(-\frac{1}{3}R_g^2q^2)$) \cite{ehm2021intrinsically, bernado2012analysis}, with an excellent agreement (Figs. \ref{fig:Si-Ex-fit},\ref{fig:Si-Guinier-deviation} \ref{fig:Si-I0} and \ref{fig:Si-Rg}). 
\added {While the exact $R_g$ value may vary for different methods of analysis, the overall trend is identical for all methods.}

\begin{figure}
\centering
\includegraphics[width=0.9\linewidth]{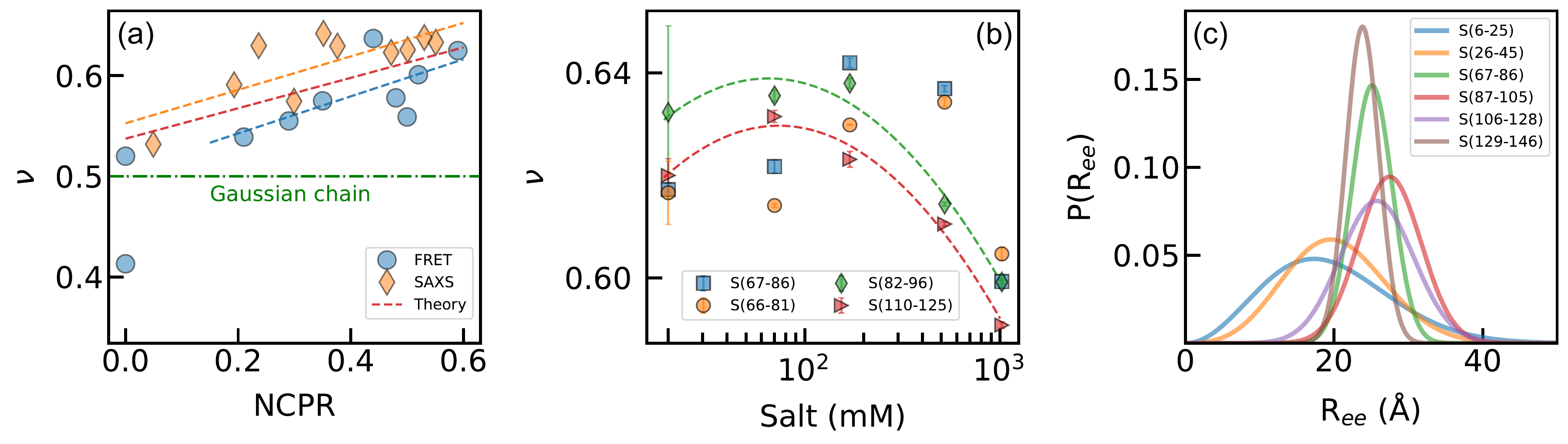}
\caption{Sub segments structural heterogeneity. (a) trFRET and SAXS were measured at 20 mM Tris buffer pH 8.0 and 150mM NaCl. Blue circles and orange diamonds are scaling exponents from trFRET and SAXS measurements, respectively. Dashed orange and blue lines are linear fits to the data. The dashed red line is a linear fit to the theoretical calculation of $\nu$ based on Zheng et al. \cite{zheng2020hydropathy} (theoretical $\nu$ values are listed in table \ref{tab:S_NCPR_KAPPA}). (b) Scaling exponent from SAXS measurements at different salt (NaCl) concentrations. Charged segments such as S(82-98) (green diamonds) expand until 70 mM salt and then collapse at higher salt concentrations. While less charged segments, such as S(67-86) (blue squares), expand until 170 mM salt. Uncertainty calculation was performed according to Eq. \ref{eq:delta_nu} (c) End-to-end distance distribution from trFRET of S segments with equal chain length. Uncharged segments, S(6-25) and S(26-45) (blue and orange lines), are more contracted compared to more charged segments, S(87-105) and S(106-128) (red and purple lines), showing more expanded structure. The distributions are calculated using the values of table \ref{tab:S_FRET_Ree} at 150mM NaCl. }
\label{fig:3}
\end{figure}

\subsubsection*{Scaling parameter correlation with NCPR}
To evaluate structural heterogeneity for segments of different lengths, it is convenient to discuss the segments' compactness Flory parameter $\nu$ as an intensive parameter. Following, the most striking structural heterogeneity is between the uncharged segments of NCPR$\approx0$, showing a more collapsed profile with $0.41<\nu<0.52$, and the charged segments that show more stretched conformations with $0.53<\nu<0.65$ (Fig. \ref{fig:3}a \added{and Table \ref{tab:table1}}). Interestingly, both SAXS and trFRET measurements show a linear correlation between NCPR and $\nu$ (Fig. \ref{fig:3}a), which is more pronounced at physiologic salinity (Fig. \ref{fig:S_segment_Nu}, Tables \ref{tab:S_FRET_Nu} and \ref{tab:SAXS1}). The NCPR is expected to influence the ensemble polymeric structure \cite{das2013conformations} since charged residues electrostatically repel each other and avoid collapse. In fact, in contrast to simple polyelectrolytes~\cite{PEscaling} it is known that the size-scaling of ideally random polyampholytes generally changes with charge density and salt concentration~\cite{Joanny}. Nonetheless, recent theoretical work implies strong sequence-specific effects on the scaling behavior~\cite{PAscaling}.  

Interestingly, the data fits well the Zheng {\it et al.} \cite{zheng2020hydropathy} estimation to $\nu$ which involves both charge and hydrophobicity that in our cases balance each other and produce an effective linear trend between $\nu$ and NCPR (Tables \ref{tab:S_Sequence}, \ref{tab:S_NCPR_KAPPA} and red dashed line in Fig.~\ref{fig:3}a). The linear correlation of $\nu$ with NCPR is reduced upon salt increment as the overall value of $\nu$ decreases for the charged peptides (Fig. \ref{fig:S_segment_Nu}, Table \ref{tab:slops}). Also, SAXS measurements of segments with NCPR$>0.4$ at low salinity (20mM Tris buffer), include strong electrostatic correlations that are reflected via scattering pattern variation at different peptide concentrations (Fig. \ref{fig:SI-Repulsion_concentration}). Given the inter-molecular interaction present at high peptide concentrations, we obtain a relatively high uncertainty for $\nu$  for low salinity (Fig. \ref{fig:3}b).
Adding 3M GdnHCl to the solution has canceled-out the $\nu$ correlation with NCPR shown with trFRET (Fig. \ref{fig:S_segment_Nu}d). Here, the denaturated IDP becomes less sequence specific with $\nu \approx 0.6$.

\begin{table}
\centering
    \resizebox{0.58\textwidth}{!}{%
\begin{tabular}{|l|c|c|cc|cc|}
\hline
\multicolumn{1}{|c|}{\multirow{2}{*}{Name}} & \multirow{2}{*}{Sequence} & \multirow{2}{*}{NCPR} & \multicolumn{2}{c|}{trFRET}                                     & \multicolumn{2}{c|}{SAXS}                                         \\ \cline{4-7} 
\multicolumn{1}{|c|}{}                      &                           &                       & \multicolumn{1}{c|}{$\langle R_{ee}\rangle$} & $\nu$ & \multicolumn{1}{c|}{$\langle R_{g, ex.}\rangle$} & $\nu$ \\ \hline
S(6-25)                                     &  
 \textcolor{white}{----}
$\color{red}\pmb{\otimes}$\textcolor{red}{A}FTSVGSITSGYSQSSQVFGR\textcolor{blue}{AA}$\color{blue}\pmb{\otimes}$  & 0.05                  & \multicolumn{1}{c|}{18.4{\scriptsize$\pm$1.5}}                               & 0.41  & \multicolumn{1}{c|}{X}                                    & X     \\ \hline

P(6-23)                                     & \pmb{...}$\color{red}\pmb{\odot}$\textcolor{red}{(C/}F)TSVGSITSGYSQSSQV(F\textcolor{blue}{/W)}\pmb{...}    \textcolor{white}{---}  & 0.00                     & \multicolumn{1}{c|}{34.0{\scriptsize$\pm$1.0}}                               & 0.63  & \multicolumn{1}{c|}{X}                                    & X     \\ \hline

S(26-45)                                    & 
\textcolor{white}{-----}
$\color{red}\pmb{\otimes}$\textcolor{red}{A}SAYSGLQSSSYLMSARSFPA\textcolor{blue}{AA}$\color{blue}\pmb{\otimes}$  & 0.05                  & \multicolumn{1}{c|}{21.9{\scriptsize$\pm$0.9}}                               & 0.52  & \multicolumn{1}{c|}{11.0{\scriptsize$\pm$0.08}}                                 & 0.54  \\ \hline

P(23-43)                                    & \pmb{...}\textcolor{blue}{(W/}F)GRSAYSGLQSSSYLMSARS(F\textcolor{red}{/C)$\color{red}\pmb{\odot}$}\pmb{...} 
\textcolor{white}{----}
& 0.10                   & \multicolumn{1}{c|}{41.9{\scriptsize$\pm$0.8}}                               & 0.64  & \multicolumn{1}{c|}{X}                                    & X     \\ \hline

S(45-64)                                    &
\textcolor{white}{-----}
$\color{red}\pmb{\otimes}$\textcolor{red}{A}YYTSHVQEEQTEVEETIEAT\textcolor{blue}{AA}$\color{blue}\pmb{\otimes}$  & 0.30                   & \multicolumn{1}{c|}{X}                                  & X     & \multicolumn{1}{c|}{12.1{\scriptsize$\pm$0.19}}                                 & 0.58  \\ \hline

P(43-64)                                    & \pmb{...}$\color{red}\pmb{\odot}$\textcolor{red}{(C/}F)PAYYTSHVQEEQTEVEETIE(A\textcolor{blue}{/W)}\pmb{...} 
\textcolor{white}{----}
& 0.27                  & \multicolumn{1}{c|}{41.6{\scriptsize$\pm$2.5}}                               & 0.68  & \multicolumn{1}{c|}{X}                                    & X     \\ \hline

S(67-86)                                    &
\textcolor{white}{--------}
$\color{red}\pmb{\otimes}$\textcolor{red}{A}KAEEAKDEPPSEGEAEEEEK\textcolor{blue}{AA}$\color{blue}\pmb{\otimes}$  & 0.35                  & \multicolumn{1}{c|}{25.3{\scriptsize$\pm$0.2}}                               & 0.58  & \multicolumn{1}{c|}{14.1{\scriptsize$\pm$0.03}}                                 & 0.64  \\ \hline

S(67-86)c                                    &
\textcolor{white}{--------}
$\color{red}\pmb{\otimes}$\textcolor{red}{A}KAEEAKDEGGSEGEAEEEEK\textcolor{blue}{AA}$\color{blue}\pmb{\otimes}$  & 0.35                  & \multicolumn{1}{c|}{24.0{\scriptsize$\pm$0.2}}                               & 0.54  & \multicolumn{1}{c|}{X}                                 & X  \\ \hline

S(66-81)                                    &
$\color{red}\pmb{\otimes}$ATKAEEAKDEPPSEGEA$\color{blue}\pmb{\otimes}$   
\textcolor{white}{------}
& 0.24                  & \multicolumn{1}{c|}{22.1{\scriptsize$\pm$0.2}}                                 & 0.56  & \multicolumn{1}{c|}{12.3{\scriptsize$\pm$0.00}}                                 & 0.63  \\ \hline

P(64-80)                                    & \pmb{...}\textcolor{blue}{(W/}A)WTKAEEAKDEPPSEGE(A\textcolor{red}{/C)$\color{red}\pmb{\odot}$}\pmb{...}   
\textcolor{white}{-------}
& 0.24                  & \multicolumn{1}{c|}{36.6{\scriptsize$\pm$0.9}}                                 & 0.71  & \multicolumn{1}{c|}{X}                                    & X     \\ \hline

S(87-105)                                   & $\color{red}\pmb{\otimes}$\textcolor{red}{A}EKEEGEEEEGAEEEEAAKDE\textcolor{blue}{AA}$\color{blue}\pmb{\otimes}$  & 0.55                  & \multicolumn{1}{c|}{27.6{\scriptsize$\pm$0.2}}                               & 0.60   & \multicolumn{1}{c|}{13.8{\scriptsize$\pm$0.11}}                                 & 0.63  \\ \hline

S(82-96)                                    & $\color{red}\pmb{\otimes}$AEEEEKEKEEGEEEEGA$\color{blue}\pmb{\otimes}$     & 0.53                  & \multicolumn{1}{c|}{25.1{\scriptsize$\pm$0.3}}                               & 0.63  & \multicolumn{1}{c|}{12.5{\scriptsize$\pm$0.03}}                                 & 0.64  \\ \hline

P(80-96)                                    & 
\textcolor{white}{-}
\pmb{...}\textcolor{blue}{(W/}A)EEEEKEKEEGEEEEG(A\textcolor{red}{/C)$\color{red}\pmb{\odot}$}\pmb{...}     & 0.53                  & \multicolumn{1}{c|}{39.8{\scriptsize$\pm$1.2}}                               & 0.73   & \multicolumn{1}{c|}{X}                                    & X     \\ \hline

S(106-128)                                  & $\color{red}\pmb{\otimes}$\textcolor{red}{A}SEDTKEEEEGGEGEEEDTKE\textcolor{blue}{AA}$\color{blue}\pmb{\otimes}$ 
\textcolor{white}{---}
& 0.50                   & \multicolumn{1}{c|}{25.8{\scriptsize$\pm$0.2}}                               & 0.58   & \multicolumn{1}{c|}{13.6{\scriptsize$\pm$0.08}}                                 & 0.63  \\ \hline

S(110-125)                                  & 
\textcolor{white}{-}
$\color{red}\pmb{\otimes}$\textcolor{red}{(A}/T)KEEEEGGEGEEEDTKE(S/\textcolor{blue}{A)}$\color{blue}\pmb{\otimes}$    & 0.47                  & \multicolumn{1}{c|}{22.4{\scriptsize$\pm$0.2}}                               & 0.56  & \multicolumn{1}{c|}{12.6{\scriptsize$\pm$0.04}}                                 & 0.62  \\ \hline

P(109-126)                                  & \pmb{...}$\color{red}\pmb{\odot}$\textcolor{red}{(C/}T)KEEEEGGEGEEEDTKE(S\textcolor{blue}{/W)}\pmb{...}    & 0.44                  & \multicolumn{1}{c|}{36.3{\scriptsize$\pm$1.0}}                               & 0.69  & \multicolumn{1}{c|}{X}                                    & X     \\ \hline

S(129-146)                                  &
\textcolor{white}{-----}
$\color{red}\pmb{\otimes}$\textcolor{red}{A}SEEEEKKEESAGEEQVAKKKD\textcolor{blue}{AA}$\color{blue}\pmb{\otimes}$ & 0.19                  & \multicolumn{1}{c|}{24.2{\scriptsize$\pm$0.2}}                               & 0.54  & \multicolumn{1}{c|}{12.9{\scriptsize$\pm$0.13}}                                 & 0.59   \\ \hline

S(129-146)c                                  &
\textcolor{white}{-----}
$\color{red}\pmb{\otimes}$\textcolor{red}{A}SEEEEKKEESAGEEQVAGGGD\textcolor{blue}{AA}$\color{blue}\pmb{\otimes}$ & 0.33                  & \multicolumn{1}{c|}{24.6{\scriptsize$\pm$0.2}}                               & 0.56  & \multicolumn{1}{c|}{X}                                 & X   \\ \hline

P(126-141)                                  & \pmb{...}\textcolor{blue}{(W/}S)EEEEKKEESAGEEQ(V\textcolor{red}{/C)$\color{red}\pmb{\odot}$}\pmb{...} 
\textcolor{white}{----}
& 0.44                  & \multicolumn{1}{c|}{36.3{\scriptsize$\pm$1.4}}                               & 0.69  & \multicolumn{1}{c|}{X}                                    & X     \\ \hline

S(130-143)                                  & $\color{red}\pmb{\otimes}$\textcolor{red}{(A}/S)EEEEKKEESAGEEQ(V/\textcolor{blue}{A)}$\color{blue}\pmb{\otimes}$ 
\textcolor{white}{------}
& 0.38                  & \multicolumn{1}{c|}{24.8{\scriptsize$\pm$0.5}}                               & 0.64  & \multicolumn{1}{c|}{11.8{\scriptsize$\pm$0.03}}                                 & 0.63  \\ \hline

\end{tabular}}
\caption{Summary of key structural heterogeneity statistics of the S and P segments at 150mM NaCl. S segments were measured as free segments without the protein context, and P segments were measured as a part of the full NFLt protein. For trFRET measurements, S segments include modification marks in blue and red where $\color{blue}\pmb{\otimes}$ and $\color{red}\pmb{\otimes}$ denote the Naphtyl and Dansyl FRET dye, respectively. SAXS measurements are done without any modification or labeling. For the trFRET measurements, P segments include modifications where parentheses mark amino acid replacements, and $\color{red}\pmb{\odot}$ denotes the coumarin dye. TrFRET $\langle R_{ee}\rangle$ and $\nu$ are extracted via radial Gaussian and SAW models, respectively. SAXS $\langle R_{g, ex.}\rangle$ and $\nu$  are extracted via extended Guinier analysis. Errors for $\nu$ are smaller than $3\%$. NCPR is calculated for sequence without modification. Further details are given in the supplementary Tables 
\ref{tab:S_Sequence},\ref{tab:P_Sequence},\ref{tab:S_FRET_Nu},\ref{tab:SAXS1},\ref{tab:S_FRET_Ree},\ref{tab:P_FRET_Ree} and \ref{tab:P_FRET_Nu}.}
\label{tab:table1}
\end{table}

\subsubsection*{Salinity effect}
With an aim to determine the source of heterogeneity, we perform trFRET and SAXS experiments in various solution conditions. Using SAXS, segments S(67-86), S(66-81), S(82-96) and S(110-125) show an increase \added {of 0.02} in the scaling parameter until 150 mM NaCl followed by a decrease \added {of 0.04} up to 1M NaCl (Fig. \ref{fig:3}b). At low salinity, we relate the increase in the scaling parameter to inter-molecular charge screening which enables intra-molecular stretching (Fig. \ref{fig:SI-InerRepulsion_Salt}), in line with theoretical work~\cite{Joanny}. The following decrease in higher salinity can be explained by intra-molecular charge screening of the sequences with net-charge~\cite{Joanny} as well as salting-out effects \cite{vancraenenbroeck2019polymer, wiggers2021diffusion}. 

In contrast, Segments S(26-45), S(45-64), S(87-105), S(129-146) and S(130-143) show insensitivity to added monovalent salt (Fig. \ref{fig:SI-independentSalt}). On the contrary, the scaling parameter shows a smaller change with salinity when analysing the trFRET with Eq. \ref{eqEt} (supplementary table \ref{tab:S_FRET_Nu}). We attribute it to distance distribution averaging while using the mean fluorescence life-time. As we will demonstrate below, an alternative analysis to the entire fluorescence decay will show that salinity plays an unconventional tuning parameter to the $R_{ee}$ distance distribution. 

\subsubsection*{FRET distance distribution}
As mentioned, the SAW-FRET analysis in Eq. \ref{eqEt}, averages out the details that might be present in the fluorescence decay curves. Following, we can fit the entire fluorescence decay curves using an empirical intramolecular end-to-end radial Gaussian distribution model \cite{edwards1965statistical} : 
\begin{equation}
P(r)=K r^{2} \operatorname{exp}\left(-b(r-a)^2\right).
\label{eqGaussFRET}
\end{equation}
Here, $a$ and $b$ are free parameters that determine the mean $R_{ee}$ and full width half max (FWHM) of the distance distribution and $K$ is a normalization factor ensuring that $\int_{0}^{\infty} P(r) dr=1$. Following, the time-resolved fluorescence data of segments containing donor only (DO) and segment containing donor and acceptor (DA) can be directly fitted to extract $a$ and $b$ (see methods). \added{The radial Gaussian model highlights a segment's deviation from polymer theory when the distribution width is narrow compared to the SAW model (Fig. \ref{fig:SI-Pr-SAW_VS_Gauss}).}

The radial Gaussian distribution model clearly reflects the alternative segments have distinguished structures (Fig. \ref{fig:3}c). For example, segments S(67-86), S(66-81),\added{ S(106-128), S(110-125) and S(129-146)} show a deviation from the SAW polymer theory with a narrow distance distribution width \added{ranging between 5.8 to 8.2 {\AA}}. In contrast, segments S(87-105), S(82-96) and S(130-143) show a wider \deleted{width of}\added{distribution between 13.9 - 15.2 {\AA}}.\deleted{ and rather a large mean distance distribution} (supplementary table \ref{tab:S_FRET_Ree}). Apparently, \deleted{these} segments \added{with narrow width} fluctuate rather mildly about a specific mean $R_{ee}$ distance. \deleted{We postulate that segments S(67-86) and S(66-81) are fluctuating around a "V" shape configuration due to di-Proline at positions 74-75 (Fig.~\ref{fig:Structural_Constraints}a). Di-Proline has been known to constrain the possible conformational space.} Moreover, segments S(67-86), S(66-81) \added{and S(110-125)} show no response to increasing salt concentration (Fig. \ref{fig:GaussMeanSalt}), indicating that this constraint is not influenced by electrostatic interaction.

\subsubsection*{Ionic bridge constrain}
\begin{figure}
\centering
\includegraphics[width=0.4\linewidth]{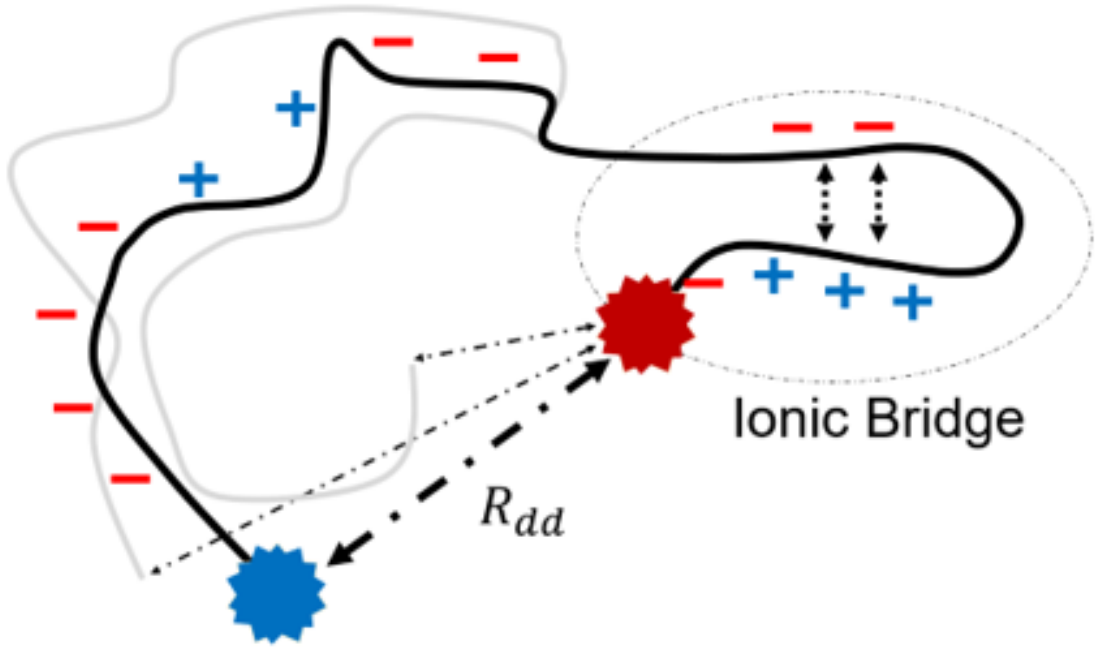}
\caption{Illustration of the structural constraints of segment\deleted{s S(67-86) (panel a) and S(129-146) (panel b). The trFRET fluorescent labels are colored with blue and red for donor and acceptor dyes. The peptides' net-charge arrangement is denoted with ``$+$" and ``$-$". Grey lines represent additional conformations of the peptides. (a) Segment S(67-86) is fluctuating around a "V" shape configuration due to di-Proline at positions 74-75.} S(129-146). The positive charged C-terminal tip forms a transient loop, resulting in constrained conformations with narrowed distance distributions.}
\label{fig:Structural_Constraints}
\end{figure}

\deleted{Similarly,} The carboxy tail domain of segment S(129-146), also shows constrained conformations with narrowed distance distribution width \added{of 5.8 {\AA}} (Fig. \ref{fig:3}c). Here, we propose the formation of a transient loop resulting from the positively charged C-terminal tip (Fig.~\ref{fig:Structural_Constraints}), previously identified to electrostatically interact with the remaining negatively charged residue along NFLt \cite{kornreich2016neurofilaments}. Supporting this claim is the $P(r)$ of segment S(130-143), which shares a similar sequence to S(129-146) apart from the positive tip. Indeed, we find that \added{while S(130-143) is 4 amino acid shorter, it still shows an identical mean value ($\sim$ 24 {\AA}) to S(126-146) (Fig. \ref{fig:7_7p}). Additionally, as  a control, we synthesized segment S(126-146)c, where three Lys at the carboxy tail are replaced with three Gly. Segment S(126-146)c shows a higher mean value ($\sim$ 26 {\AA}), as expected, because of less possible ionic paring; however, it still has a rather narrow distribution (Fig. \ref{fig:7_7p}).}

\deleted{S(130-143) shows a wider $P(r)$  with identical mean value to S(126-146) (Fig. \ref{fig:7_7p}). Additionally, using SAXS, S(126-146) and S(130-143) (with and without the charged tip, respectively) do not show any salt dependency.} 

Moreover, while segments S(6-25), S(26-45), S(87-105), S(82-96),S(106-128) and S(130-143), show \added{$\sim$3 {\AA}} decrease in $R_{ee}$ as a response to increasing salt concentration (Fig. \ref{fig:GaussMeanSalt}), segment S(129-146) is the only one that shows \added{$\sim$ 1 {\AA}} increase in $R_{ee}$. \added{In contrast, after mutating into S(126-146)c, the {$R_{ee}$} decreased by {$\sim$}2{\AA} after the addition of salt because of screening of the now dominating electrostatic repulsion.}

These results further emphasize the breaking of ionic bridges that form the loop. \added{Using SAXS, S(126-146) and S(130-143) (with and without the charged tip, respectively) do not show any salt dependency.}\deleted{In addition,} It further demonstrates apparent inconsistencies between SAXS and FRET that we discuss and explain later. 

\subsubsection*{Evaluating the persistence length by WLC model}

\deleted{Further support for our hypothesis to the structural anomalies of S(67-86) and S(129-146)} 
\added{Further support for structural heterogeneity} 
\replaced{is given}{are validated} by evaluating the persistence length ($l_p$) from the trFRET data. Using a Worm Like Chain (WLC) model we fit the fluorescence decay and find that while most segments have $l_p = 1.4\pm 0.2$ nm (Supplementary table \ref{tab:S_FRET_WLC}), segments S(67-86) ,S(129-146) \added{and S(106-128)} result with much higher $l_p$ values larger than $2.0$ nm. Importantly, \added{since it is possible that apparent short end-to-end distances (compared to the contour length) will have high persistence length, we set the contour length as a free parameter in the analysis. This results in an apparent contour length which is close to the value of the persistence length. This effect is highlighted in the segments mentioned above and is shown in supplementary table \ref{tab:S_FRET_WLC}.}\deleted {the fit converges only when the contour length $l_c$ for those specific segments is left as a free parameter showing a value smaller than the actual segments' length (Supplementary table \ref{tab:S_FRET_WLC}).} \deleted{The discrepancies of $l_p$ and $l_c$ in comparison to the other segments are well inline with the `V'-shaped and loop formation presented above.} These results show again the existence of structural constraints, regardless of the model used to analyze the data. 



\subsubsection*{SAXS distance distribution by EOM}
For additional corroboration to the $R_g$ values obtained from extended Guinier analysis (Eq. \ref{Eq-extendedGunier}), we implement the ensemble optimization method (EOM) as an independent approach \cite{bernado2007structural,tria2015advanced}. Here, the ensemble of peptide structures whose average scattering curve fits the experimental data is selected from a pool of possible sequence-dependent conformations. The EOM consistently overestimates the $R_g$ values by $\sim$9\% in comparison to the extended Guinier analysis (Eq.~\ref{Eq-extendedGunier}). However, the two approaches have a satisfying correlation (Fig.~\ref{fig:SI-Rg-estimation}). 

The EOM sheds light on the different conformations that each segment adopts upon salt addition (Figs.~\ref{fig:SI-EOM-salt}, \ref{fig:SI-EOM-R_flex}) where the most influenced segments are S(45-64), \replaced{S(67-86),}{S(67-81)} and S(129-146). In those segments the distribution's width expands upon salt addition, hence, adopting more expended structures (Figs. \ref{fig:SI-EOM-salt},  \ref{fig:SI-EOM-R_flex}). Notably, this is in contrast to the experimental trFRET results showing for segments \deleted{segments S(45-64)} S(67-81) and S(67-86) insensitivity to salt variation, excluding S(129-146) that expands with the addition of salt.

Furthermore, using the EOM, we can estimate the $R_{ee}$ probability distributions and compare \replaced{them}{it} to the trFRET empirical model distribution (Eq. \ref{eqGaussFRET}). Although SAXS is less sensitive to the peptides' ends fluctuations, we observe similarities between some of the distributions analyzed by both experimental methods  (Fig. \ref{fig:SI-EOM-vs-FRET}). Importantly, EOM analysis shows wider distributions than the Gaussian distribution from trFRET. We note that the largest deviation between the two techniques belongs to segments showing intra-molecular constraints in the trFRET.

\subsection*{Structural heterogeneity and expansion in the NFLt protein's context}

The structural heterogeneity between peptide sequences is expected since different amino-acid interactions constrain the ensemble differently. In fact, this relation is key in determining the structure of folded proteins. The lack of structure in IDPs, on the other hand, is expected to smear out some of the molecular heterogeneity, at least for large enough sequences as in the context of a protein.

\begin{figure}
\centering
\includegraphics[width=0.5\linewidth]{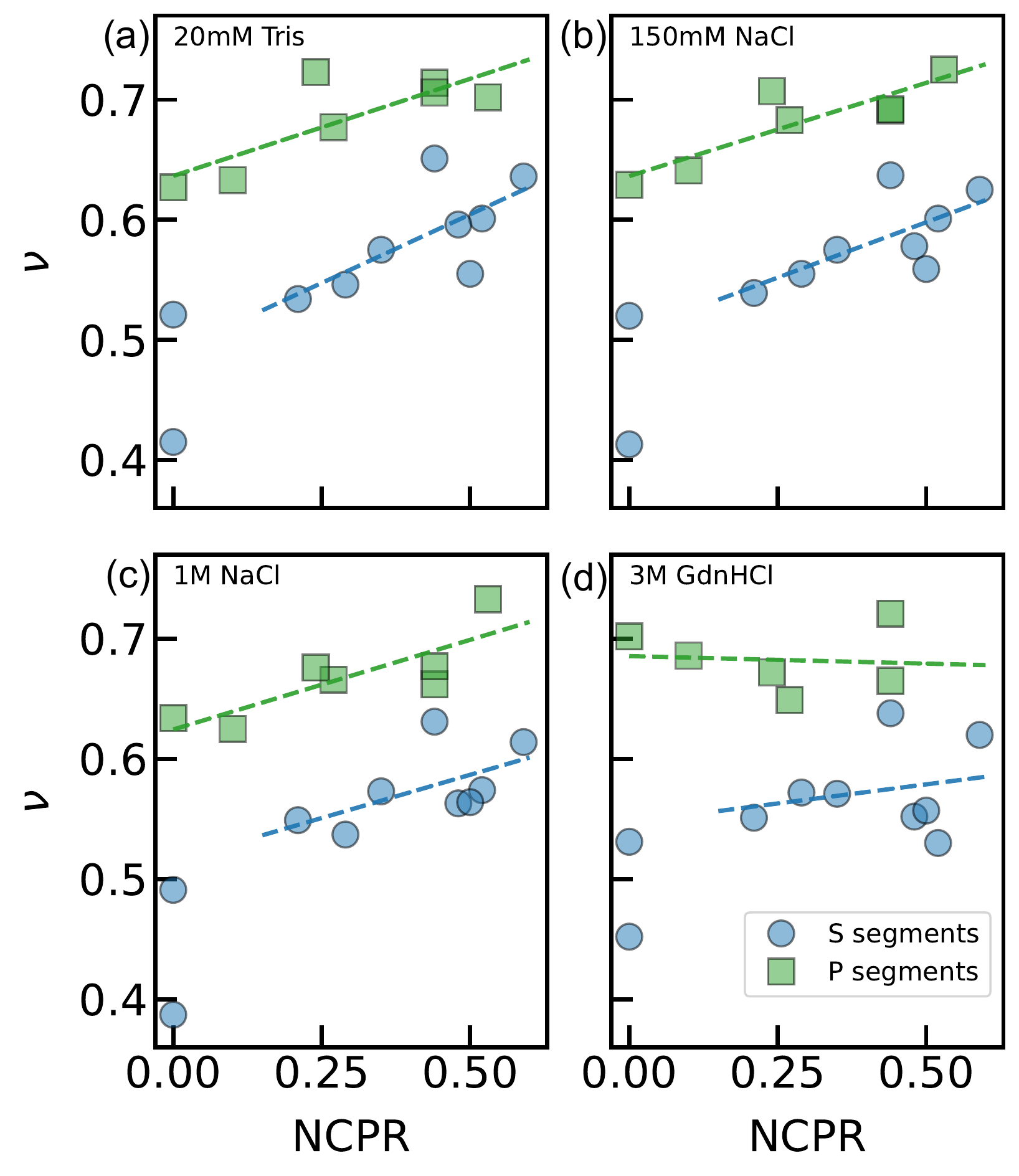}
\caption{Scaling exponent for all trFRET measurements at different conditions. S segments are measured free in solution, while P segments are measured in the context of the full NFLt model IDP. \deleted{Black arrow points on segment P(64-80) that deviate from the linear trend presumably due to inherent di-proline constrain.} All segments at all conditions show that P segments expand with larger scaling exponents than S segments. Dashed lines are linear fits to the data. The lines' slops decrease upon salt addition as detailed in Table \ref{tab:slops}. All $\nu$ relative uncertainties are below 1\%.}
\label{fig:protein_nu}
\end{figure}

We further raise the question of how the structural properties of each isolated segment change in the context of the entire NFLt. As mentioned above, P segments are fully expressed NFLt with labels for trFRET for each segment (More details in supplementary table \ref{tab:P_Sequence}). The SAW model fits well to the fluorescence decay, and here, we denote the mean dye-to-dye distance as $R_{dd}$. However, for all P segments, the $\nu$ values (and consequently the $R_{dd}$ values) are larger than those measured for the corresponding S segments (Fig. \ref{fig:protein_nu} and tables \ref{tab:P_FRET_Ree}, \ref{tab:P_FRET_Nu}). Then again, similarly to the S segments, we find a linear relation between $\nu$ and the NCPR, including the non-charged segments (Fig. \ref{fig:protein_nu}).

\subsubsection*{Salinity effect in the protein's context}
For all P segments, at all studied salt concentrations (0-1M), we find an expanded polymeric state with $\nu>0.6$. In fact, the highly charged P(80-96) shows an impressive expansion with $\nu=0.73$ at 150mM NaCl. Again, as with the S segments, the linear correlation between $\nu$ and NCPR is moderated with increasing salt and absent in denaturating condition (3M GdnHCl) (Fig. \ref{fig:protein_nu}). 

The relation between the expansion and salt implies towards electrostatic interaction. Interestingly, a radial Gaussian distance distribution analysis for segments P(6-23) and P(23-43) shows an \added{$\sim$2 {\AA} and $\sim$ 4 {\AA}} increasing mean $R_{dd}$ with salt concentration (Fig. \ref{fig:P_GaussMeanSalt}). This increase in the mean value is completely reversed to that found for the isolated segment S(6-25) and S(26-46) \added{showing decreases of $\sim$3 {\AA} and $\sim$2 {\AA}, respectively}. \deleted{Moreover, the constraint that was previously exposed in segments S(66-81) is still noticeable, as P(64-80) deviates from the overall linear trend between $\nu$ and NCPR (Fig. \ref{fig:protein_nu}) \added{with $\nu=0.72$}.} 

The data shows that regardless of the specific detail of each segment sequence, its structural dimension (i.e., $R_{ee}$) increases while tethered to the rest of the protein. This phenomenon is unexpected for long and self-similar homogeneous polymers, although documented in the formation of tertiary structures for folded proteins \cite{guo2008denaturant}. Next, we demonstrate that the expansion can be eliminated by replacing the protein context with inert tethers.

\subsection*{Structural heterogeneity without an expansion in the presence of tethered self avoiding walk polymers}

\subsubsection*{Non-interacting tethering design}
The expansion of all P segments compared to their S segment counterparts raises the question of how unique this phenomenon is and whether it depends on the sequence of the tethered contextual protein. Given the Flory parameters ($\nu\geq 0.5$) of both S and P segments, it is tempting to consider them as non-interacting or even repelling. \deleted{However, for polymers in $\theta$ solvent ($\nu=0.5$),  the context of additional tethered chains should not expand the structure of the segment \cite{ghosh2022rules}.}
\added{However, for perfectly ideal (Gaussian) polymers, the context of additional tethered chains should not expand the structure of the segment \cite{rubinstein2003polymer}.}


To evaluate the role of the tethered sequence in the expansion of P segments, we study the structural properties of S(6-25) and S(106-128) in the context of extensions of SAW chains. To mimic SAW chains, we include the repeats of three amino acids, Proline, Alanine and Serine (PAS) to both ends of the S segments. PAS repeats have been shown to adopt a random coil-like structure in an aqueous solution \cite{breibeck2018polypeptide, gebauer2018prospects}. Since PAS repeats are uncharged and relatively hydrophilic, we expect it not to have a strong attractive interaction with the tethered segment. 

Uncharged segment S(6-25) and charged segments S(106-128) are fluorescently labeled as before but each with additional 3 or 7 PAS repeats at both ends of the chain (table \ref{tab:PAS_Sequence}, and Fig. \ref{fig:PAS}a). While not as long as the native tethers, the lack of size dependence of the two tested PAS tethers indicates that length is unlikely to be a major factor. As \replaced{a}{another} control, we also measured a segment of (PAS)$_7$ labeled at its ends while adding the same (PAS)$_3$ or (PAS)$_7$ from each terminal of the chain. 

\subsubsection*{Tethering effect on the scaling parameter}
The PAS-only repeats show a scaling factor between random and real polymer with $\nu=0.58$ by SAXS and $\nu=0.54$ by trFRET. For most cases, the scaling value is rather homogeneous across different PAS polymer lengths and salinities. Unlike the previous results, using trFRET we find that the segments' $\nu$ and $R_{dd}$ (Eqs. \ref{eqSAW}, \ref{eqGaussFRET}) are relatively insensitive to the addition PAS repeats chains (Figs. \ref{fig:PAS}a,b, \ref{fig:SI-PAS-FRET-decay}). This result, in combination with the difference between S and P segments, demonstrates that the segments' structure is sensitive to amino-acid of nearby chains. When including as much as possible SAW polymer, such as PAS, the average structure remains as of a free segment (S). However, the NFLt context interacts with each segment leading to expansion. 

\subsubsection*{Tethering simulation}
\deleted{Furthermore,} We performed reference Monte-Carlo (MC) simulations of analogously tethering simple homopolymers in good solvent conditions (see Methods). The simulations indicate that the influence of tethering for simple homopolymer on swelling is small (< 7\%).Beyond 10 residues long tethers, the simulations show that the tethers' length alone does not play a large role in modulating the segment size (Fig.~\ref{fig:MC}).

\subsubsection*{Salinity effect in the Tethering context} 
\begin{figure}
\centering
\includegraphics[width=0.5\linewidth]{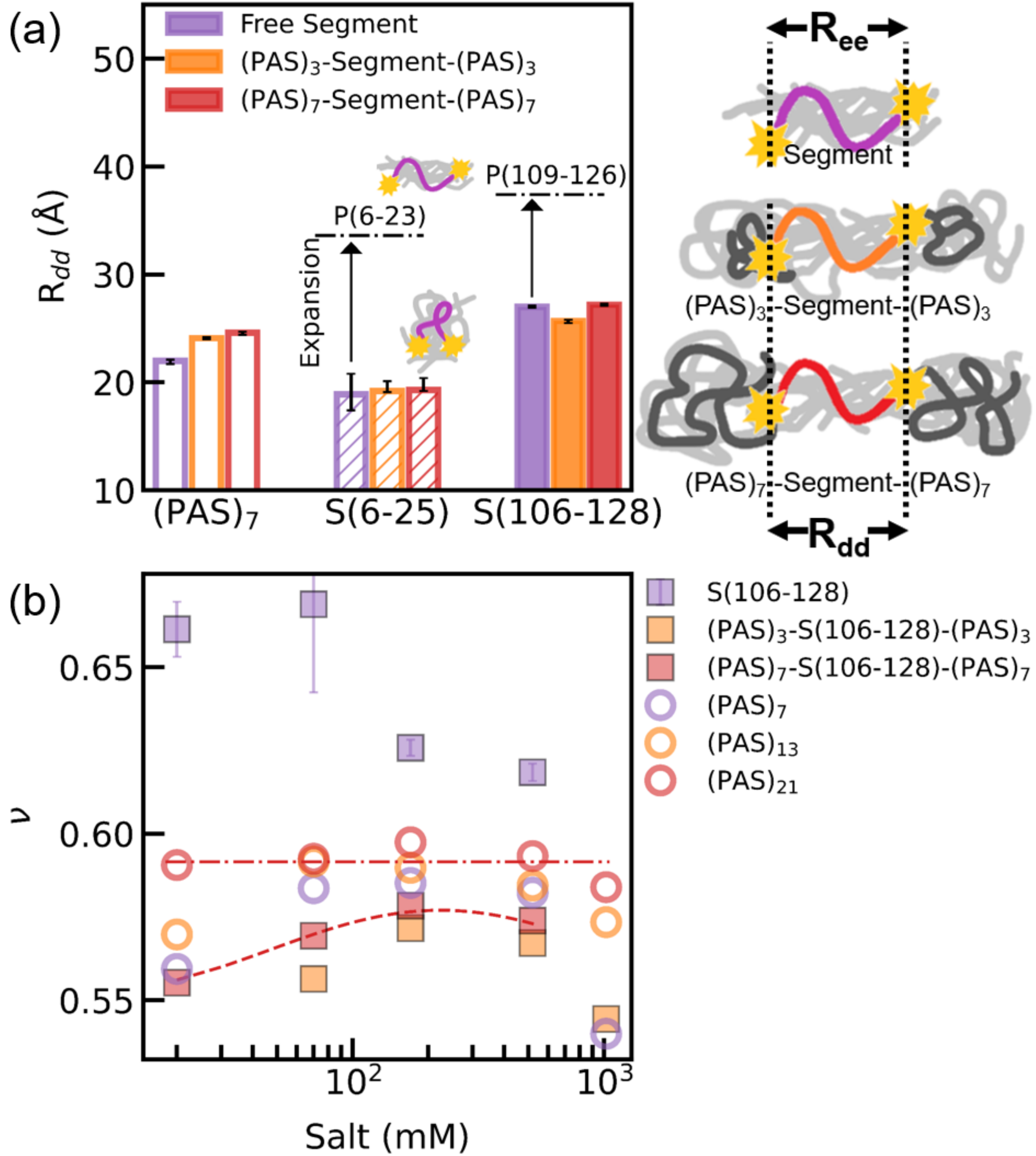}
\caption{(a) Dye to dye distance of segments (PAS)$_7$, S(6-25) and S(106-128) as free segments (purple), tethered from each side with (PAS)$_3$ (orange), and with (PAS)$_7$ (red). Increasing PAS chain length results in a very small change in the $R_{dd}$ (trFRET Fluorescent decays presented in Fig.  \ref{fig:SI-PAS-FRET-decay}). Dashed-dotted lines indicate the $R_{dd}$ value of the corresponding P segments. On the right schematic illustration of the result from PAS trFRET measurements. Tethered PAS repeats from each side of segment S(106-128) did not change the $R_{ee}$. (b) Scaling exponent from SAXS measurements at different salt concentrations. PAS-only segments are less dependent on salt concentration and behave as SAW polymers.}
\label{fig:PAS}
\end{figure}

In contrast to trFRET measurement, the $\nu$ value obtained from SAXS, represents the scaling of the entire construct that includes segment S(106-128) and the additional PAS repeats. Since, the inclusion of the PAS repeats lowers the NCPR from 0.5 to below 0.3, the resulting $\nu$ is decreased as well \added{from $\nu \sim 0.66$} to $\nu \sim 0.57$ (Fig. \ref{fig:PAS}c). This is also consistent with the SAXS linear pattern on Fig. \ref{fig:3}a.

Nonetheless, for charged segment S(106-128) including the PAS repeats, we still observe first subtle increase \added {from $\nu=0.56$ to $\nu=0.58$} and then decrease in $\nu$, \added{to $\nu=0.57$,} upon added salt (dashed red line in Fig. \ref{fig:PAS}c). 
For short only-PAS repeats similar hump-like behavior in $\nu$ is observed, although these sequences are uncharged (Fig. \ref{fig:PAS}c). We note that to keep the peptide chains concentration constant, the peptide molarity is increasing for smaller PAS repeats.  

A similar trend is observed with trFRET (Supplementary table \ref{tab:PAS_FRET_Ree}). There, a decrease in $R_{dd}$ \deleted{and $\nu$} is milder with the addition of PAS chains for both S(6-25) and S(106-128). We note that since PAS is charge neutral, its inclusion lowers the polymer's charge fraction. 
 
\section*{Discussion}

The interplay of intramolecular interactions that affect structural heterogeneity \added{have been studied but still lacking a comprehensive} \deleted{has eluded} experimental investigation in the context of IDPs. Here we use polymer physics models combined with SAXS and trFRET to characterize the structural heterogeneity of a model IDP. Overall, the result by both SAXS and trFRET shows that the NFLt consists of intramolecular structural heterogeneity - different domains within the IDP occupy space differently, although all display disorder properties. While NFLt can be defined as a strong polyampholyte (FCR = 0.45 and NCPR = 0.24)\cite{mao2010net}, introspection of sub-segments shows sequence heterogeneity that ranges from polymer description as weak polyampholyte to strong polyelectrolyte. Consistently, we find that Flory's exponent is scaling from chain in theta solvent with $\nu=0.5$ to chain in good solvent with $\nu=0.6$ and further stretching up to $\nu=0.7$ in the IDP context. While the segments' expansion correlates well with the net charge at low ionic strength \cite{muller2010charge} (Fig. \ref{fig:protein_nu}), and the order parameter that combines the sequence hydropathy and charge decoration \cite{zheng2020hydropathy} (Fig. \ref{fig:3}), the additional expansion in the context of the protein is beyond current modeling. Nonetheless, such scalings promote the ability to predict the ensemble structure of IDPs segments from their sequence. Interestingly, the scaling correlation with NCPR is tuned down with increasing salinity pointing towards dominant electrostatic interaction.

Additionally, denaturation of IDP homogenizes their disordered structure. In such cases, the IDP segments, regardless of their sequence, have a very similar structural scaling to chemically unfolded structured protein ($R_g=1.927 N^{0.598}$ \AA) \cite{kohn2004random}. On average, our SAXS measurements at 3M GdnHcl show a scaling behaviour of $R_g=(1.99 \pm 0.06) N^{0.61\pm 0.02}$ \AA. We emphasize that both trFRET and SAXS show expansion with denaturing, which is consistent with recent works \cite{borgia2016consistent}.

The main difference we find with the scaling exponent derived from SAXS data being consistently larger than that from trFRET data. This discrepancy is in line with other documented examples of results originating from the two experimental techniques \cite{ruff2017saxs,best2020emerging,yoo2012small,fuertes2017decoupling,riback2017innovative}. The discrepancies have been suggested to emerge due to the inference strategy utilized to convert mean FRET transfer efficiencies, measured in smFRET experiments, into estimates of $R_g$. The latter can then be directly compared to $R_g$ measured from SAXS \cite{song2017conformational}. This is consistent with the idea that the origin of the discrepancy is in the ability to decouple measures of specific pairwise distances ($R_{ee}$) from the averaging over the square of all pairwise distances ($R_g^2$) \cite{fuertes2017decoupling}. Moreover, it was suggested that $R_{ee}/R_g$ is not constant and depends on the scaling parameter \cite{zheng2018inferring}. Indeed, for different salinity, $R_{ee}/R_g$ ratio changes (Fig. \ref{fig:ReeRg}). Last, the discrepancies can originate from the hydrophobic nature of the fluorophore in FRET experiments \cite{riback2019commonly,reinartz2020fret}. However, when \replaced{comparing}{measuring} fluorescently labeled \added{and unlabeled} segments with SAXS, we do not observe a significant decrease in $\nu$ (supplementary Fig. \ref{fig:SI-un_and_labeled-S5p}). Moreover, as mentioned, FRET is only sensitive to the positions and fluctuations of the fluorophores, while SAXS to the bulk mass\cite{fuertes2017decoupling}. Therefore, hydrophobic fluorophores in the context of a hydrophilic segment might slightly collapse the ensemble, resulting in with smaller $\nu$ (supplementary Fig. \ref{fig:S_segment_Nu}).      

Another discrepancy between SAXS and FRET relates to the scaling exponent versus salt concentration showing a maximum in SAXS data, particularly for the charged segments (Fig. \ref{fig:3}b). This maximum is not present in the FRET data. We note that these highly charged segments, particularly at low salinity, show inter-molecular repulsion evident by a deep in the scattering at $q\rightarrow 0$ (Fig. \ref{fig:SI-InerRepulsion_Salt},  \ref{fig:SI-Repulsion_concentration}). Our SAXS data, even at the lowest segment concentration (1 mg/ml) can not role out inter-molecular interference (Fig. \ref{fig:SI-Repulsion_concentration}). Therefore, at 20mM Tris, polyelectrolyte segments at mM concentrations are less expanded than in $\mu$M concentrations, due to a stronger (peptide concentration-dependent) repulsion\cite{muthukumar2012counterion}. Hence, the expansion seen in fig. \ref{fig:3}b upon salt addition
can be attributed to electrostatic inter-molecular interactions. In most trFRET data, $R_{ee}$ does decrease with increasing salt concentration (Fig. \ref{fig:GaussMeanSalt}). This is consistent with electrostatic screening to intra-molecular interactions. Combining both intra- and inter-molecular interaction the entire SAXS data at different salinity can be understood. 

 
The scaling exponent of most segments correlates with increasing net charge. However, the most significant deviation is found in the N-terminal regime, where the net charge is minimal and hydrophobic\added{ity} is maximal. The deviation persists even when compared to current IDP theory \cite{zheng2020hydropathy, devarajan2022effect}. Using trFRET, we find S(6-25) to be more collapsed than expected with $\nu=0.42$. In contrast segment S(26-46) shares a similar NCPR but shows $\nu=0.52$. Unfortunately, we were unable to measure S(6-25) with SAXS due to aggregation (Fig. \ref{fig:SI-Pep1}). The additional contraction of S(6-25) could result from higher Gly and Ser content \cite{rahamim2015resolution}.

Additional deviation from polymer physics is found with \added{several }segments \deleted{S(67-86) and S(129-146)} showing specific structural constraints. There, the full fluorescence decay analysis shows a relatively narrow $R_{ee}$ distribution width to its mean that can not be modeled with standard polymer statistics. \added{Interestingly, we find an empirical correlation between the relative distribution width and $l_c/\mathcal{P}$ (Fig. \ref{fig:SI-ClusterCorrelation}). Here, $l_c$ is the segments'  contour length and $\mathcal{P}$ counts the number of negatively charged clusters, with at least two neighbouring amino-acids, in the segment. Presumably, shorter segments are less likely to form long-range constrains and multiple charged patches can induce multiple alternative ionic bridges that will cover a wide range of distances.}

Moreover, while in most cases, we find a decrease in the $R_{ee}$ with increasing salinity, we see a counter-intuitive exception with these segments, which shows \deleted{apathy or increasing}\added{insensitivity or even increase} in the mean distance. By examination of segments' sequence, we relate the deviation \added{from conventional polyelectrolyte theories} to structural constraints.\deleted{originating from di-Proline restriction or ionic bridge formation} \added{For example, }in S(129-146), \added{we relate} the increase in $R_{ee}$ \added{with added salinity} \deleted{is caused by the} \added{to} breaking \deleted{of a the loop formation due to} \added{the loop formed by the} three Lys in a row at the C-terminal. Revealing these constraints from the trFRET data is manifested by careful analysis of the entire fluorescence decay. By introducing effective contour length for WLC analysis, and radial Gaussian distribution, we are able to identify the segments that deviate from expected polymer statistics. The effect of molecular constraints can also be visible in the SAXS data with expansion upon truncation of the positive tip in S(129-146) \added{from $\nu=0.61$ to $\nu=0.63$} \deleted{and the highest $\nu$ value for S(67-86)} (supplementary Table \ref{tab:SAXS1}).

\deleted{Most surprisingly,} We find that all segments considerably expand when measuring them in the context of the entire NFLt ($R_{dd}/R_{ee}\approx 1.2$). \added{An expansion is expected for SAW chain but the magnitude of the expansion is unexpected}. For non-interacting polymers ($\nu = 0.5$), trivially no modification by the tethered ends is expected, while our MC simulations of simple homopolymers in theta solvents show only very small deswelling effects (<6\%), see Fig.~\ref{fig:MC}. On the contrary, the MC simulations only show a mild swelling effect (<7\%) for SAW tethered chains. Here, the swelling increases as the tethered chains' length increase and saturates at a length of 20 residues. For the S versus P segments experiments (Fig. \ref{fig:protein_nu}), we demonstrated that regardless of the segment sequence, there are significant expansions. We explain this expansion by the multiple transient contacts between distant amino acids. These contacts compete with local intra-molecular interactions within the segments. Since local interaction within the segment necessarily compacts the structure, competing contacts from amino acids located far away along the chain effectively expand the segments' structure. Including tethered SAW chains of PAS repeats supports this long-ranged contact model. Here, when the tethered chain is inert, the dye-to-dye distance is indifferent to the added chains (Fig. \ref{fig:PAS}), as in the reference MC simulations. 

Interestingly, only the P segments, measured in the context of the entire protein, \deleted{did }show\added{ed} SAW statistics with variable $\nu$ that fit the entire fluorescence decays. The P segment data do show structural heterogeneity with correlation to NCPR. Such heterogeneity for charged IDPs has been simulated before \cite{baul2019sequence}. Similarly, small IDP segments of 23 amino acid length and NCPR = 0.09 in $\alpha$-synuclein \cite{grupi2011segmental}, result in $R_{dd}=33.6$ {\AA}. Here, P(23-43), having similar chain length and net charge, agrees with the above data with $R_{dd}=33.4$ {\AA}.  

Moreover, the structural heterogeneity is in agreement with the inclusion of the structural constraints that were evident for the short segments and as part of the entire protein. The constraints have milder effect due to the above mentioned abundant long-ranged contacts present in the protein context. Surprisingly, these competing forces result in SAW statistics for all seven measured segments. \replaced{That said}{At the same time}, we can not rule out that the statistics are a coincidence or a special case shown in NFLt, although the fact that all segments scale similarly does point to generality.          

In conclusion, we find intra-molecular structural heterogeneity in \replaced{a}{our} model system IDP. The ensemble structure correlates with NCPR; however, local structural constraints of specific sequence motifs must be taken into account. Moreover, we found that long-range contacts are key determinants in NFLt. These contacts effectively compete with the short-ranged intra-molecular interactions, thus expanding the structure and hiding the local constraints. As demonstrated here, future functional investigations of IDP should take into account the structural heterogeneity resulting from the primary sequence and the environmental context.

\section*{Acknowledgements}
The synchrotron SAXS data was collected at beamline P12 operated by EMBL Hamburg at the PETRA III storage ring (DESY, Hamburg, Germany), and at beamline B21 at Diamond Light Source. We would like to thank Daniel Franke (Desy), Nathan Cowieson, Charlotte Edwards-Gayle and Katsuaki Inoue (all three of them Diamond Light Source) for the assistance in using the beamlines.

This work has been supported by the National Science Foundation under Grant No. MCB-2113302, the United States-Israel Bi-national Science Foundation under Grant No. 2020787, the Israel Science Foundation under Grants No. 1454/20, the Israel Science Foundation under Grant No. 1435/19, and by iNEXT-Discovery, project No. 871037, funded by the Horizon 2020 program of the European Commission.

We thank Robert Best for kindly sharing his FRET analysis code and for valuable discussions about the results. We also acknowledge fruitful discussions and assistance in the experiments with Haim Diamant, Dan Amir, Yacov Kantor, Joshua Riback, Tobin Sosnick, Vaishali Sethi, Valentina Alberti, Elvira Haimov, Boris Redko, and ChatGPT.

\bibliography{lib}

 \newpage
\renewcommand{\thefigure}{S\arabic{figure}}
\renewcommand{\thetable}{S\arabic{table}}
\renewcommand{\theequation}{S\arabic{equation}}
\setcounter{figure}{0} 
\setcounter{table}{0}
\setcounter{equation}{0}

\newpage
\begin{center}
\title{SI Appendix}
\label{Appendix}
\end{center}

\section*{Methods}
\subsection*{Peptide preparation}
Synthesis and purification of peptides were performed by the Blavatnik Center of Drag Discovery at Tel-Aviv University and LifeTein LLC (Hillsborough, Nj, USA). Peptides were identified using mass spectrometry and were purified by over 95\% via high-performance liquid chromatography (HPLC). Labeled peptides were designed to have Alanine residue next to the FRET dye to achieve an equal spectral environment for all of the peptides. 

\subsection*{Protein purification and labeling } 
\begin{figure}[h]
\centering
\includegraphics[width=170mm,scale=1.0]{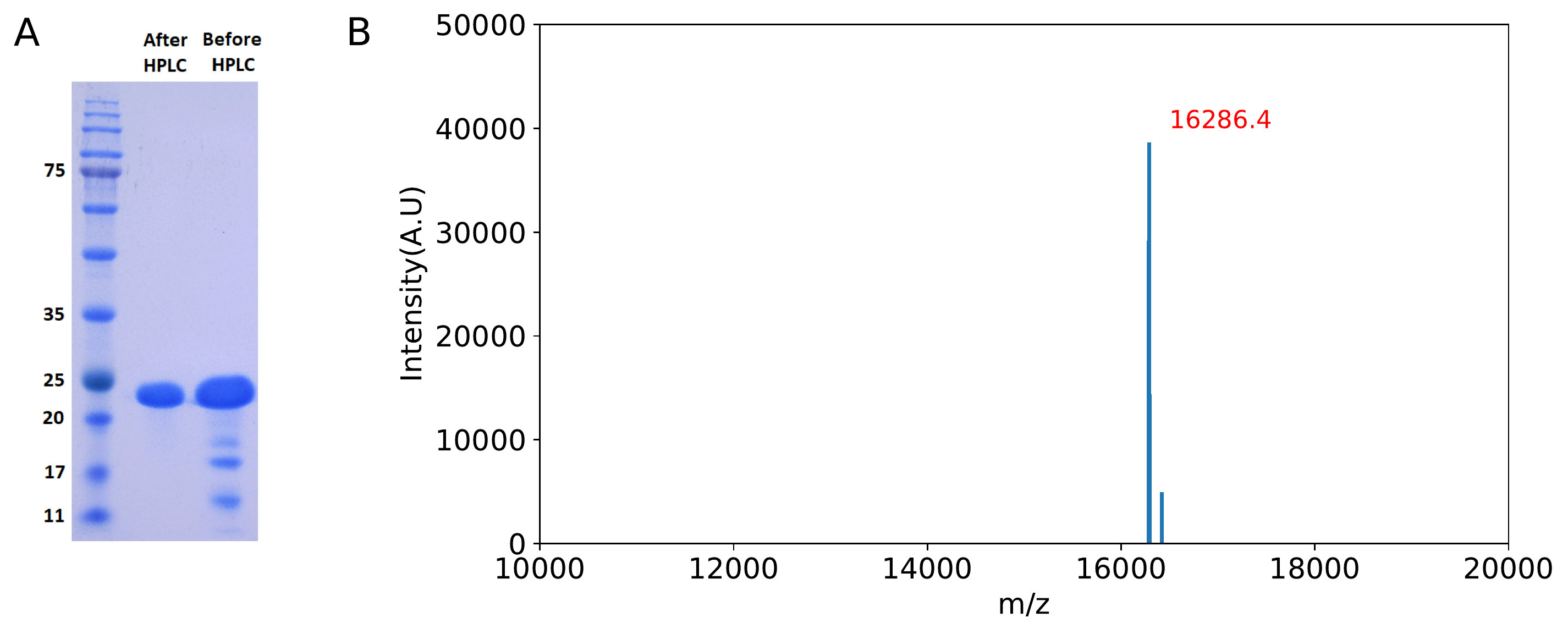}
\caption{A. SDS-PAGE Tris-Glycine 15\% of NFLt model protein showing purity above 95\%. The molecular weight is 16.2kDa but shown as 24 in the gel which is common for IDPs. \added{B. Deconvoluted ESI-TOF MS spectra of NFLt 64w. Expected mass is 16348 Da.}}
\label{fig:Gel}
\end{figure}

Protein purification followed Morgan et al. \cite{morgan2020glassy} with several modifications. The gene encoding NFLt (supplementary) ligated into a pET vector with a PagP fusion protein on the C-terminus. Mutagenesis was done using supreme NZYProof DNA polymerase  (NZYTech, Portugal). To confirm the mutations, the full length PagP fusion protein and the NFLt domain gene were sequences for each mutant vector. Competent Escherichia coli BL21(DE3) Rosetta was transformed with the modified pET vector and subsequently plated on agar plates containing $100 \mu g/ml$ ampicillin and $30 \mu g/ml$ chloramphenicol. A single colony was picked for starting cultures and grown overnight in 50 ml LB containing $100 \mu g/ml$ ampicillin and $30 \mu g/ml$ chloramphenicol. Cells were palletized, resuspended, and transformed to 1 L  Terrific broth containing $100 \mu g/ml$ ampicillin and $30 \mu g/ml$ chloramphenicol. Expression cultures were grown in a baffled Erlenmeyer flask in a shaking incubator at 37{\degree}C at 280rpm for 3-5 hr until the optical density at 600 nm reached 0.7-1.0. Protein expression was induced by the addition of Isopropyl b-D-1-thiogalactopyranoside to a final concentration of 0.5 mM. The cultures were grown for 4-6 hr before harvesting. Cells were palletized and stored at -80{\degree}C for later use. For purification of proteins, cell pellets were resuspended in a 10 ml lysis buffer for each 1 g bacterial pellet. The lysis buffer contained 20 mM Tris buffer pH 8.0, 0.1\% 2-Mercaptoethanol, 1\% Triton, and 0.5mg/ml Lysozyme. The solution was incubated at 25{\degree}C for 20 min. Next, the solution was added with 10 mM MgSo4 and 1k units of Benzonase nuclease for 20 min at 25{\degree}C. The solution was centrifuged at 18,500 g for 30 min at 4{\degree}C. Next, the pellet was homogenized in a washing buffer containing 20 mM Tris pH 8.0, 6 M Guanidine HCL, 20 mM imidazole and 0.1\% 2-Mercaptoethanol. After centrifugation at 18,500 g for 30 min at 4{\degree}C the supernatant was loaded on a 10 ml home-packed nickel affinity column that was equilibrated with a washing buffer, at a rate of 1 mL/min. After washing with 100 ml, the protein solution was eluted with an elution buffer containing 20 mM Tris buffer pH 8.0, 0.5M Imidazole, and 0.1\% 2-Mercaptoethanol. The solution was then dialyzed overnight against 1 L of 20 mM Tris pH 8.0, followed by another dialysis against 1 L of 50 mM 3-(N-morpholino)propanesulfonic acid (MOPS) buffer pH 8.5. After dialysis, a cleavage reaction was initiated by 5 mM NiSO4 in the presence of 6 M Guanidine and incubated for 20 hr at 50{\degree}C. Cleavage reaction was stopped by 50 mM EDTA and followed by adding 0.1\% 2-Mercaptoethanol. The cleaved protein was dialyzed twice overnight against 1 L 20 mM Tris pH 8.0, 2 mM EDTA, and 0.1\% 2-Mercaptoethanol. Cleaved PagP precipitate was centrifuge at 18,500 g for 30 min and discarded. The protein was adjusted to 6 M Guanidine and was loaded on a 100 mL size-exclusion column (HiPrep 16/60 Sephacryl S-200 HR) at a rate of 1 mL/min pre-equilibrated with washing buffer containing 20 mM Tris pH 8.0, 1M Guanidine, 2 mM EDTA and 0.1\% 2-Mercaptoethanol. Eluted NFLt was dialysis against 1 L of 20 mM Tris pH 8.0 and 0.1\% 2-Mercaptoethanol. For labeling preparation, NFLt containing Trp and Cys mutation was first reduced by 50 mM 2-Mercaptoethanol. Next, 2-Mercaptoethanol was washed using several concentration and dilution cycles with 50 mM HEPES buffer pH 7.2 using an Amicon ultra centrifugal filter unit of MWCO 10kDa. At the labeling reaction, the protein concentration was $\sim$3mg/ml. Fluorescence dye, 7-Iodoacetamidocoumarin-4-carboxylic acid (coumarin) (Chem Cruz), was dissolved in DMSO and was added to the protein solution at a molar ratio of 1:20. Reaction solution was kept under dark and slow stirring for 5 hr. The labeling reaction was stopped by adding 50 mM 2-Mercaptoethanol followed by overnight dialysis against 1 L of 20 mM Tris at pH 8.0 with 0.1\% 2-Mercaptoethanol. Final purification was done by HPLC using a semi-preparative Vydac C18 column. The column was pre-equilibrated with 0.1\% trifluoroacetic acid (TFA). Labeled protein was loaded and eluted by a linear gradient from 0\% to 50\% acetonitrile and 0.1\% TFA for 30 min at a rate of 2 ml/min. Final purity was $>$95\% as determined by SDS-PAGE \added{and protein identity was confirmed by mass spectrometry} (Fig.~\ref{fig:Gel}). Typically growth was of 4 L cultures, and the yield of pure labeled protein was 5-10 mg.

As a control for the labeling efficiency, the reduced state of the labeled protein was reacted with 5,5-dithio-bis-(2-nitrobezoic acid). This control was important since unlabeled protein can increase the apparent distance shown by FRET. Results show less than 5\% unlabeled fraction of protein in the labeled protein solution. This result is not changing the conclusion of this paper.

\subsection*{trFRET measurements}
The time-correlated single photon counting (TCSPC) method was used at the fluorescence center of Bar-Ilan University, Israel, and Tel-Aviv University, Israel. Most of the data were collected at Bar-Ilan, where the excitation source was a femtosecond Ti sapphire laser (Chameleon, Coherent). The laser output was frequency tripled by a flexible second and third harmonics generator (A.P.E). A pulse selector (A.P.E.) was used to reduce the basic 80 MHz pulse rate to 4.0 MHz and 8.0 MHz for S and P segments, respectively. The excitation was at 290 nm and 295 nm for S and P segments, respectively. The emission wavelength was selected by a double $\frac{1}{8}$ m subtractive monochromator (DIGIKROMCM112, Albuquerque, NM) and directed to the surface of a fast photomultiplier (Hamamatsu, R988OU-210) biased at -1100 V. The donor emission (Naphthyl for S segment and Tryptophan for P segment) was collected at 350 nm (emission bandwidth 20 nm). A single-photon counting board (SPC 630; Backer and Hickl GmbH) fed via a preamplifier (HFAC-26DB 0.1UA, Brookline MA) and triggered by a photodiode (PHD-400N) was used for data collection. The response of the system yielded a pulse of full-width at half-maximum (FWHM) of 200 ps. The emission was collected with a polarizer at the magic angle (54.7{\degree}) relative to the excitation polarization. The reference impulse response function (IRF) profile used for deconvolution of the experimental decay curves was a scattered light pulse generated by placing a glass in the cell. Both S and P segments were measured at a concentration of 10 $\mu$M. Samples were routinely magnetically stirred during the measurements.

At Tel-Aviv university, TCSPC measurements were made using a Horiba FluoroHub-B with a pulsed LED source (Horiba) operating at 1 MHz with a wavelength of 284 nm. Emission was measured at 330 nm with a 15 nm bandwidth. Segments were measured at a concentration of 100 $\mu$M. The IRF was measured at 290 nm using 5 $\mu$m SiO2 beads (1\% in water). 

All measurements were done at 25{\degree}C in a 20 mM TRIS buffer at pH = 8.0. The background emission was routinely subtracted from the corresponding fluorescence decay curve. \added{Represented measurements were repeated at least twice and analyzed separately, resulting in  relative errors (standard deviation /average value) of $R_{ee}$ and $R_{dd}$ smaller than 0.05, 0.07 for the S and P segments, respectively. }  

\subsection*{trFRET analysis}

The analysis of a specific distribution model is done by two different methods. The first method is to initially extract the fluorescence exponential average lifetime for the calculation of the mean energy transfer $\langle E\rangle=1-\frac{<\tau_{DA}>}{<\tau_{DO}>}$ where $<\tau_{DA}>$ and $<\tau_{DO}>$ are the average donor lifetime with and without the presence of an acceptor, respectively. For the S segments, a single exponential fluorescence lifetime, is extracted from the DO measurements, and an average of three lifetimes is extracted from the DA measurements. Next, $\langle E\rangle$ is used via Eq. \ref{eqEt} to fit the SAW model (Eq. \ref{eqSAW}).

The second analysis method is fitting the entire fluorescence decay to: 

\begin{equation}
I(t)=A \sum_{i=1}^n\int_{0}^{l_{c}} P(r) \exp \left(-\frac{t}{\tau_{DO_i}}\left[1+\left(\frac{R_{0}}{r}\right)^6\right]\right) d r.
    \label{eq_Full_Decay_fitting}
\end{equation}

\begin{figure}[h]
\centering
\includegraphics[width=0.7\linewidth]{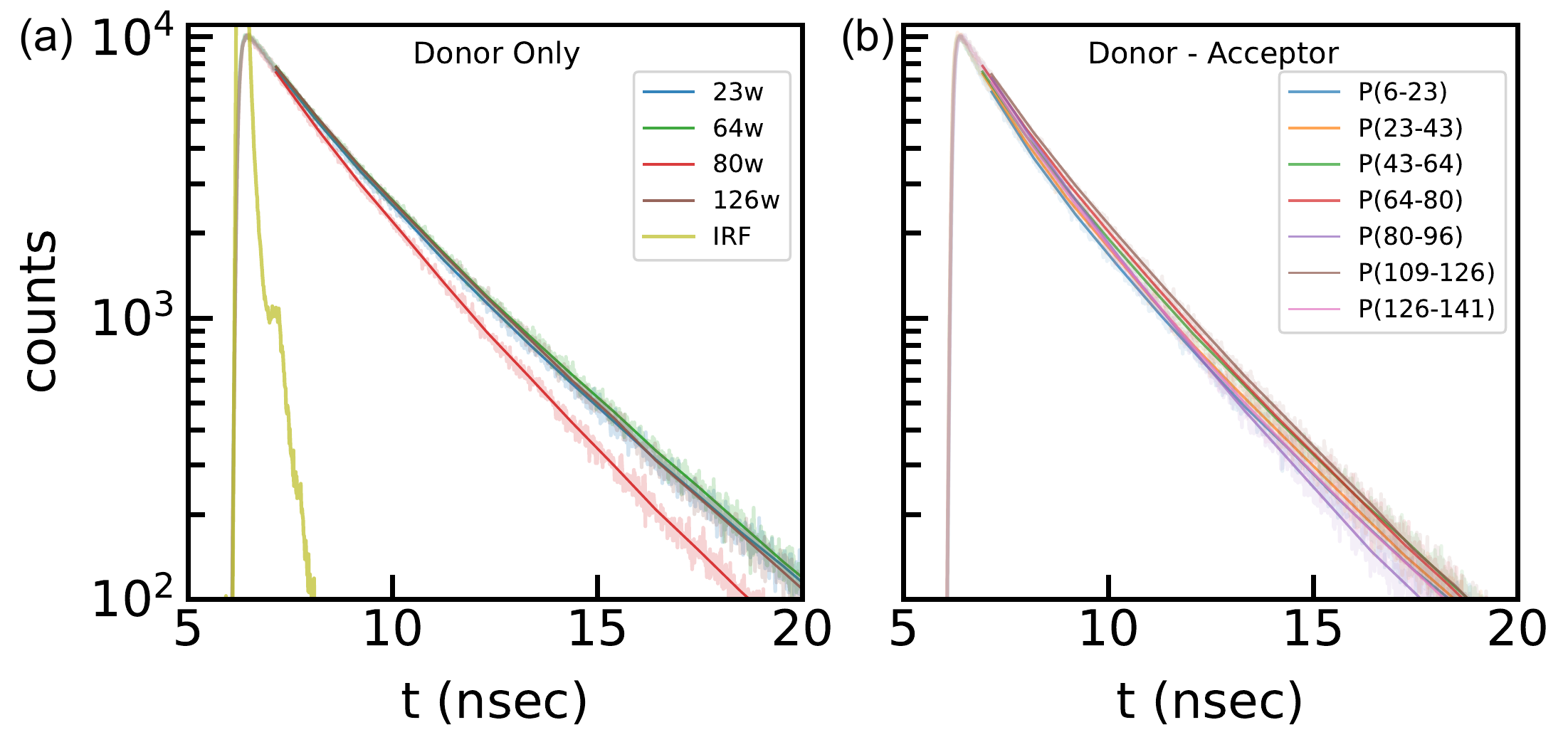}
\caption{Time-resolved fluorescent decays of segments with only a donor (DO) (a) and donor in the presence of an acceptor (DA) (b) in the context of the entaire NFLt (P segments). }
\label{fig:Protein_Decays}
\end{figure}

Here, $l_c$ is the segment contour length, $R_0$ is the F{\"o}rster distance and $A$ is a proportion parameter. For the S segments, fitting to the radial Gaussian model (Eq. \ref{eqGaussFRET}) was done via Eq. \ref{eq_Full_Decay_fitting} since the SAW model fit was insufficient. For the P segments, a multi-exponential fluorescence with three lifetimes (i.e., $n=3$ in eq. \ref{eq_Full_Decay_fitting}), $\tau_{DO_i}$, is extracted from the DO measurements (Fig. \ref{fig:Protein_Decays}) followed by fitting to SAW and radial Gaussian model via Eq. \ref{eq_Full_Decay_fitting}. Eq. \ref{eq_Full_Decay_fitting} was convolved with the IRF function, and optimal parameters were obtained with $\chi^2$ minimization with respect to the experimental fluorescent decay, assuming the lack of any diffusion effect. Confidence intervals for the mean of the distribution were calculated by applying a rigorous error analysis procedure.

\subsection*{Circular Dichroism }
Far-UV circular dichroism spectra of S segments and NFLt were measured at a concentration of 30 $\mu$M in a 5mM sodium phosphate buffer, 1mm cuvette at 25{\degree}C,  using Chirascan Circular Dichroism Spectrometer (Applied Photophysics, Leatherhead, Surrey, UK). Each scan was recorded over the range of 190-260 nm, bandwidth of 1 nm, and an average time of 0.5 s per point. For each sample, three scans were measured and averaged. 

\subsection*{SAXS measurements}
All S segments powders were diluted by buffer and concentration with 3 kDa Amicon filters. To ensure proper pH and buffer condition the dilution and concentration were repeated at least five times. S segments with additional PAS repeats were diluted and dialyzed overnight and measured with Nanodrop 2000 spectorphotometer (Thermo Scientific) for concentration determination. Buffers were prepared with 1 mM of TCEP to reduce radiation damage. Samples' final concentration was 4 mg/ml, and a series of 3 dilutions (1, 2, 3 mg/ml) was prepared to determine of inter-molecular interactions. 

\begin{figure}[h]
\centering
\includegraphics[width=0.7\linewidth]{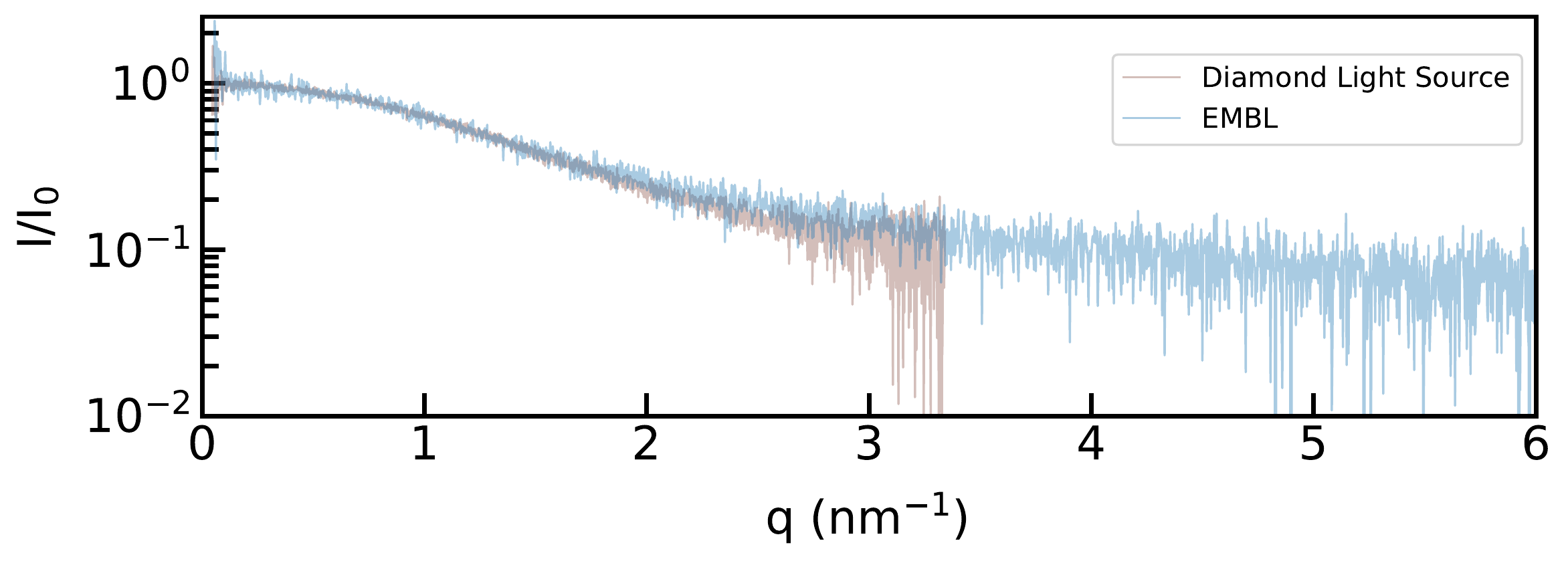}
\caption{\added{Example of SAXS data reproducibility. Brown and blue line colors are SAXS measurements of the same peptide at Diamond Light Source and EMBL respectively.}}
\label{fig:SI-EMBL-vs-DLS}
\end{figure}

Preliminary measurements were measured at Tel-Aviv University with a Xenocs GeniX Low Divergence CuK$\alpha$ radiation source setup with scatterless slits \cite{li2008scatterless} and a Pilatus 300K detector. All samples were measured at two synchrotron facilities: beamline B21, Diamond Light Source, Didcot, UK, and beamline P12, EMBL, DESY, Hamburg, Germany \cite{blanchet2015versatile}. \added{Data is reproducible, see Fig.~\ref{fig:SI-EMBL-vs-DLS}.}

\begin{figure}[h]
\centering
\includegraphics[width=0.5\linewidth]{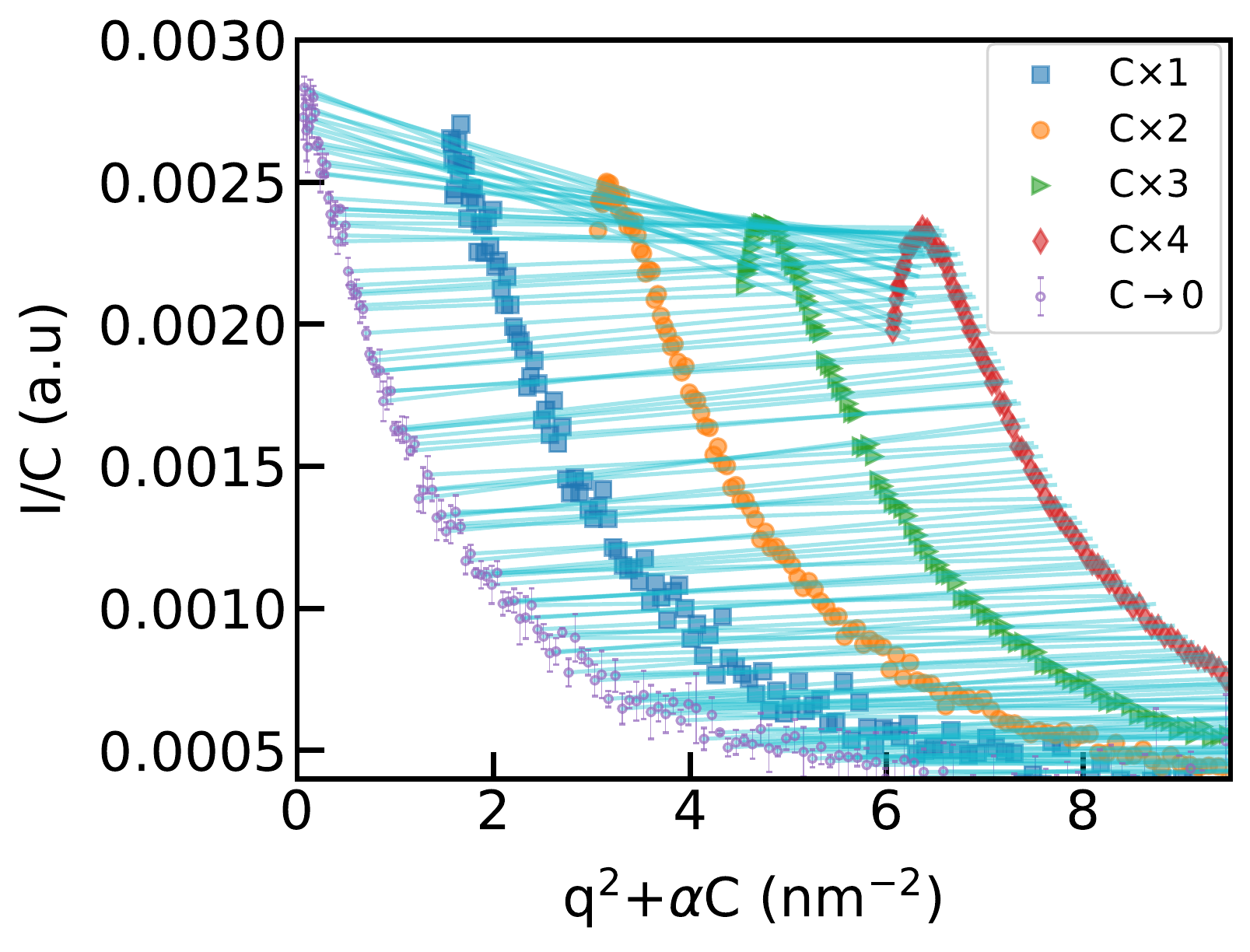}
\caption{An example of linear extrapolation of scattering intensity to zero concentration with a typical Zimm plot. SAXS measured intensity normalized by peptide concentration (C) versus $q^2+\alpha C$ of segment S(110-125) at 20 mM Tris buffer (without additional NaCl) and at four peptide concentrations multiples (1, 2, 3 and 4 mg/ml). Here $\alpha=1.5$ nm$^{-2}$ml/mg is an arbitrary constant that is chosen for plotting purposes. Data was rebinned for clarity. Cyan solid lines are linear fits for constant values of $q^2$ and different concentration. The purple dots are the linear concentration extrapolation to zero.}
\label{fig:SI-extrapolation}
\end{figure}

AUTORG was used for Guinier analysis \cite{petoukhov2007atsas}. Extended Guinier analysis was done with ``curve\_fit" function from the scipy python library. To extract $R_g$ and $\nu$, extended Guinier analysis were conducted for $0.15<qR_g<2$. 
Error estimation of the Flory exponent ($\nu$) was calculated as follows:
\begin{equation}
    \Delta \nu=\sqrt{\Delta \nu_{fit}^2+\Delta \nu_{std}^2},
\label{eq:delta_nu}
\end{equation}
where $\Delta \nu_{fit}$ is the fit uncertainty and $\Delta \nu_{std}$ is the standard deviation of about 300 extended Guinier fits taking alternative ranges of the maximum wave-vector $q_{max}$ such that $1.85<q_{max}R_g<2.1$. In most cases the relative errors are below 1\%. However, the exceptional were segments with NCPR $>0.4$ at low salinity, in which we encountered large variations in the scattering patterns at small angles for different peptide concentrations (Fig. \ref{fig:SI-Repulsion_concentration}). The difference results from high correlations between the peptides due to strong electrostatic interaction \cite{muthukumar2012counterion}. For these cases, we did the extended Guinier analysis on the linear extrapolation of the scattering intensity to zero peptide concentration (Fig. \ref{fig:SI-extrapolation}).

\subsection*{EOM}
\begin{figure}[h]
\centering
\includegraphics[width=1.0\linewidth]{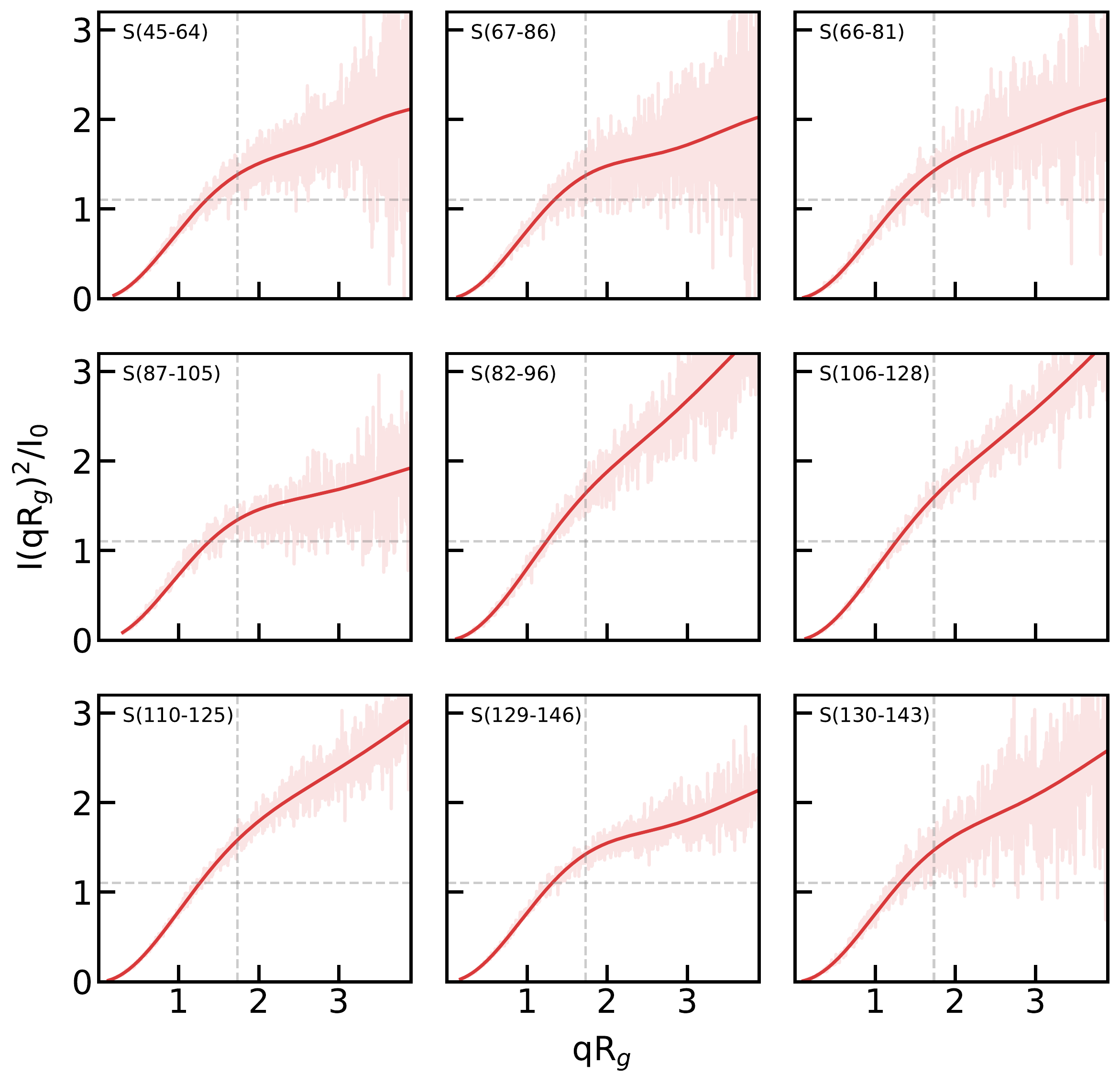}
\caption{EOM fits to the segments at 20 mM Tris, presented on a normalized Kratky plot.}
\label{fig:SI-EOM-Kratky-Tris}
\end{figure}

\begin{figure}[h]
\centering
\includegraphics[width=1.0\linewidth]{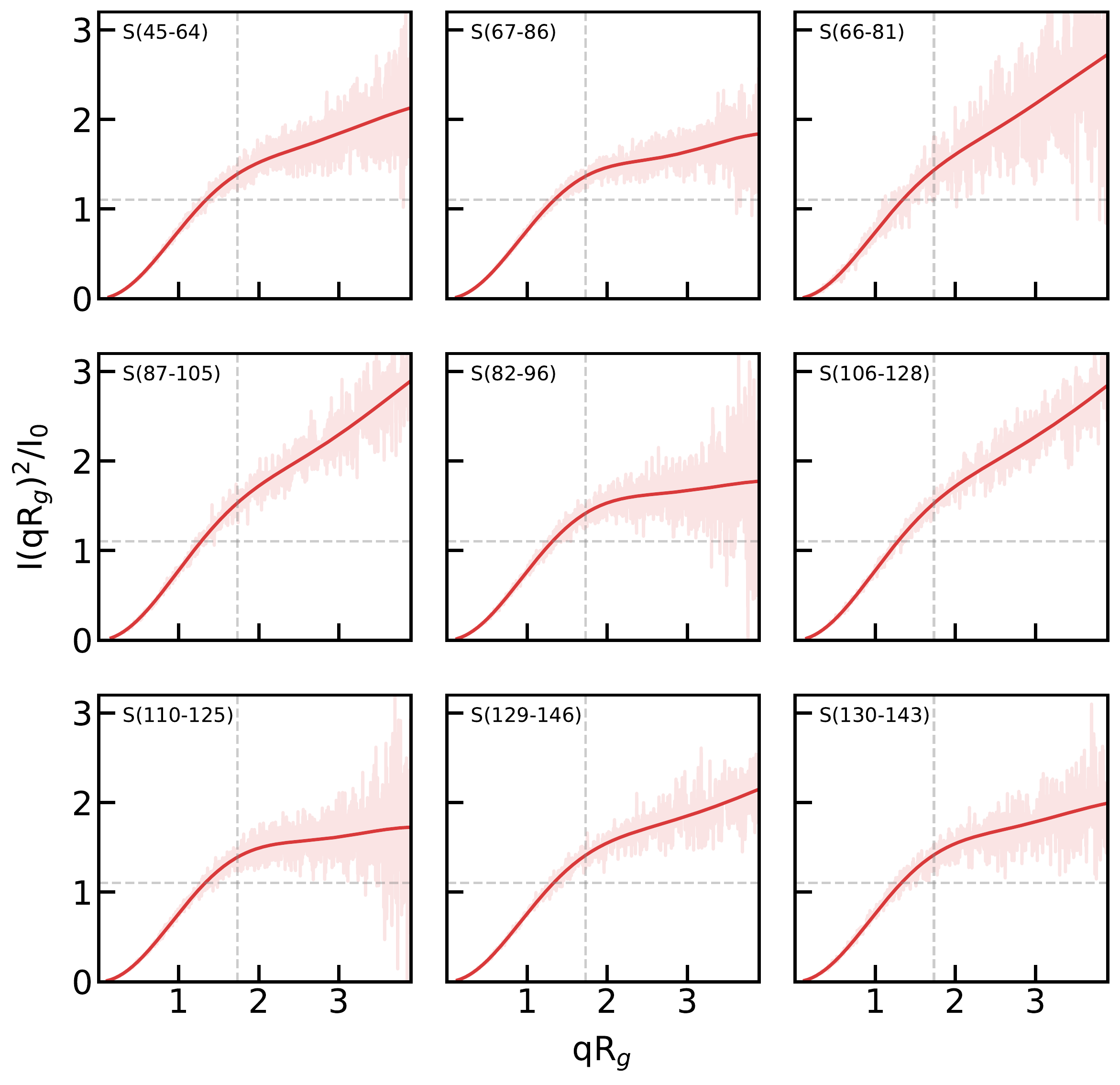}
\caption{EOM fits to the segments at 20 mM Tris and 50 mM NaCl, presented on a normalized Kratky plot.}
\label{fig:SI-EOM-Kratky-50mM}
\end{figure}

\begin{figure}[h]
\centering
\includegraphics[width=1.0\linewidth]{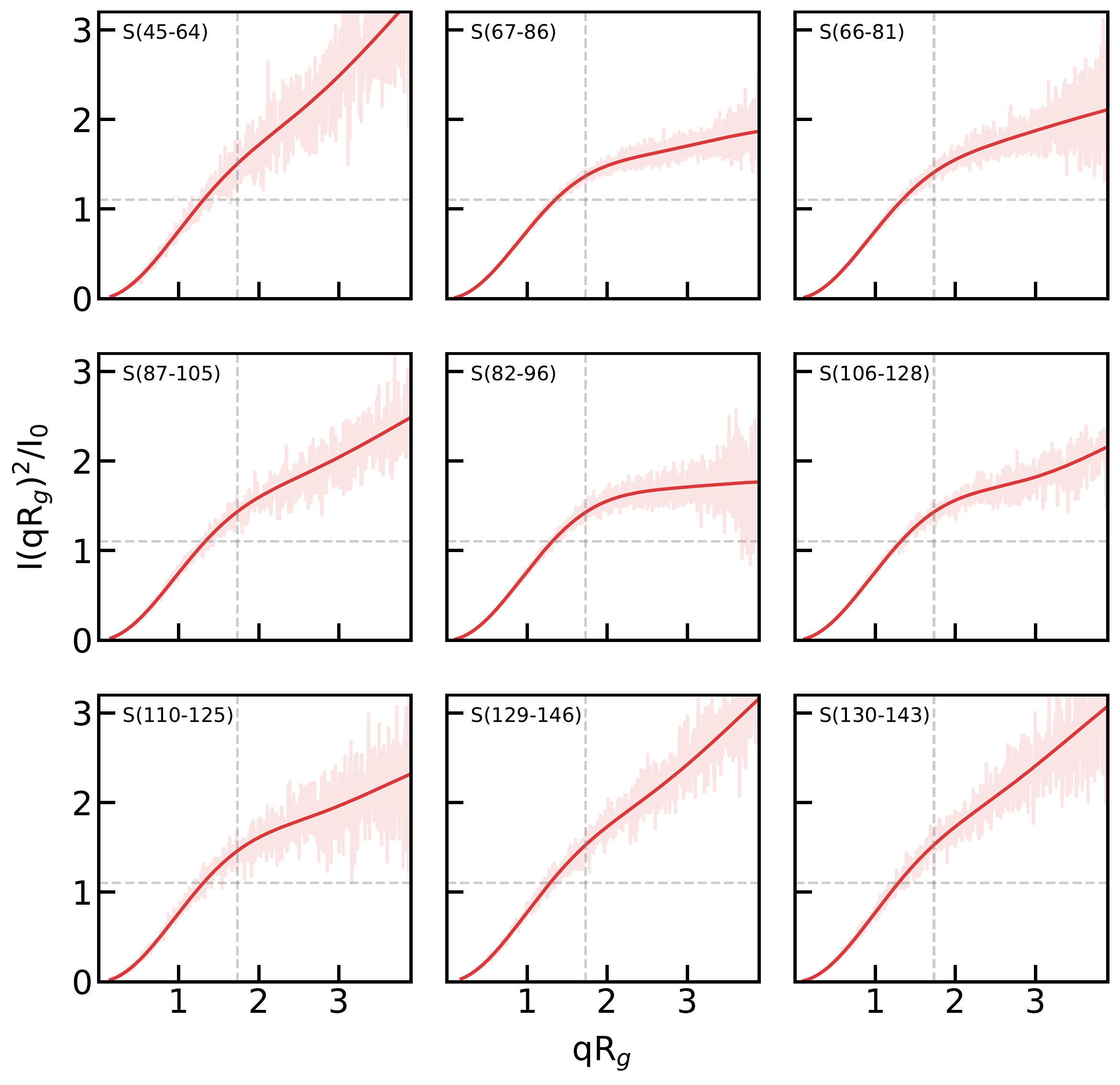}
\caption{EOM fits to the segments at physiological conditions (20 mM Tris and 150 mM NaCl), presented on a normalized Kratky plot.}
\label{fig:SI-EOM-Kratky-150mM}
\end{figure}

\begin{figure}[h]
\centering
\includegraphics[width=1.0\linewidth]{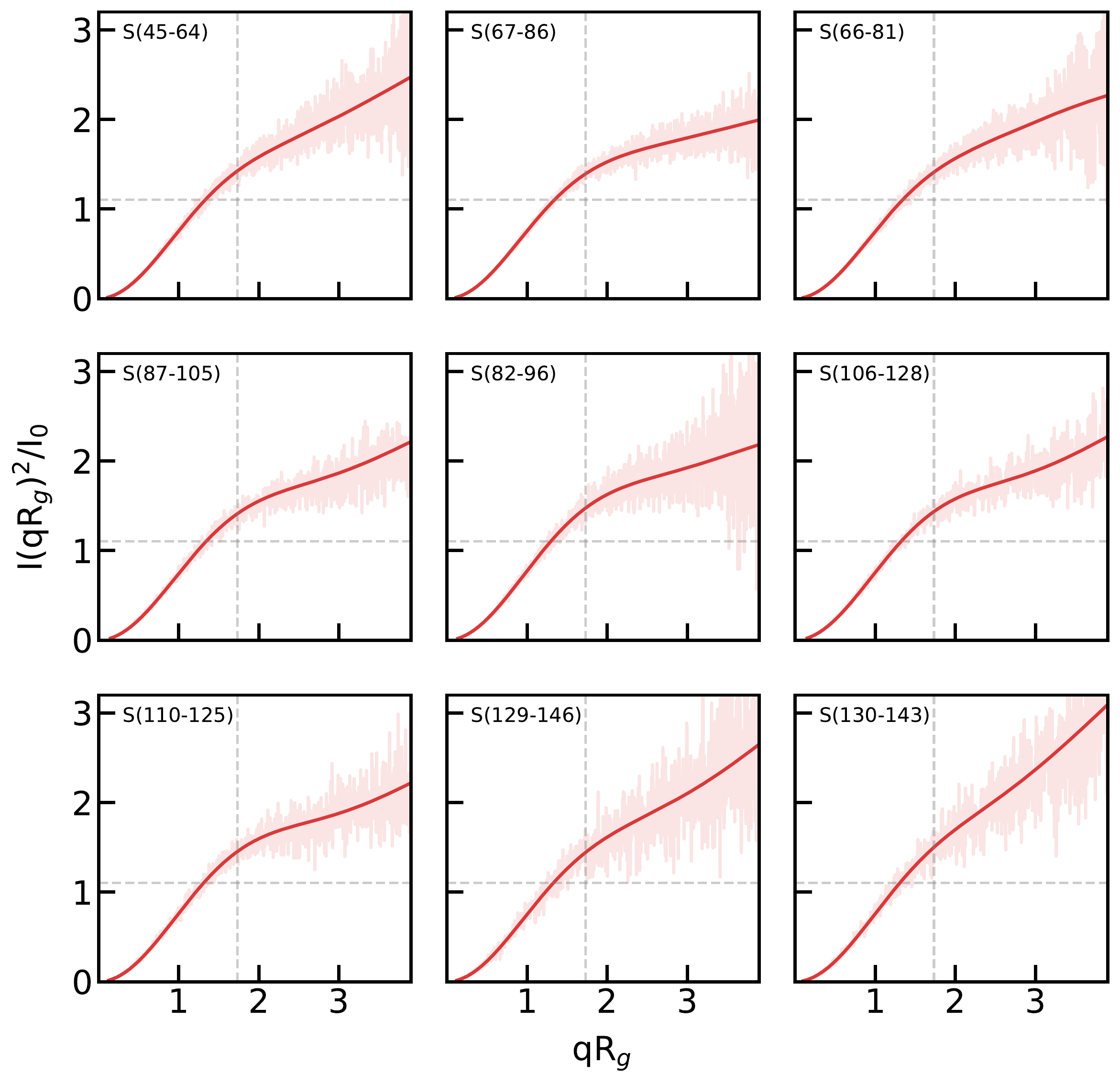}
\caption{EOM fits to the segments at 20 mM Tris and 500 mM NaCl, presented on a normalized Kratky plot.}
\label{fig:SI-EOM-Kratky-500mM}
\end{figure}

In the ensemble optimization method (EOM), a pool containing 10,000 possible conformations based on peptide sequence is generated with RANCH program. The scattering amplitudes of each conformation are calculated with \added{the program CRYSOL \cite{svergun1995crysol} explicitly taking into account the hydration shell around the peptides.} \deleted{FFMAKER program.} \cite{tria2015advanced} Then, the ensemble (sub-group of pool conformations) whose combined theoretical scattering intensity best describes the experimental SAXS data is selected with a genetic algorithm (GAJOE program) \cite{tria2015advanced, sagar2021comment}. \added{Examples of EOM fits are presented in Figs. \ref{fig:SI-EOM-Kratky-50mM}, \ref{fig:SI-EOM-Kratky-150mM}, \ref{fig:SI-EOM-Kratky-500mM}, \ref{fig:SI-EOM-Kratky-Tris}.}

Since IDPs are characterized by an enormous number of conformations, we did not apply the option of ensemble size optimization. Instead, we fixed the ensemble size on 50 conformers (the maximal conformers allowed). This way, we also avoided the result of bimodal $R_g$, $R_{ee}$ and size distributions \cite{tria2015advanced, sagar2021comment}. \added{Additionally, we note that the EOM analysis has been done on the highest peptide-concentration data that did not manifest strong signs of inter-peptide interactions at the lower $q$ regime. It includes the highly charged segments S(67-86), S(66-81), S(82-96), and S(110-125), where we did not use the extrapolated data for EOM analysis.}

\subsection*{Worm like chain (WLC) analysis}

For trFRET, the WLC model for distance distribution probability function \cite{o2009accurate} was used to analyze the S segment with two free parameters, the persistence length, $l_p$, and the contour length, $l_c$. 

\begin{equation}
P(r)=\frac{4 \pi\left(r / l_{c}\right)^{2} f\left(l_{p^{\prime}} l_{c}\right)}{l_{c}\left(1-\left(r / l_{c}\right)^{2}\right)^{9 / 2}} \exp \left[\frac{-3 l_{c}}{4 l_{p}\left(1-\left(r / l_{c}\right)^{2}\right)}\right]
\label{S3}
\end{equation}

with:
\begin{equation}
f\left(l_{p}, l_{c}\right)=\frac{1}{\pi^{3 / 2} \exp (-\alpha) \alpha^{-3 / 2}\left(1+3 / \alpha+15 /\left(4 \alpha^{2}\right)\right)}
\end{equation}

where $\alpha=\frac{3l_c}{4l_p}$

\subsection*{Charge parameters}
The NCPR is the absolute net charge value (at pH 8) divided by number of amino acids ($N$). The SCD was calculated as follows:
\begin{equation}
    SCD=\frac{1}{N}\sum_{i=1}^{N-1}\sum_{j=i+1}^N q_iq_j\sqrt{j-i},
\end{equation}
where $q_x$ is the net charge of an amino acid at position $x$ in the sequence. The SHD was calculated as follows:
\begin{equation}
    SHD =  \frac{1}{N}\sum_{i=1}^{N-1}\sum_{j=i+1}^N \frac{\lambda_j+\lambda_i}{j-i}. 
\end{equation}
Here, $\lambda_x$ is a hydropathy value (between 0 to 1) of an amino acid at position $x$. More hydrophobic amino acids' hydropathy is closer to 1.

\begin{table}[h]
\centering
    \resizebox{\textwidth}{!}{%
\begin{tabular}{lllllllllllllllllllll}
\hline
\multicolumn{1}{|c|}{Amino acid} & \multicolumn{1}{c|}{R} & \multicolumn{1}{c|}{H}     & \multicolumn{1}{c|}{K}     & \multicolumn{1}{c|}{D}     & \multicolumn{1}{c|}{E}     & \multicolumn{1}{c|}{S}     & \multicolumn{1}{c|}{T}     & \multicolumn{1}{c|}{N}     & \multicolumn{1}{c|}{Q}     & \multicolumn{1}{c|}{C}     & \multicolumn{1}{c|}{G}     & \multicolumn{1}{c|}{P} & \multicolumn{1}{c|}{A}    & \multicolumn{1}{c|}{I}     & \multicolumn{1}{c|}{L}     & \multicolumn{1}{c|}{M}     & \multicolumn{1}{c|}{F} & \multicolumn{1}{c|}{W}     & \multicolumn{1}{c|}{Y}     & \multicolumn{1}{c|}{V}     \\ \hline
\multicolumn{1}{|c|}{$\lambda$}  & \multicolumn{1}{c|}{0} & \multicolumn{1}{c|}{0.514} & \multicolumn{1}{c|}{0.514} & \multicolumn{1}{c|}{0.378} & \multicolumn{1}{c|}{0.459} & \multicolumn{1}{c|}{0.595} & \multicolumn{1}{c|}{0.676} & \multicolumn{1}{c|}{0.432} & \multicolumn{1}{c|}{0.514} & \multicolumn{1}{c|}{0.595} & \multicolumn{1}{c|}{0.649} & \multicolumn{1}{c|}{1} & \multicolumn{1}{c|}{0.73} & \multicolumn{1}{c|}{0.973} & \multicolumn{1}{c|}{0.973} & \multicolumn{1}{c|}{0.838} & \multicolumn{1}{c|}{1} & \multicolumn{1}{c|}{0.946} & \multicolumn{1}{c|}{0.865} & \multicolumn{1}{c|}{0.892} \\ \hline
                                 &                        &                            &                            &                            &                            &                            &                            &                            &                            &                            &                            &                        &                           &                            &                            &                            &                        &                            &                            &                            \\
                                 &                        &                            &                            &                            &                            &                            &                            &                            &                            &                            &                            &                        &                           &                            &                            &                            &                        &                            &                            &                           
\end{tabular}}
\caption{Hydropathy values of amino acids. Values were taken from \cite{zheng2020hydropathy}.}
\end{table}

\subsection*{Theoretical $\nu$ value} The red dashed line values in Fig. \ref{fig:3}a were calculated according to Zheng et al. $\nu_{cal}=-0.0423\times SHD + 0.0074 \times SCD + 0.701$ \cite{zheng2020hydropathy}.

\subsection*{Monte-Carlo simulations of tethering effects on homopolymers swelling in different solvent qualities} 

We employ standard Metropolis Monte-Carlo (MC) simulations \cite{at} to study tethering effects on simple homopolymers for different solvent qualities. The homopolymer is modeled as a simple bead-spring chain with harmonic bonds between neighboring monomers $i$ and $j$ at positions $\vec r_i$ and $\vec r_j$, respectively, and the Lennard-Jones (LJ) interaction between all monomers. The spring constant of the harmonic potential, $k(|{\vec r_i} -  {\vec r_j}|-b)^2/2$ is chosen as $k=20~k_BT/\sigma^2$ in units of the thermal energy $k_BT$ and the monomer LJ size $\sigma$ which equals the bond length $b$. The LJ energy $\epsilon$ defines the solvent quality \cite{Bley}. For $\epsilon=0$ we have non-interacting ideal chains, for  $\epsilon=0.1~k_BT$ we have good solvent (SAW) behavior, while for $\epsilon=0.44~k_BT$ we have chains in a theta-solvent. We now consider a piece of 'A' chain of $N_A=20$ monomers and tether symmetrically two pieces of same length $N_B$ of a 'B'-chain to A, and average the end-to-end distance $R_{ee}$ of A as a function of thethering length $N_B$. A and B have the same interaction parameters, hence the whole chain is a homopolymer.  

\begin{figure}[h]
\centering
\includegraphics[width=0.7\linewidth]{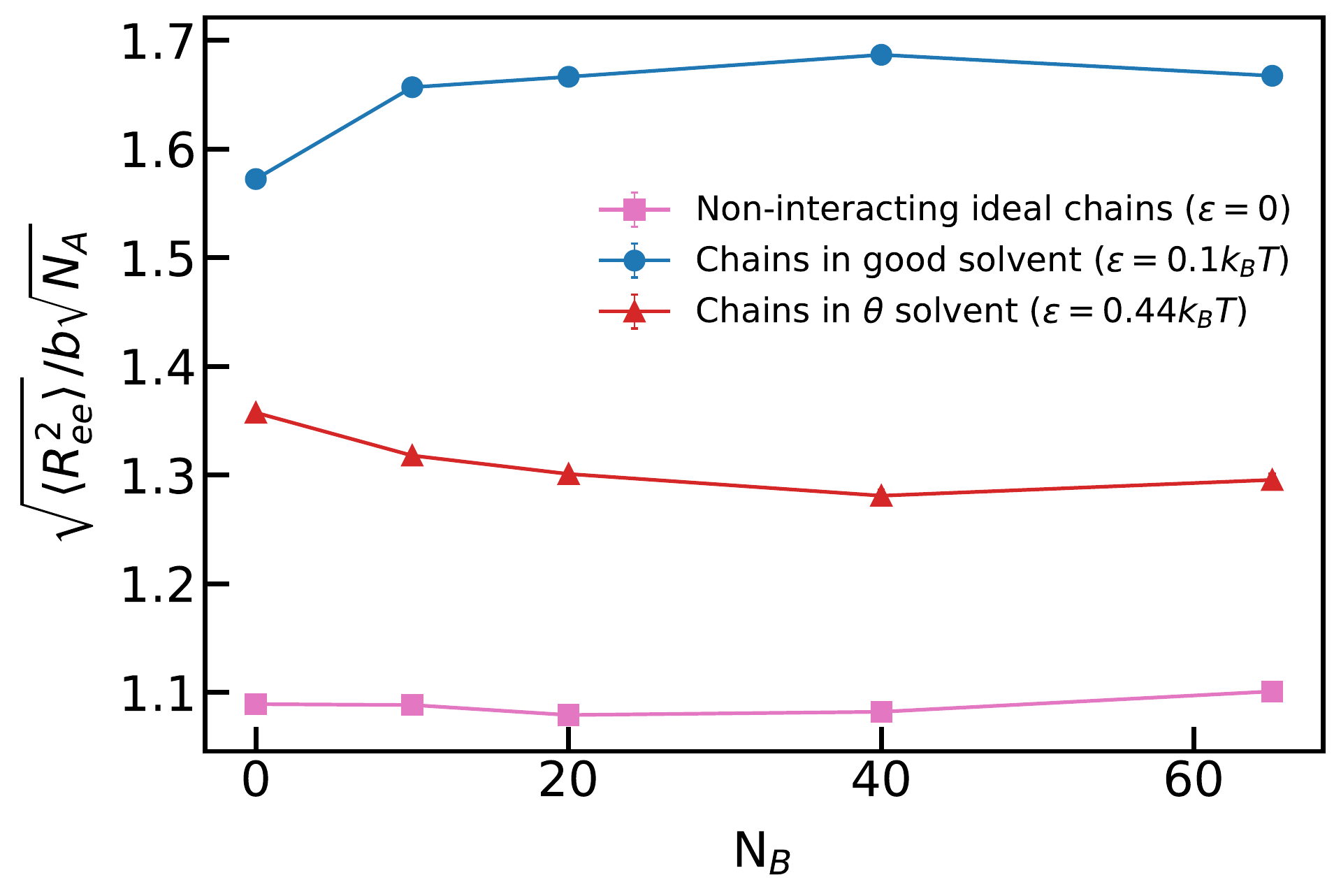}
\caption{Monte-Carlo simulation results of the effect of symmetric tethering of outer B-chains to the end-to-end distance $R_{ee}(N_B)$ a central A-chain of monomer length $N_A=20$ as a function of tethering length $N_B$. Chains A and B have the same interaction parameters and are homopolymers. For Lennard-Jones energy $\epsilon=0$ we have non-interacting ideal chains, for  $\epsilon=0.1~k_BT$ we have good solvent (SAW) behavior, while for $\epsilon=0.44~k_BT$ we have chains in a theta-solvent~\cite{Bley}. Some swelling and shrinking effects are visible for good and theta solvent, respectively, while relatively small, less than ca. 7\%. Size units are bond length equal the Lennard-Jonas monomer size, $b=\sigma$.}
\label{fig:MC}
\end{figure}

After equilibration of $10^5$ MC steps, averages are gathered in $2\cdot 10^7$ steps. The maximum MC translational step width is chosen such that the MC step acceptance ratio is about 40\%. The results are shown in Fig.~\ref{fig:MC}. Some significant swelling and shrinking effects are visible for good and theta solvent, respectively, while relatively small, less than ca. 7\%. For perfectly ideal chains we observe as expected no effect of tethering. 

 \newpage
\newpage

\section*{Supporting information}

\begin{figure}[h]
\centering
\includegraphics[width=1.0\linewidth]{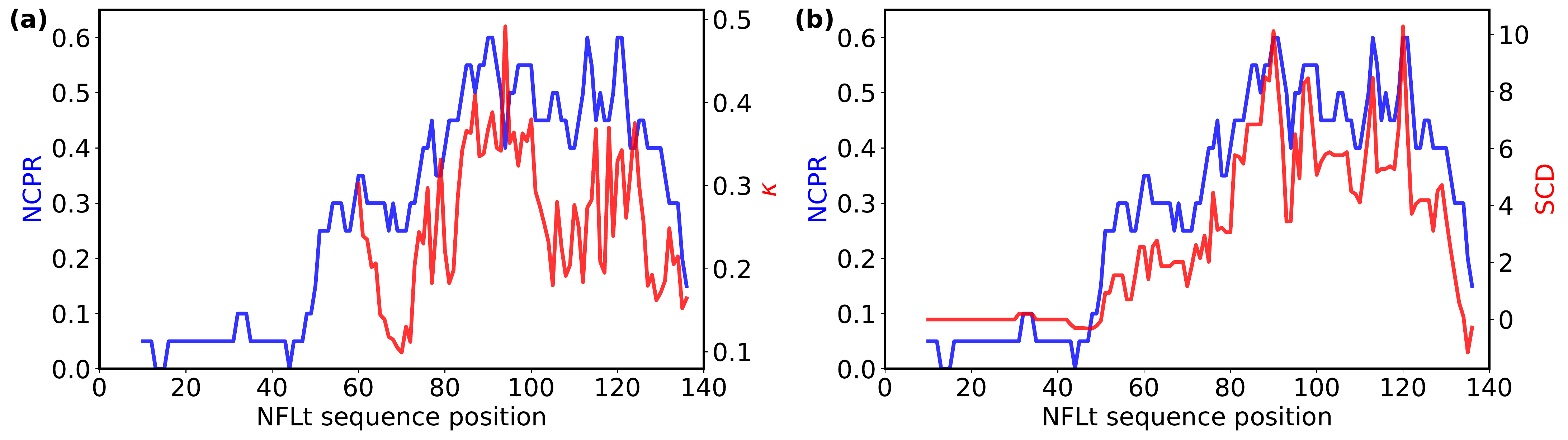}
\caption{NCPR and $\kappa$ were calculated as described here \cite{das2013conformations}. SCD was calculated as described in the method. Each value represent the calculation result of 20 amino acid segment length. $\kappa$ was calculate only for polyampholyte segments. }
\label{fig:NCPR_VS_sequence}
\end{figure}

\begin{figure}[h]
\centering
\includegraphics[width=0.6\linewidth]{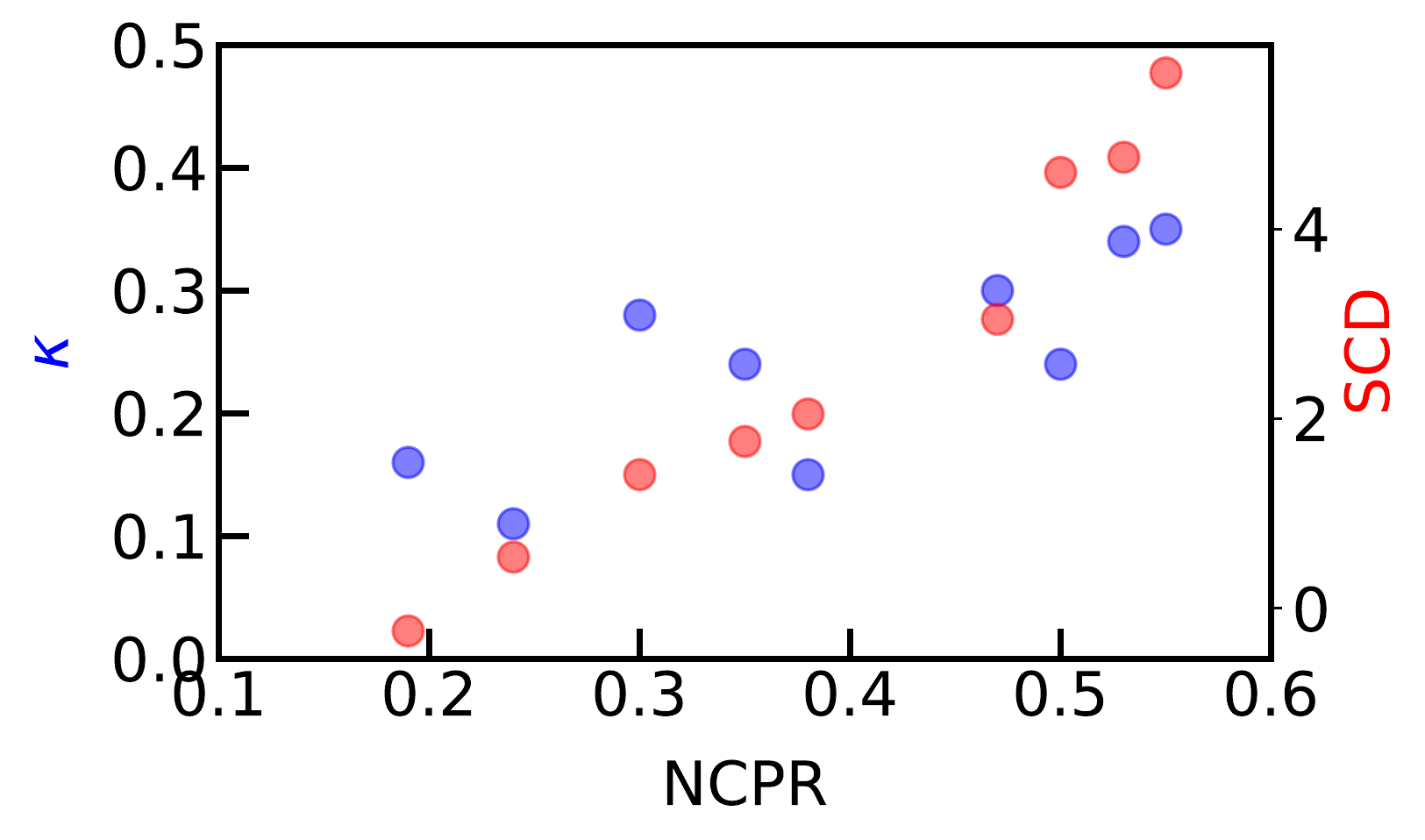}
\caption{NCPR and $\kappa$ were calculate as describe here \cite{das2013conformations}. SCD was calculated as describe is the method. Positive correlation of $\kappa$ and SCD with NCPR for different segments along the NFLt sequence. }
\label{fig:Kappa_VS_NCPR}
\end{figure}

\begin{table}[h]
    \centering
    \resizebox{\textwidth}{!}{%
    \begin{tabular}{|l|l|l|l|l|l|l|}
        \hline Name & sequence for FRET & Length & Mw & sequence for SAXS & Length & Mw \\
        \hline S(6-25) & (DNS)AFTSVGSITSGYSQSSQVFGRA(NaphA) & 23 & 2667.94 &  FTSVGSITSGYSQSSQVFGR & 20 & 2095.25 \\
         \hline S(26-45) & (DNS)ASAYSGLQSSSYLMSARSFPAA(NaphA) & 23 & 2683.01 & SAYSGLQSSSYLMSARSFPA & 20 & 2110.33 \\
         \hline S(45-64) & (DNS)AYYTSHVQEEQTEVEETIEATA(NaphA) & 23 & 2959.15 & YYTSHVQEEQTEVEETIEAT & 20 & 2386.46 \\
        \hline S(67-86) & (DNS)AKAEEAKDEPPSEGEAEEEEKA(NaphA) & 23 & 2803.94 & KAEEAKDEPPSEGEAEEEEK & 20 & 2231.27 \\
        \hline S(66-81) & (DNS)ATKAEEAKDEPPSEGE(NaphA) & 17 & 2118.26 & ATKAEEAKDEPPSEGEA & 17 & 1758.81 \\
        \hline S(87-105) & (DNS)AEKEEGEEEEGAEEEEAAKDEA(NaphA) & 23 & 2838.85 & EKEEGEEEEGAEEEEAAKDE & 20 & 2266.18 \\
        \hline S(82-96) & (DNS)AEEEEKEKEEGEEEEG(NaphA) & 17 & 2310.34 & AEEEEEKEKEEGEEEEGA & 17 & 2080.01 \\
        \hline S(106-128) & (DNS)ASEDTKEEEEGGEGEEEDTKEA(NaphA) & 23 & 2828.82 & SEDTKEEEEGGEGEEEDTKE & 20 & 2256.14 \\
        \hline S(110-125) & (DNS)AKEEEEGGEGEEEDTKE(NaphA) & 18 & 2325.36 & TKEEEEGGEGEEEDTKES & 18 & 2011.94 \\
        \hline S(129-146) & (DNS)ASEEEEKKEESAGEEQVAKKKDA(NaphA) & 24 & 2980.21 & SEEEEKKEESAGEEQVAKKKD & 21 & 2407.53 \\
        \hline S(130-143) & (DNS)AEEEEKKEESAGEEQ(NaphA) & 16 & 2152.23 & SEEEEKKEESAGEEQV & 16 & 1836.84 \\
        \hline
        
    \end{tabular}
    }
    \caption {\textbf {S Segments sequence used for trFRET and SAXS.} Two Alanine residues were added to the FRET sequence ends to insure equal spectral for the NaphtyleAlanine donor (NaphA) and the Dansyl acceptor (DNS) }
    \label{tab:S_Sequence}
\end{table}

\begin{table}[h]
    \centering
    \resizebox{0.7\textwidth}{!}{%
    \begin{tabular}{|l|l|l|}
        \hline Name & sequence & N$^a$ \\
         \hline NFLt (146 AA) & MTRLSFTSVGSITSGYSQSSQVFGRSAYSGLQSSSYLMSARSFPAY & \\
         & Y\textbf{N}SHVQEEQTEVEETIEATKAEEAKDEPPSEGEAEEEEKEKEEGEEE & \\
         & EGAEEEEAAKDESEDTKEEEEGGEGEEEDTKESEEEEKKEESAGEE & \\
         & QVAKKKD &  \\
        \hline P(6-23)  & MTRLS\textcolor{red}{CTSVGSITSGYSQSSQVW}GRSAYSGLQSSSYLMSARSFPAY &  18  \\
         & YTSHVQEEQTEVEETIEATKAEEAKDEPPSEGEAEEEEKEKEEGEEE &  \\
         & EGAEEEEAAKDESEDTKEEEEGGEGEEEDTKESEEEEKKEESAGEE &  \\
         & QVAKKKD &  \\
        \hline P(23-43)  & MTRLSFTSVGSITSGYSQSSQV\textcolor{red}{WGRSAYSGLQSSSYLMSARSC}PAY & 21  \\
         & YTSHVQEEQTEVEETIEATKAEEAKDEPPSEGEAEEEEKEKEEGEEE &  \\ 
         & EGAEEEEAAKDESEDTKEEEEGGEGEEEDTKESEEEEKKEESAGEE &  \\ 
         & QVAKKKD &  \\
        \hline P(43-64)  & MTRLSFTSVGSITSGYSQSSQVFGRSAYSGLQSSSYLMSARS\textcolor{red}{CPAYY} & 22  \\
         & \textcolor{red}{TSHVQEEQTEVEETIEW}TKAEEAKDEPPSEGEAEEEEKEKEEGEEE &  \\
         & EGAEEEEAAKDESEDTKEEEEGGEGEEEDTKESEEEEKKEESAGEE &  \\
         & QVAKKKD &  \\
        \hline P(64-80)  & MTRLSFTSVGSITSGYSQSSQVFGRSAYSGLQSSSYLMSARSFPAYY &  17  \\
         & TSHVQEEQTEVEETIE\textcolor{red}{WTKAEEAKDEPPSEGEC}EEEEKEKEEGEEE &  \\
         & EGAEEEEAAKDESEDTKEEEEGGEGEEEDTKESEEEEKKEESAGEE &  \\
         & QVAKKKD &  \\
        \hline P(80-96) & MTRLSFTSVGSITSGYSQSSQVFGRSAYSGLQSSSYLMSARSFPAYY &  17  \\
         & TSHVQEEQTEVEETIEATKAEEAKDEPPSEGE\textcolor{red}{WEEEEKEKEEGEEE} &  \\
         & \textcolor{red}{EGC}EEEEAAKDESEDTKEEEEGGEGEEEDTKESEEEEKKEESAGEE &  \\
         & QVAKKKD &  \\
        \hline P(109-126) & MTRLSFTSVGSITSGYSQSSQVFGRSAYSGLQSSSYLMSARSFPAYY &  18  \\
         & TSHVQEEQTEVEETIEATKAEEAKDEPPSEGEAEEEEKEKEEGEEEE &  \\
         & GAEEEEAAKDESED\textcolor{red}{CKEEEEGGEGEEEDTKEW}EEEEKKEESAGEE &  \\
         & QVAKKKD &  \\
        \hline P(126-141) & MTRLSFTSVGSITSGYSQSSQVFGRSAYSGLQSSSYLMSARSFPAYY &  16  \\
         & TSHVQEEQTEVEETIEATKAEEAKDEPPSEGEAEEEEKEKEEGEEEE &  \\
         & GAEEEEAAKDESEDTKEEEEGGEGEEEDTKE\textcolor{red}{WEEEEKKEESAGEE} &  \\
         & \textcolor{red}{QC}AKKKD &  \\
        \hline
    \end{tabular}
    }
    \caption{\textbf {Sequence of full NFLt and P-Segments used for trFRET.}  NFLt domain sequence is drived from 
    Neurofilament light (UniProtKB P08551) positions 
     (398 - 543). For improving purification yield we introduce a single mutation T47N. Marked in red are the segments which end's residue were replaced with Tryptophan donor and a Cysteine residue that was additionally coupled with Coumarin dye.
     $N^a$ is the number of amino acid between donor and acceptor sites.}
     
    \label{tab:P_Sequence}
\end{table}

\begin{table}[h]
    \centering
    \begin{tabular}{|l|l|l|l|l|l|}
        \hline Name & N & FCR & NCPR & $\kappa$ & $\nu_{cal}$ \\
        \hline S(6-25) & 23/20 & 0.05/0.05 & 0/0.05 &   & 0.54\\
         \hline S(26-46) & 23/20 & 0.05/0.05 & 0/0.05 &  & 0.54\\
         \hline S(45-64) & 20/20 & 0.27/0.30 & 0.3/0.3 & 0.32/0.28 & 0.57\\
        \hline S(67-86) & 23/20 & 0.59/0.65 & 0.35/0.35 & 0.24/0.24 & 0.58\\
        \hline S(66-81) & 17/17 & 0.50/0.47 & 0.29/0.24 & 0.10/0.11 & 0.58\\
        \hline S(87-105) & 23/20 & 0.68/0.75 & 0.52/0.55 & 0.38/0.35 & 0.62\\
        \hline S(82-96) & 17/17 & 0.81/0.78 & 0.59/0.53 & 0.35/0.34 & 0.63\\
        \hline S(106-128) & 23/20 & 0.64/0.70 & 0.48/0.50 & 0.26/0.24 & 0.61\\
        \hline S(110-125) & 18/18 & 0.71/0.67 & 0.50/0.47 & 0.31/0.30 & 0.61\\
        \hline S(129-146) & 24/21 & 0.61/0.67 & 0.21/0.19 & 0.20/0.16 & 0.57\\
        \hline S(130-143) & 16/16 & 0.67/0.63 & 0.44/0.38 & 0.15/0.15 & 0.61\\
        \hline Full NFLt  & 146 & 0.45 & 0.24 & 0.25  & 0.59\\
        \hline
        
    \end{tabular}
    \caption {\textbf {S segments sequence characteristics}. Values in each column are calculate for trFRET (left) and SAXS (right) segments. $\kappa$ is calculated only for polyampholytes, $\nu_{cal}=-0.0423\cdot SHD+0.0074\cdot SCD+0.701$ according to ref \cite{zheng2020hydropathy}}
    \label{tab:S_NCPR_KAPPA}
\end{table}

\begin{table}[h]
    \centering
    \begin{tabular}{|l|l|l|l|l|}
        \hline Name & N & FCR & NCPR & $\kappa$ \\
        \hline P(6-23) & 18 & 0.0 & 0.0 &   \\
         \hline P(23-43) & 21 & 0.1 & 0.1 &  \\
         \hline P(43-64) & 22 & 0.27 & 0.27 & 0.40 \\
        \hline P(64-80) & 17 & 0.47 & 0.24 & 0.12 \\
        \hline P(80-96) & 17 & 0.76 & 0.53 & 0.35 \\
        \hline P(109-126) & 18 & 0.67 & 0.44 & 0.33\\
        \hline P(126-141) & 16 & 0.69 & 0.44 & 0.15\\
        
        \hline
        
    \end{tabular}

    \caption {\textbf {P segments sequence characteristics}.  $\kappa$ is calculated only for polyampholytes} 
    \label{tab:P_NCPR_KAPPA}
\end{table}

\begin{figure}[h]
\centering
\includegraphics[width=0.9\linewidth]{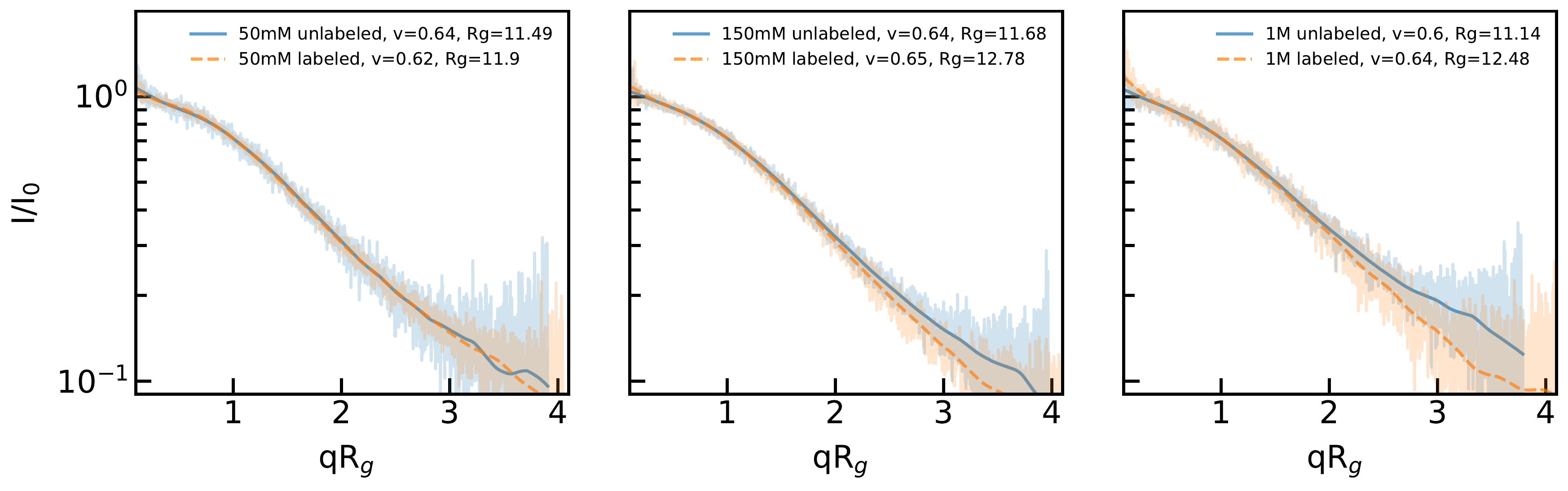}
\caption{SAXS comparison of segment S(82-96), labeled and unlabeled. \replaced{From left to right panels}{(a), (b) and (c)} at 50, 150, and 1000 mM NaCl respectively. The labeling adds to the $R_g$ about $\sim$ 1  {\AA}. Besides, the data has a similar pattern until $qR_g\geq2$, i.e., in the relevant SAXS analysis range.}
\label{fig:SI-un_and_labeled-S5p}
\end{figure}

\begin{figure}[h]
\centering
\includegraphics[width=0.8\linewidth]{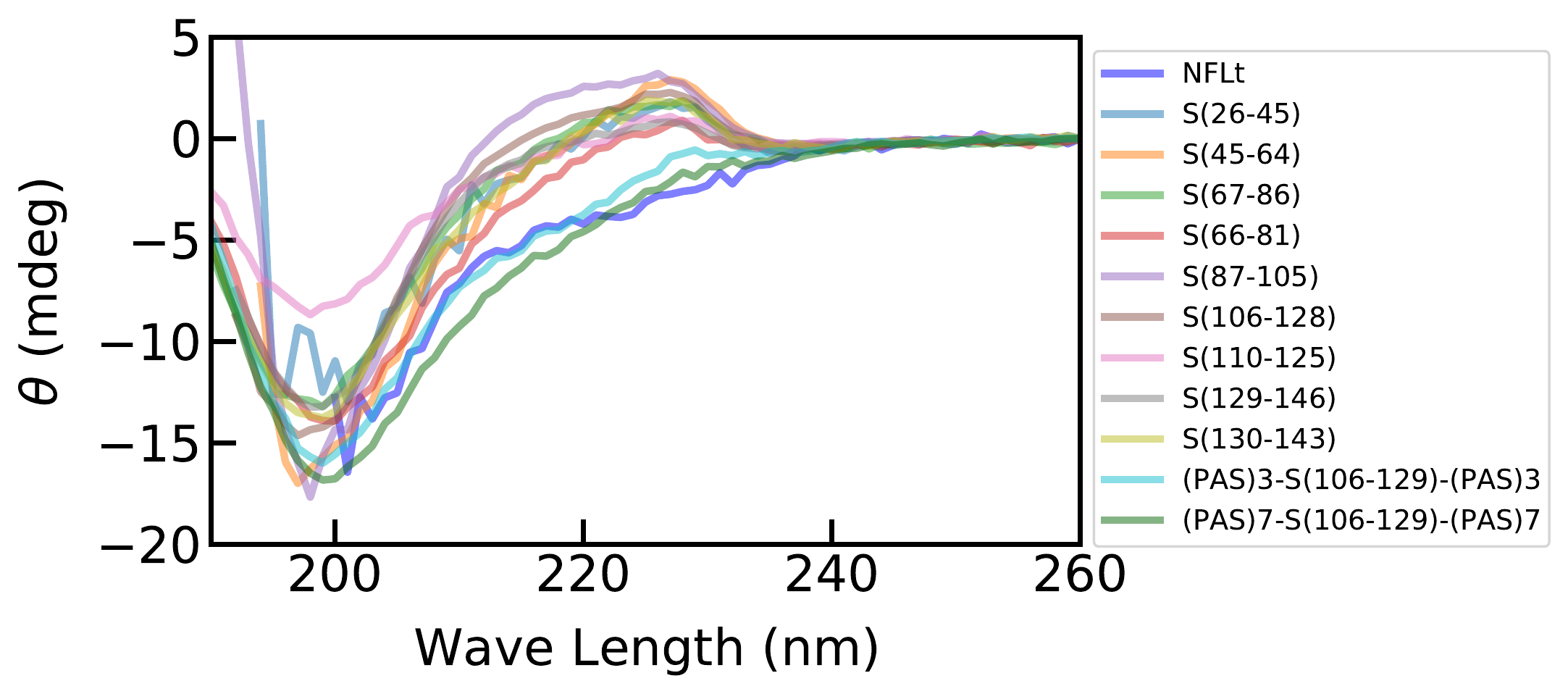}
\caption{Circular dichroism (CD) measurements of segments and full NFLt in 5mM NaPO4 buffer at pH 7.0. CD signal present random coil spectrum for all segments and full protein which indicates no secondary structure.}
\label{fig:CD}
\end{figure}

\begin{figure}[h]
\centering
\includegraphics[width=1\linewidth]{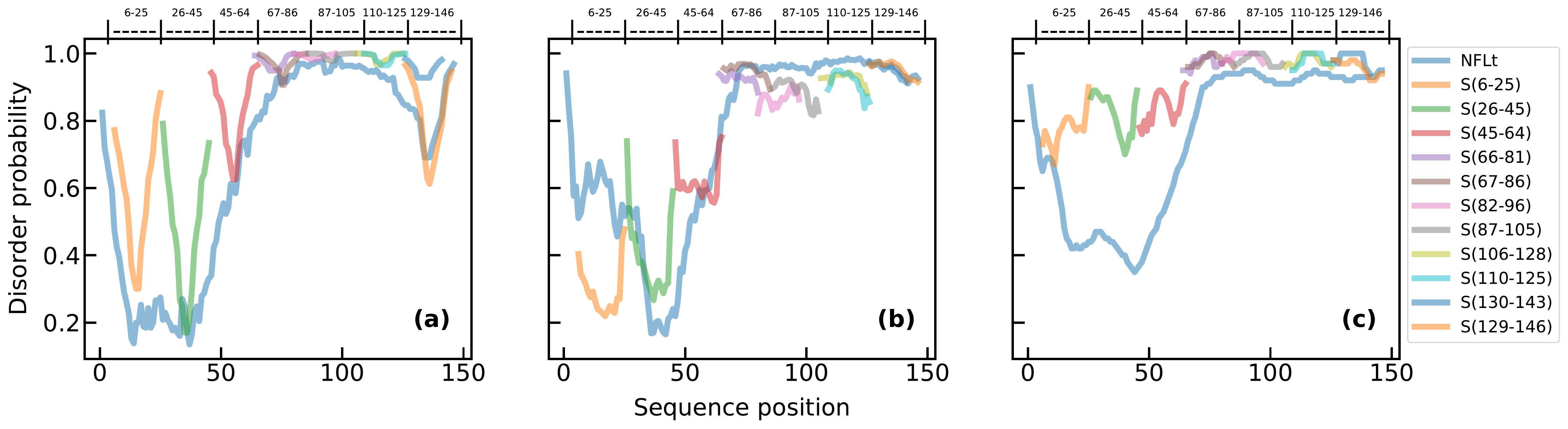}
\caption{Result of software disorder prediction, (a) NetSurfP-3.0 (b) IUPred2A (c) PrDOS. The output showing a high probability for disorder of all S segments and full NFLt protein after 60 position. Default parameters were used.}
\label{fig:PrDOS}
\end{figure}

\begin{figure}[h]
\centering
\includegraphics[width=0.5\linewidth]{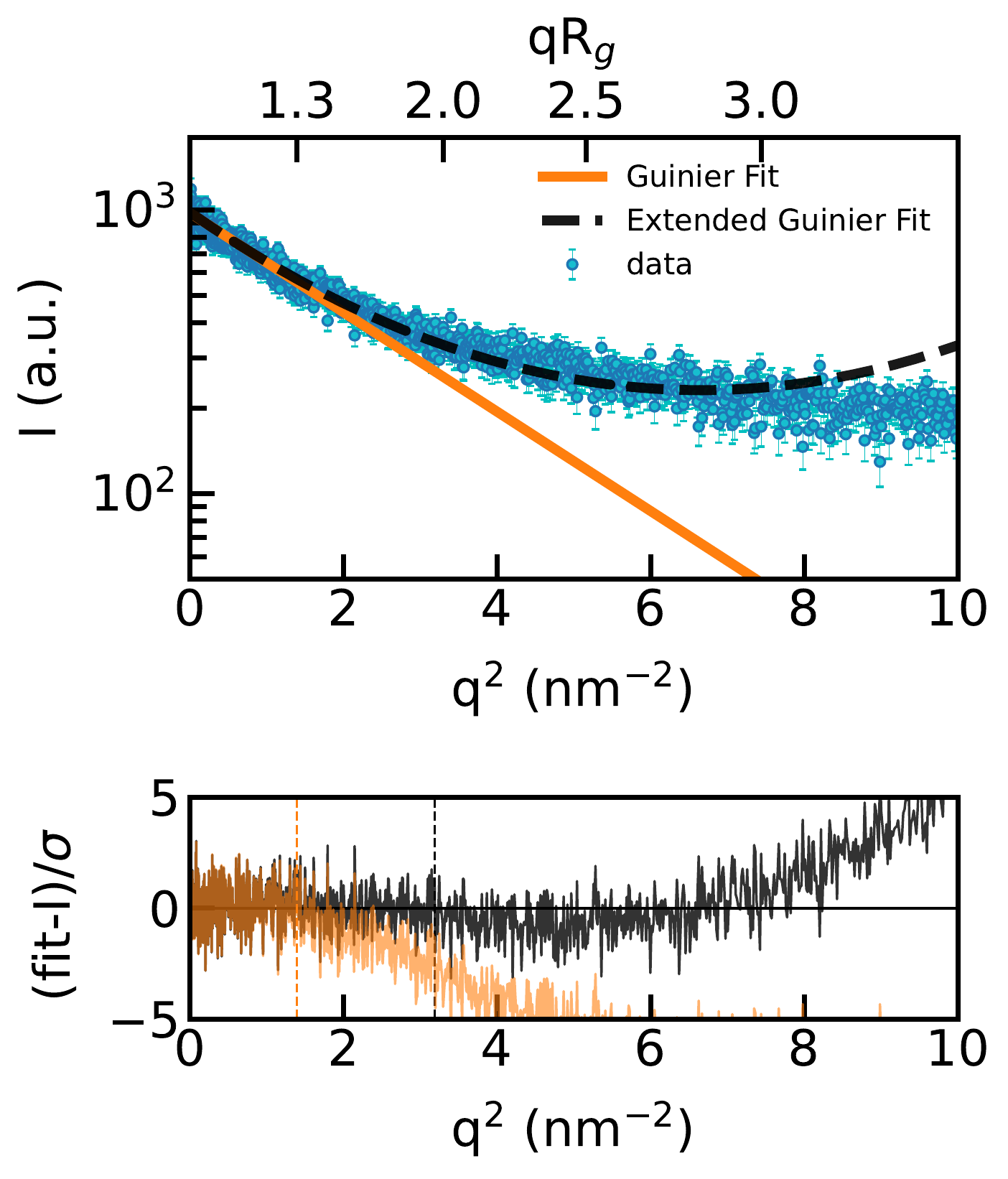}
\caption{Example for an Extended Guinier fit (black dashed line) vs. the regular Guinier (orange solid line). In the residual plot the vertical dashed lines mark the maximal range of the each fit (orange for regular Guinier and black for extended). The traditional Guinier fit, deviated at $qR_g=1.3$ while the extended fit at $qR_g=2$.}
\label{fig:Si-Ex-fit}
\end{figure}

\begin{figure}[h]
\centering
\includegraphics[width=1.0\linewidth]{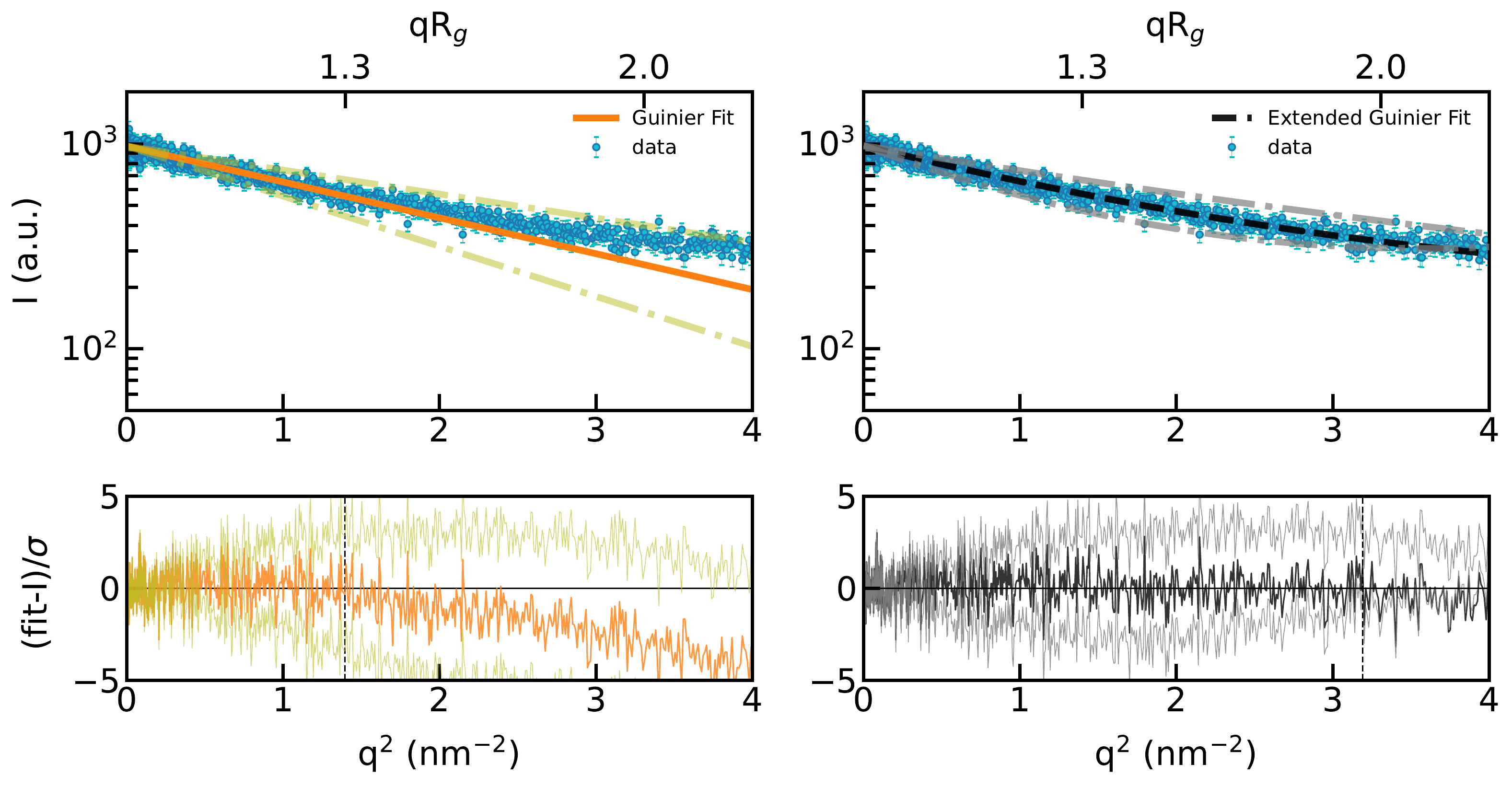}
\caption{\added{Demonstration of Guinier (left) and extended Guinier (right) fits' sensitivity (same data as in Fig.~\ref{fig:Si-Ex-fit}). The left panels demonstrate a deviation of 2 {\AA} in the $R_g$ (yellow lines) for the regular Guinier fit, and, the right panels demonstrate a deviation of 0.1 in the Flory exponent ($\nu$) (gray lines).}}
\label{fig:Si-Guinier-deviation}
\end{figure}

\begin{figure}[h]
\centering
\includegraphics[width=0.5\linewidth]{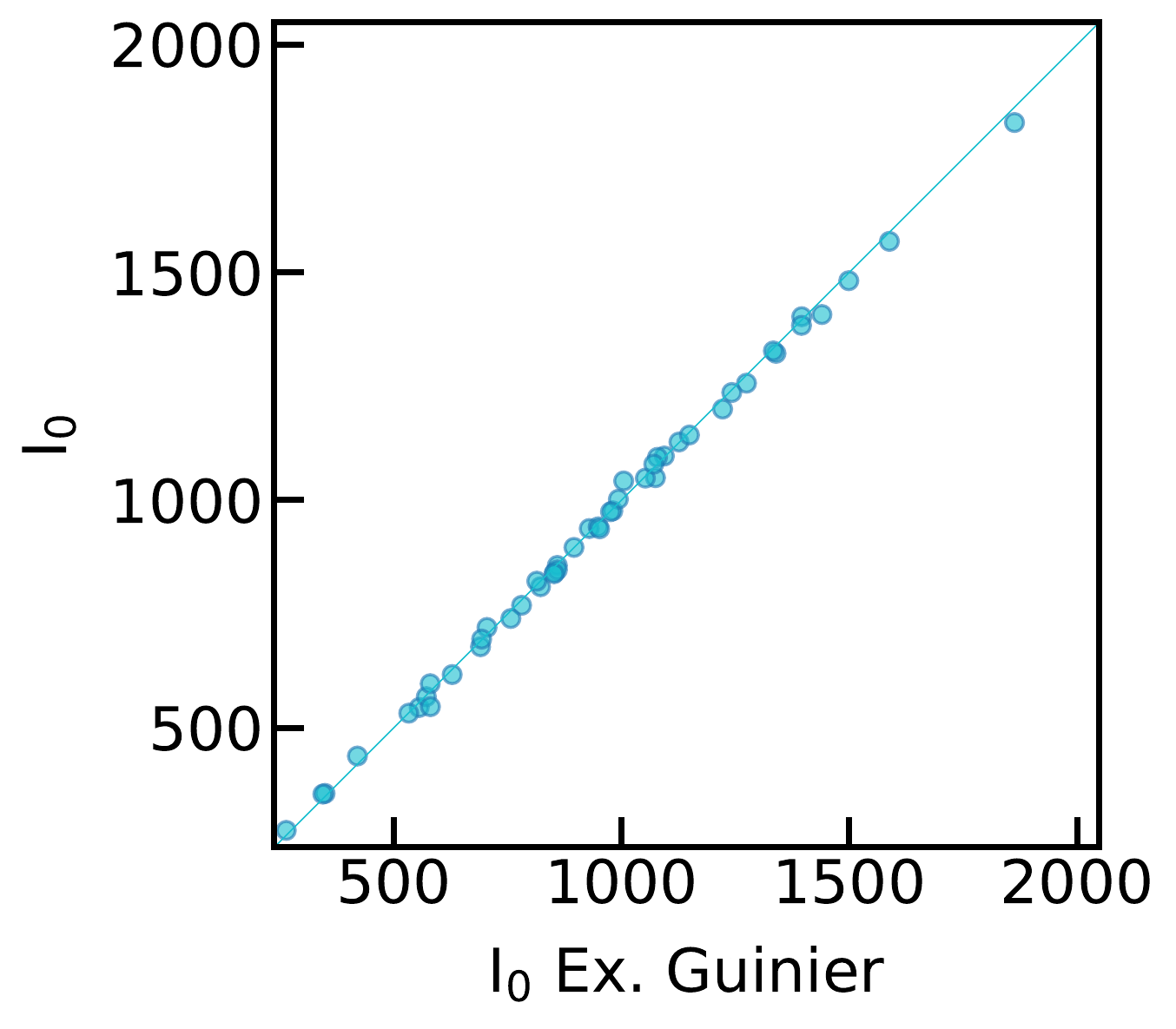}
\caption{$I_0$ obtained from the regular Guinier analysis vs. the obtained from the extended fit. The solid cyan line is the plot of $y=x$. There is an excellent agreement between $I_0$ that was extracted from the extended Guinier to the $I_0$ extracted from the regular Guinier.}
\label{fig:Si-I0}
\end{figure}

\begin{figure}[h]
\centering
\includegraphics[width=0.5\linewidth]{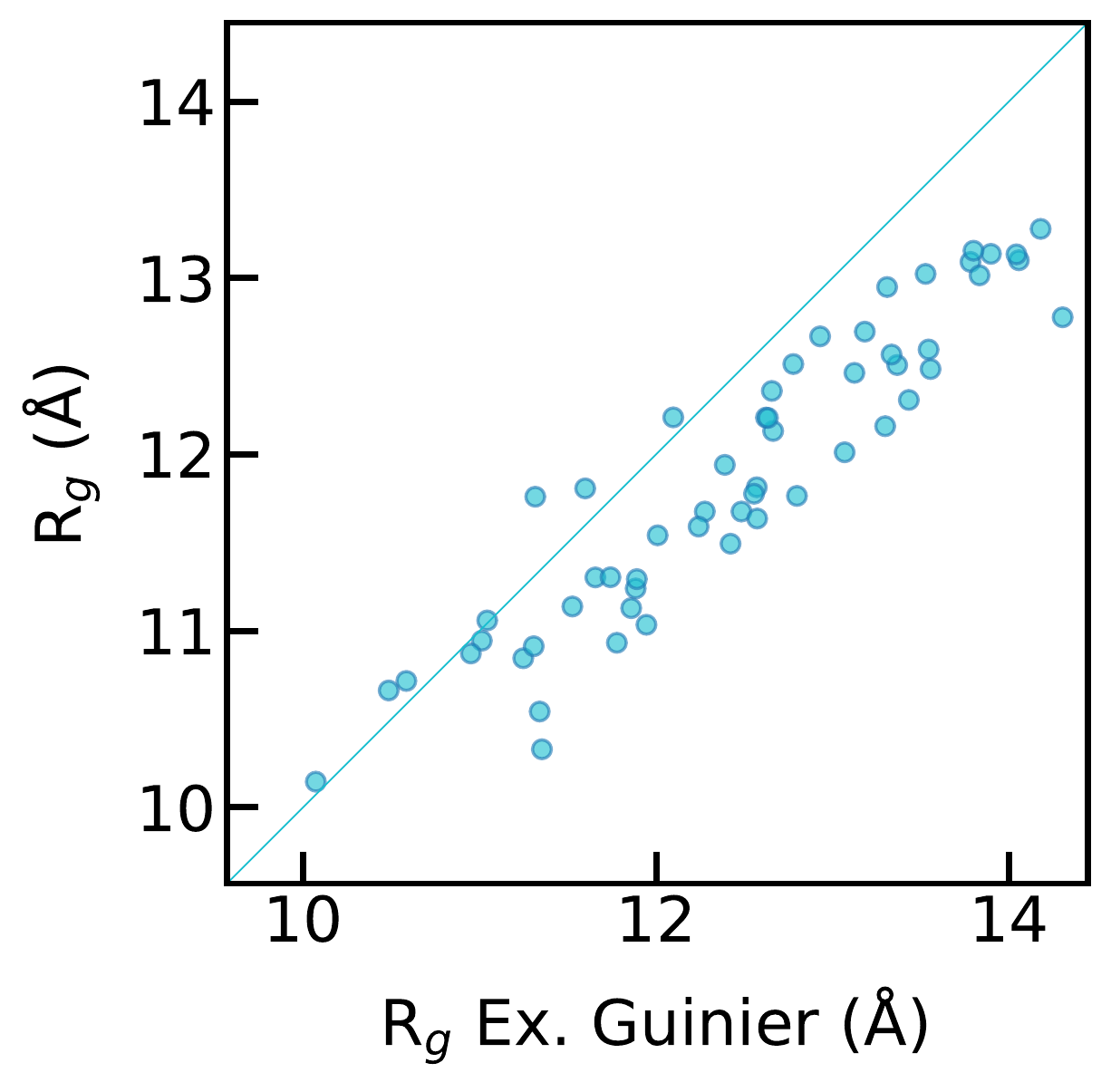}
\caption{Radius of gyration obtained from the regular Guinier analysis vs. the obtained from the extended fit. The solid cyan line is the plot of $y=x$. The regular analysis, underestimates the $R_g$. The deviation in $R_g$ is discussed in the original paper of Zheng et al. \cite{zheng2018extended}.}
\label{fig:Si-Rg}
\end{figure}

\begin{figure}[h]
\centering
\includegraphics[width=0.9\linewidth]{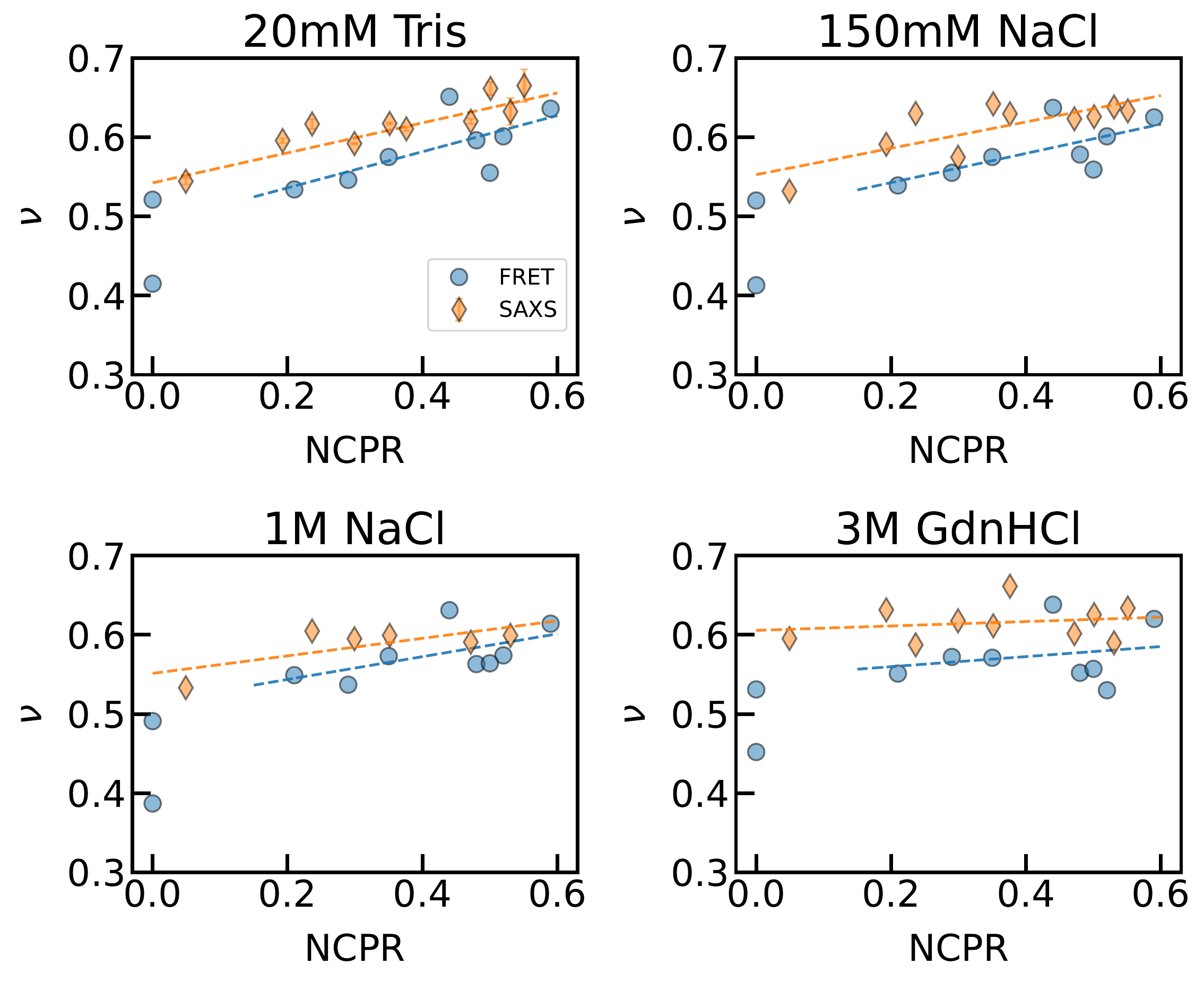}
\caption{Scaling exponent for all S segments at different conditions. Dashed line are linear fit to the data. }
\label{fig:S_segment_Nu}
\end{figure}

\begin{figure}[h]
\centering
\includegraphics[width=0.5\linewidth]{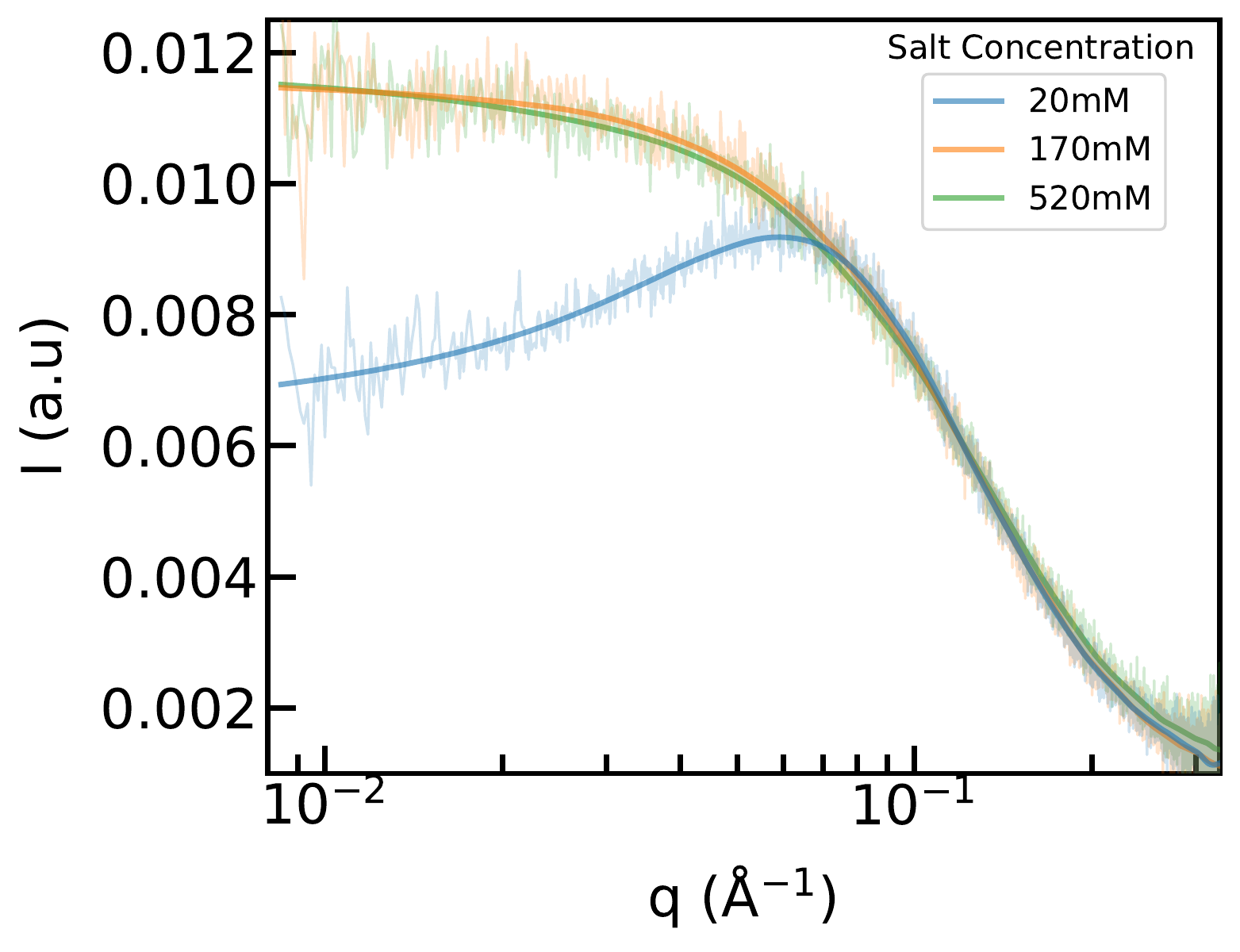}
\caption{SAXS measured intensity versus $q$ of segment S(110-125) at different salinity and peptide concentration of 4 mg/ml. Solid thick lines are moving averages. It can be seen that the peptide solution at 20 mM salt, scatters less X-ray at small angles (blue line), compared to higher salinity (orange and green lines). The shape of the blue curve is a result of inter-molecular repulsion that is screened upon salt addition. The minimal repulsion range is $2\pi/q_{peak}=$ 11 nm where $q_{peak}$ is the $q$ value of the blue curve's peak.}
\label{fig:SI-InerRepulsion_Salt}
\end{figure}

\begin{figure}[h]
\centering
\includegraphics[width=0.5\linewidth]{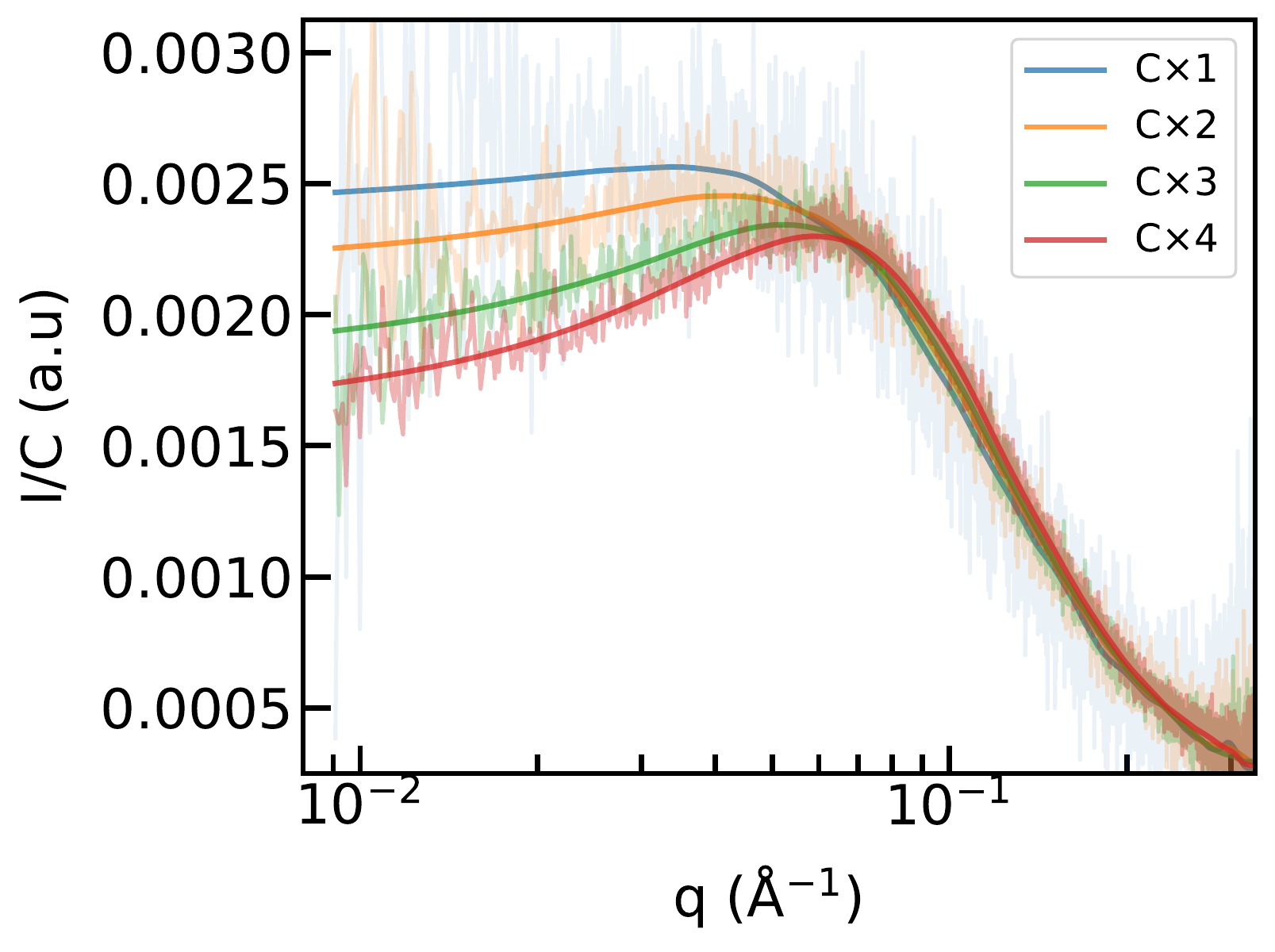}
\caption{SAXS measured intensity normalized by peptide concentration (C) versus $q$ of segment S(110-125) at 20 mM Tris buffer (without additional NaCl) and at four peptide concentrations multiples (1, 2, 3 and 4 mg/ml). Solid thick lines are moving averages. Higher peptide concentration solutions scatters less X-ray at small angles (red line), compared to lower peptide concentration (blue line). The shape of the curves is a result of inter-molecular repulsion and the peak of each curve indicates the minimal repulsion range. This repulsion range increases with higher peptide concentrations as the low angle peak shifts to higher $q$-values with higher concentrations.}
\label{fig:SI-Repulsion_concentration}
\end{figure}

\begin{figure}[h]
\centering
\includegraphics[width=0.5\linewidth]{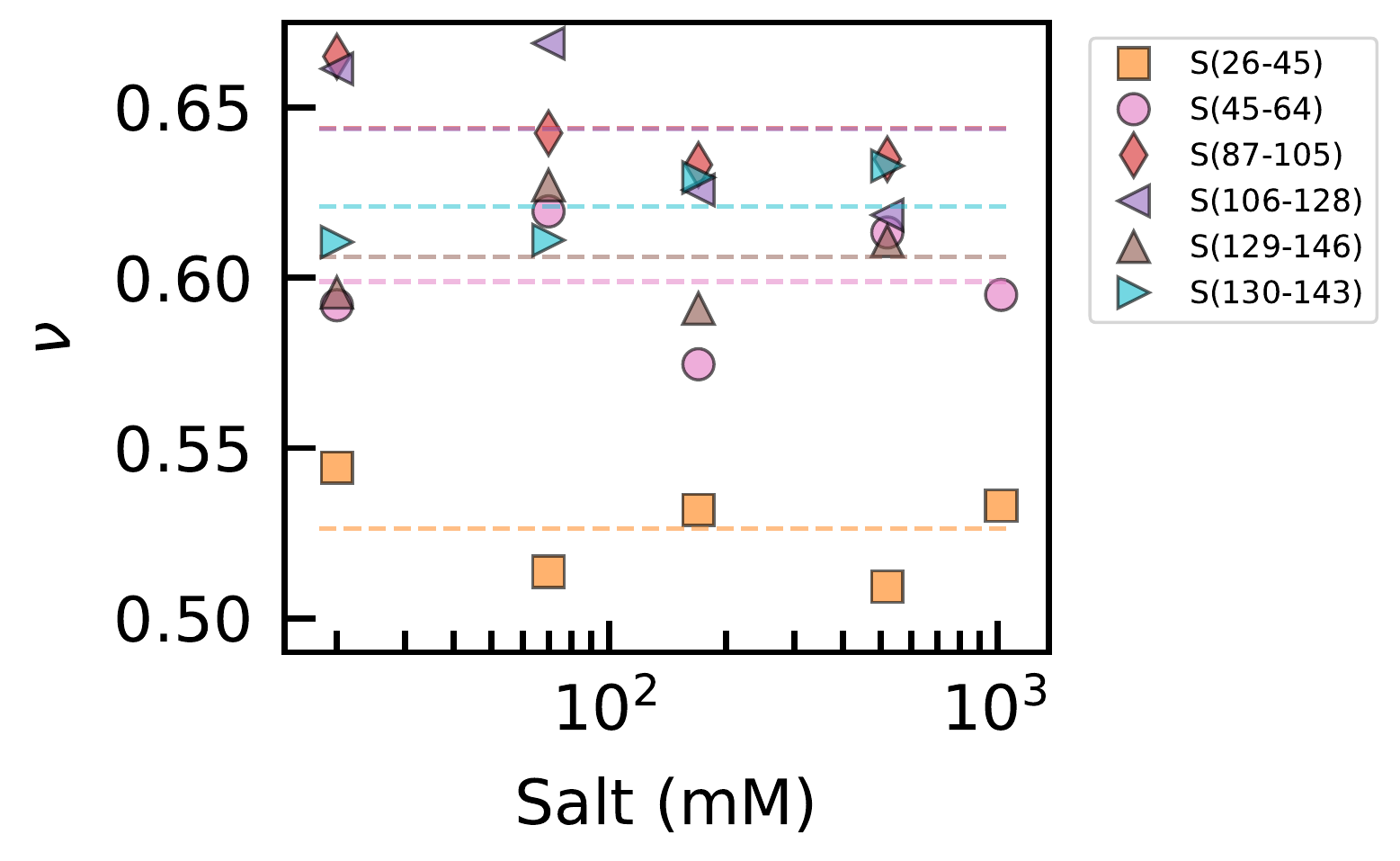}
\caption{Salt-independent S segments: Flory exponent vs. salt concentration. Dashed lines are the average $\nu$ value.}
\label{fig:SI-independentSalt}
\end{figure}

\begin{figure}[h]
\centering
\includegraphics[width=1.0\linewidth]{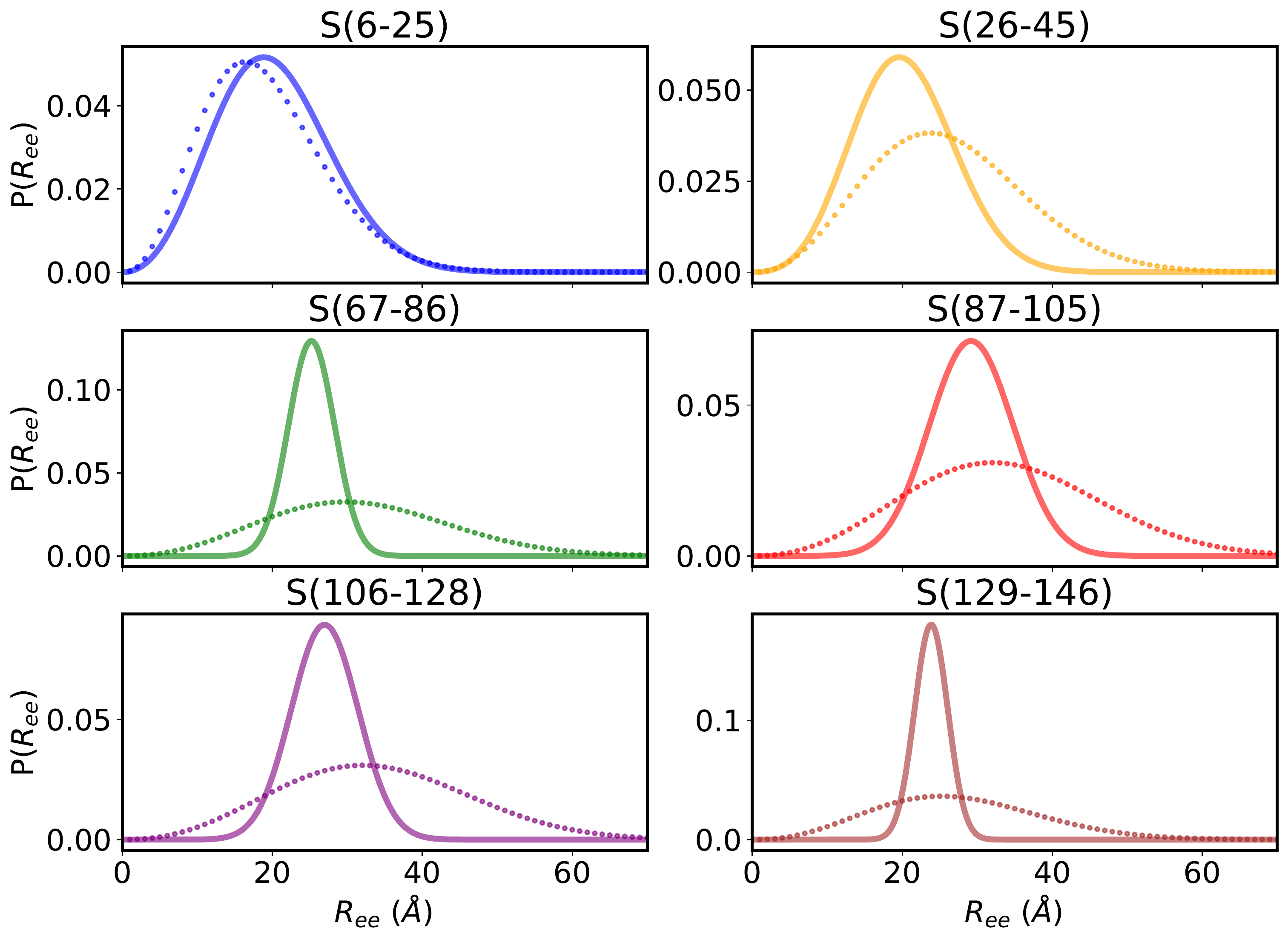}
\caption{Comparison of trFRET distance distribution obtain from the radial Gaussian model (line) and the SAW model (dot). Segments are color as in (Fig.~3c)}
\label{fig:SI-Pr-SAW_VS_Gauss}
\end{figure}

\begin{table}[h]
\resizebox{\textwidth}{!}{%
\begin{tabular}{|c|cccc|cccc|cccc|cccc|}
\hline
           & \multicolumn{4}{c|}{0M NaCl}                                                                                      & \multicolumn{4}{c|}{0.15M NaCl}                                                                                   & \multicolumn{4}{c|}{1M NaCl}                                                                                      & \multicolumn{4}{c|}{3M GdnHcl}                                                                                    \\ \hline
Segment    & \multicolumn{1}{c|}{$\tau_{DO}$} & \multicolumn{1}{c|}{$\tau_{DA}$} & \multicolumn{1}{c|}{\textless{}ET\textgreater{}} & $\nu$   & \multicolumn{1}{c|}{$\tau_{DO}$} & \multicolumn{1}{c|}{$\tau_{DA}$} & \multicolumn{1}{c|}{\textless{}ET\textgreater{}} & $\nu$   & \multicolumn{1}{c|}{$\tau_{DO}$} & \multicolumn{1}{c|}{$\tau_{DA}$} & \multicolumn{1}{c|}{\textless{}ET\textgreater{}} & $\nu$   & \multicolumn{1}{c|}{$\tau_{DO}$} & \multicolumn{1}{c|}{$\tau_{DA}$} & \multicolumn{1}{c|}{\textless{}ET\textgreater{}} & $\nu$   \\ \hline
S(6-25)    & \multicolumn{1}{c|}{21.2}  & \multicolumn{1}{c|}{5.5}   & \multicolumn{1}{c|}{74.2}                        & 0.42 & \multicolumn{1}{c|}{22.0}  & \multicolumn{1}{c|}{5.6}   & \multicolumn{1}{c|}{74.6}                        & 0.41 & \multicolumn{1}{c|}{18.7}  & \multicolumn{1}{c|}{3.8}   & \multicolumn{1}{c|}{79.7}                        & 0.39 & \multicolumn{1}{c|}{20.1}  & \multicolumn{1}{c|}{6.9}   & \multicolumn{1}{c|}{65.9}                        & 0.45 \\ \hline
S(26-46)   & \multicolumn{1}{c|}{20.0}  & \multicolumn{1}{c|}{10.3}  & \multicolumn{1}{c|}{48.5}                        & 0.52 & \multicolumn{1}{c|}{20.5}  & \multicolumn{1}{c|}{10.5}  & \multicolumn{1}{c|}{48.8}                        & 0.52 & \multicolumn{1}{c|}{19.0}  & \multicolumn{1}{c|}{8.3}   & \multicolumn{1}{c|}{56.3}                        & 0.49 & \multicolumn{1}{c|}{23.0}  & \multicolumn{1}{c|}{12.4}  & \multicolumn{1}{c|}{45.9}                        & 0.53 \\ \hline
S(67-86)   & \multicolumn{1}{c|}{38.3}  & \multicolumn{1}{c|}{24.9}  & \multicolumn{1}{c|}{35.0}                        & 0.58 & \multicolumn{1}{c|}{38.0}  & \multicolumn{1}{c|}{24.7}  & \multicolumn{1}{c|}{35.0}                        & 0.58 & \multicolumn{1}{c|}{36.5}  & \multicolumn{1}{c|}{23.5}  & \multicolumn{1}{c|}{35.6}                        & 0.57 & \multicolumn{1}{c|}{30.3}  & \multicolumn{1}{c|}{19.4}  & \multicolumn{1}{c|}{36.0}                        & 0.57 \\ \hline
S(66-81)   & \multicolumn{1}{c|}{38.3}  & \multicolumn{1}{c|}{16.9}  & \multicolumn{1}{c|}{55.9}                        & 0.55 & \multicolumn{1}{c|}{38.0}  & \multicolumn{1}{c|}{17.6}  & \multicolumn{1}{c|}{53.8}                        & 0.56 & \multicolumn{1}{c|}{36.5}  & \multicolumn{1}{c|}{15.3}  & \multicolumn{1}{c|}{58.0}                        & 0.54 & \multicolumn{1}{c|}{30.3}  & \multicolumn{1}{c|}{15.3}  & \multicolumn{1}{c|}{49.6}                        & 0.57 \\ \hline
S(87-105)  & \multicolumn{1}{c|}{37.3}  & \multicolumn{1}{c|}{26.4}  & \multicolumn{1}{c|}{29.1}                        & 0.60 & \multicolumn{1}{c|}{37.0}  & \multicolumn{1}{c|}{26.3}  & \multicolumn{1}{c|}{29.1}                        & 0.60 & \multicolumn{1}{c|}{35.7}  & \multicolumn{1}{c|}{23.1}  & \multicolumn{1}{c|}{35.2}                        & 0.57 & \multicolumn{1}{c|}{29.3}  & \multicolumn{1}{c|}{15.8}  & \multicolumn{1}{c|}{46.2}                        & 0.53 \\ \hline
S(82-96)   & \multicolumn{1}{c|}{37.3}  & \multicolumn{1}{c|}{24.4}  & \multicolumn{1}{c|}{34.5}                        & 0.64 & \multicolumn{1}{c|}{37.0}  & \multicolumn{1}{c|}{23.3}  & \multicolumn{1}{c|}{37.1}                        & 0.63 & \multicolumn{1}{c|}{35.7}  & \multicolumn{1}{c|}{21.5}  & \multicolumn{1}{c|}{39.7}                        & 0.61 & \multicolumn{1}{c|}{28.5}  & \multicolumn{1}{c|}{17.6}  & \multicolumn{1}{c|}{38.3}                        & 0.62 \\ \hline
S(106-128) & \multicolumn{1}{c|}{34.1}  & \multicolumn{1}{c|}{23.8}  & \multicolumn{1}{c|}{30.2}                        & 0.60 & \multicolumn{1}{c|}{33.8}  & \multicolumn{1}{c|}{22.2}  & \multicolumn{1}{c|}{34.3}                        & 0.58 & \multicolumn{1}{c|}{33.1}  & \multicolumn{1}{c|}{20.6}  & \multicolumn{1}{c|}{38.0}                        & 0.56 & \multicolumn{1}{c|}{28.1}  & \multicolumn{1}{c|}{16.7}  & \multicolumn{1}{c|}{40.7}                        & 0.55 \\ \hline
S(110-125) & \multicolumn{1}{c|}{34.1}  & \multicolumn{1}{c|}{16.7}  & \multicolumn{1}{c|}{51.1}                        & 0.56 & \multicolumn{1}{c|}{33.8}  & \multicolumn{1}{c|}{16.9}  & \multicolumn{1}{c|}{49.9}                        & 0.56 & \multicolumn{1}{c|}{33.1}  & \multicolumn{1}{c|}{17.0}  & \multicolumn{1}{c|}{48.8}                        & 0.56 & \multicolumn{1}{c|}{28.1}  & \multicolumn{1}{c|}{14.0}  & \multicolumn{1}{c|}{50.4}                        & 0.56 \\ \hline
S(129-146) & \multicolumn{1}{c|}{36.1}  & \multicolumn{1}{c|}{20.4}  & \multicolumn{1}{c|}{43.5}                        & 0.53 & \multicolumn{1}{c|}{35.7}  & \multicolumn{1}{c|}{20.6}  & \multicolumn{1}{c|}{42.3}                        & 0.54 & \multicolumn{1}{c|}{34.2}  & \multicolumn{1}{c|}{20.6}  & \multicolumn{1}{c|}{39.6}                        & 0.55 & \multicolumn{1}{c|}{28.7}  & \multicolumn{1}{c|}{17.4}  & \multicolumn{1}{c|}{39.2}                        & 0.55 \\ \hline
S(130-143) & \multicolumn{1}{c|}{36.1}  & \multicolumn{1}{c|}{23.7}  & \multicolumn{1}{c|}{34.3}                        & 0.65 & \multicolumn{1}{c|}{35.7}  & \multicolumn{1}{c|}{22.4}  & \multicolumn{1}{c|}{37.4}                        & 0.64 & \multicolumn{1}{c|}{34.2}  & \multicolumn{1}{c|}{20.9}  & \multicolumn{1}{c|}{38.8}                        & 0.63 & \multicolumn{1}{c|}{28.7}  & \multicolumn{1}{c|}{18.0}  & \multicolumn{1}{c|}{37.1}                        & 0.64 \\ \hline
\end{tabular}
}

\caption {\textbf {S segments scaling exponent determine by fitting trFRET measurements with SAW model.} $<ET>$ is calculate by $1-\tau_{DA}/\tau_{DO}$ where $\tau_{DO}$ and $\tau_{DA}$ are the average fluorescence life time of segments containing donor only or donor and acceptor , respectively. Lifetime values  presented in [nsec] } 
\label{tab:S_FRET_Nu}
\end{table}

\begin{table}[h]
\resizebox{\textwidth}{!}{%
\begin{tabular}{|llllllllllllllll|}
\hline
\multicolumn{1}{|l|}{}           & \multicolumn{5}{c|}{20 mM Tris}                                                                                                                                                                                                                                              & \multicolumn{5}{c|}{50 mM NaCl}                                                                                                                                                                                                                                              & \multicolumn{5}{c|}{150 mM NaCl}                                                                                                                                                                                                                        \\ \hline
\multicolumn{1}{|l|}{Name}       & \multicolumn{1}{l|}{$R_g$} 
& \multicolumn{1}{l|}{$R_g$ ex.} 
& \multicolumn{1}{l|}{$\nu$} & 
\multicolumn{1}{l|}{$R_g$ EOM} & \multicolumn{1}{l|}{$R_{ee}$ EOM} & \multicolumn{1}{l|}{$R_g$} & 
\multicolumn{1}{l|}{$R_g$ ex.} & 
 \multicolumn{1}{l|}{$\nu$} & 
 \multicolumn{1}{l|}{$R_g$ EOM} & \multicolumn{1}{l|}{$R_{ee}$ EOM} & \multicolumn{1}{l|}{$R_g$} & 
 \multicolumn{1}{l|}{$R_g$ ex.} & 
  \multicolumn{1}{l|}{$\nu$} & 
 \multicolumn{1}{l|}{$R_g$ EOM} & $R_{ee}$ EOM \\ \hline
\multicolumn{1}{|l|}{S(26-45)}   & \multicolumn{1}{l|}{11.76\scriptsize{$\pm$0.51}} &
\multicolumn{1}{l|}{11.31\scriptsize{$\pm$0.11}}     
& \multicolumn{1}{l|}{0.544\scriptsize{$\pm$0.004}} & 
\multicolumn{1}{l|}{12.98}     & \multicolumn{1}{l|}{33.6}         & \multicolumn{1}{l|}{10.72\scriptsize{$\pm$0.38}} & 
\multicolumn{1}{l|}{10.58\scriptsize{$\pm$0.07}}     & 
\multicolumn{1}{l|}{0.514\scriptsize{$\pm$0.003}} & 
\multicolumn{1}{l|}{12.6}      & \multicolumn{1}{l|}{31.7}         & \multicolumn{1}{l|}{10.94\scriptsize{$\pm$0.63}} & 
\multicolumn{1}{l|}{11.01\scriptsize{$\pm$0.08}}     & 
\multicolumn{1}{l|}{0.532\scriptsize{$\pm$0.003}} & 
\multicolumn{1}{l|}{12.82}     & 32.79        \\ \hline
\multicolumn{1}{|l|}{S(45-64)}   & \multicolumn{1}{l|}{11.81\scriptsize{$\pm$0.15}} & 
\multicolumn{1}{l|}{12.57\scriptsize{$\pm$0.03}}     
& \multicolumn{1}{l|}{0.592\scriptsize{$\pm$0.001}} & 
\multicolumn{1}{l|}{13.29}     & \multicolumn{1}{l|}{31.73}        & \multicolumn{1}{l|}{12.51\scriptsize{$\pm$0.70}} & 
\multicolumn{1}{l|}{13.37\scriptsize{$\pm$0.01}}     & 
\multicolumn{1}{l|}{0.620\scriptsize{$\pm$0.000}} & 
\multicolumn{1}{l|}{13.86}     & \multicolumn{1}{l|}{34.68}        & \multicolumn{1}{l|}{12.21\scriptsize{$\pm$0.72}} & 
\multicolumn{1}{l|}{12.10\scriptsize{$\pm$0.19}}     & 
\multicolumn{1}{l|}{0.575\scriptsize{$\pm$0.007}} & 
\multicolumn{1}{l|}{13.91}     & 34.94        \\ \hline
\multicolumn{1}{|l|}{S(67-86)}   & \multicolumn{1}{l|}{12.16\scriptsize{$\pm$1.27}} & 
\multicolumn{1}{l|}{13.30\scriptsize{$\pm$0.04}}     
& \multicolumn{1}{l|}{0.617\scriptsize{$\pm$0.001}} & 
\multicolumn{1}{l|}{13.87}     & \multicolumn{1}{l|}{31.27}        & \multicolumn{1}{l|}{12.31\scriptsize{$\pm$0.63}} & 
\multicolumn{1}{l|}{13.43\scriptsize{$\pm$0.04}}     & 
\multicolumn{1}{l|}{0.622\scriptsize{$\pm$0.001}} & 
\multicolumn{1}{l|}{13.77}     & \multicolumn{1}{l|}{30.86}        & \multicolumn{1}{l|}{13.10\scriptsize{$\pm$0.26}} & 
\multicolumn{1}{l|}{14.06\scriptsize{$\pm$0.03}}     & 
\multicolumn{1}{l|}{0.642\scriptsize{$\pm$0.001}} & 
\multicolumn{1}{l|}{14.34}     & 34.11        \\ \hline
\multicolumn{1}{|l|}{S(66-81)}   & \multicolumn{1}{l|}{11.03\scriptsize{$\pm$0.77}} & 
\multicolumn{1}{l|}{11.94\scriptsize{$\pm$0.16}}     
& \multicolumn{1}{l|}{0.617\scriptsize{$\pm$0.006}} & 
\multicolumn{1}{l|}{12.67}     & \multicolumn{1}{l|}{31.02}        & \multicolumn{1}{l|}{11.24\scriptsize{$\pm$0.17}} & 
\multicolumn{1}{l|}{11.88\scriptsize{$\pm$0.01}}     & 
\multicolumn{1}{l|}{0.614\scriptsize{$\pm$0.000}} & 
\multicolumn{1}{l|}{12.82}     & \multicolumn{1}{l|}{31.72}        & \multicolumn{1}{l|}{11.68\scriptsize{$\pm$0.23}} & 
\multicolumn{1}{l|}{12.28\scriptsize{$\pm$0.00}}     & 
\multicolumn{1}{l|}{0.630\scriptsize{$\pm$0.000}} & 
\multicolumn{1}{l|}{12.87}     & 32.31        \\ \hline
\multicolumn{1}{|l|}{S(87-105)}  & \multicolumn{1}{l|}{15.53\scriptsize{$\pm$1.21}} & 
\multicolumn{1}{l|}{15.13\scriptsize{$\pm$0.73}}     
& \multicolumn{1}{l|}{0.665\scriptsize{$\pm$0.021}} 
& 
\multicolumn{1}{l|}{14.4}      & \multicolumn{1}{l|}{32.52}        & \multicolumn{1}{l|}{12.21\scriptsize{$\pm$0.35}} & 
\multicolumn{1}{l|}{14.54\scriptsize{$\pm$1.12}}     & 
\multicolumn{1}{l|}{0.643\scriptsize{$\pm$0.033}} & 
\multicolumn{1}{l|}{14.42}     & \multicolumn{1}{l|}{33.35}        & \multicolumn{1}{l|}{13.09\scriptsize{$\pm$1.53}} & 
\multicolumn{1}{l|}{13.78\scriptsize{$\pm$0.11}}     & 
\multicolumn{1}{l|}{0.633\scriptsize{$\pm$0.003}} & 
\multicolumn{1}{l|}{14.58}     & 33.8         \\ \hline
\multicolumn{1}{|l|}{S(82-96)}   & \multicolumn{1}{l|}{10.32\scriptsize{$\pm$0.51}} & 
\multicolumn{1}{l|}{11.35\scriptsize{$\pm$0.46}}     
& \multicolumn{1}{l|}{0.632\scriptsize{$\pm$0.017}} & 
\multicolumn{1}{l|}{12.82}     & \multicolumn{1}{l|}{29.17}        & \multicolumn{1}{l|}{11.49\scriptsize{$\pm$0.63}} & 
\multicolumn{1}{l|}{12.42\scriptsize{$\pm$0.03}}     & 
\multicolumn{1}{l|}{0.636\scriptsize{$\pm$0.001}} & 
\multicolumn{1}{l|}{13.02}     & \multicolumn{1}{l|}{29.69}        & \multicolumn{1}{l|}{11.68\scriptsize{$\pm$0.27}} & 
\multicolumn{1}{l|}{12.48\scriptsize{$\pm$0.03}}     & 
\multicolumn{1}{l|}{0.638\scriptsize{$\pm$0.001}} & 
\multicolumn{1}{l|}{13.14}     & 30.86        \\ \hline
\multicolumn{1}{|l|}{S(106-128)} & \multicolumn{1}{l|}{13.39\scriptsize{$\pm$0.60}} & 
\multicolumn{1}{l|}{15.19\scriptsize{$\pm$0.29}}     
& \multicolumn{1}{l|}{0.661\scriptsize{$\pm$0.008}} & 
\multicolumn{1}{l|}{14.26}     & \multicolumn{1}{l|}{33.32}        & \multicolumn{1}{l|}{16.43\scriptsize{$\pm$0.93}} & 
\multicolumn{1}{l|}{15.60\scriptsize{$\pm$0.94}}     & 
\multicolumn{1}{l|}{0.669\scriptsize{$\pm$0.026}} & 
\multicolumn{1}{l|}{14.42}     & \multicolumn{1}{l|}{33.27}        & \multicolumn{1}{l|}{12.48\scriptsize{$\pm$1.25}} & 
\multicolumn{1}{l|}{13.56\scriptsize{$\pm$0.08}}     & 
\multicolumn{1}{l|}{0.626\scriptsize{$\pm$0.002}} & 
\multicolumn{1}{l|}{14.19}     & 31.7         \\ \hline
\multicolumn{1}{|l|}{S(110-125)} & \multicolumn{1}{l|}{11.40\scriptsize{$\pm$0.69}} & 
\multicolumn{1}{l|}{11.96\scriptsize{$\pm$0.09}}     
& \multicolumn{1}{l|}{0.620\scriptsize{$\pm$0.003}} & 
\multicolumn{1}{l|}{13.15}     & \multicolumn{1}{l|}{30.33}        & \multicolumn{1}{l|}{11.76\scriptsize{$\pm$0.62}} & 
\multicolumn{1}{l|}{12.80\scriptsize{$\pm$0.03}}     & 
\multicolumn{1}{l|}{0.631\scriptsize{$\pm$0.001}} & 
\multicolumn{1}{l|}{13.26}     & \multicolumn{1}{l|}{29.46}        & \multicolumn{1}{l|}{11.64\scriptsize{$\pm$0.43}} & 
\multicolumn{1}{l|}{12.57\scriptsize{$\pm$0.04}}     & 
\multicolumn{1}{l|}{0.623\scriptsize{$\pm$0.002}} & 
\multicolumn{1}{l|}{13.44}     & 30.4         \\ \hline
\multicolumn{1}{|l|}{S(129-146)} & \multicolumn{1}{l|}{12.01\scriptsize{$\pm$0.92}} & 
\multicolumn{1}{l|}{13.07\scriptsize{$\pm$0.10}}     
& \multicolumn{1}{l|}{0.596\scriptsize{$\pm$0.003}} & 
\multicolumn{1}{l|}{13.85}     & \multicolumn{1}{l|}{29.91}        & \multicolumn{1}{l|}{13.13\scriptsize{$\pm$1.08}} & 
\multicolumn{1}{l|}{14.04\scriptsize{$\pm$0.10}}     & 
\multicolumn{1}{l|}{0.627\scriptsize{$\pm$0.003}} & 
\multicolumn{1}{l|}{14.66}     & \multicolumn{1}{l|}{34.32}        & \multicolumn{1}{l|}{12.67\scriptsize{$\pm$0.85}} & 
\multicolumn{1}{l|}{12.93\scriptsize{$\pm$0.13}}     & 
\multicolumn{1}{l|}{0.591\scriptsize{$\pm$0.004}} & 
\multicolumn{1}{l|}{14.91}     & 35.85        \\ \hline
\multicolumn{1}{|l|}{S(130-143)} & \multicolumn{1}{l|}{10.54\scriptsize{$\pm$0.11}} & 
\multicolumn{1}{l|}{11.34\scriptsize{$\pm$0.06}}     
& \multicolumn{1}{l|}{0.611\scriptsize{$\pm$0.003}} & 
\multicolumn{1}{l|}{12.76}     & \multicolumn{1}{l|}{28.98}        & \multicolumn{1}{l|}{10.33\scriptsize{$\pm$0.21}} & 
\multicolumn{1}{l|}{11.35\scriptsize{$\pm$0.03}}     & 
\multicolumn{1}{l|}{0.611\scriptsize{$\pm$0.001}} & 
\multicolumn{1}{l|}{12.73}     & \multicolumn{1}{l|}{28.82}        & \multicolumn{1}{l|}{10.93\scriptsize{$\pm$0.20}} & 
\multicolumn{1}{l|}{11.78\scriptsize{$\pm$0.03}}     & 
\multicolumn{1}{l|}{0.629\scriptsize{$\pm$0.001}} & 
\multicolumn{1}{l|}{12.94}     & 30.34        \\ \hline
\\ \hline
\multicolumn{1}{|l|}{}           & \multicolumn{5}{c|}{500 mM NaCl}                                                                                                                                                                                                                                             & \multicolumn{5}{c|}{1 M NaCl}                                                                                                                                                                                                                                                & \multicolumn{5}{c|}{3 M GdnHCl}                                                                                                                                                                                                                         \\ \hline
\multicolumn{1}{|l|}{Name}       & \multicolumn{1}{l|}{$R_g$} & 
\multicolumn{1}{l|}{$R_g$ ex.} 
& \multicolumn{1}{l|}{$\nu$} & 
\multicolumn{1}{l|}{$R_g$ EOM} & \multicolumn{1}{l|}{$R_{ee}$ EOM} & \multicolumn{1}{l|}{$R_g$} & 
\multicolumn{1}{l|}{$R_g$ ex.} & 
 \multicolumn{1}{l|}{$\nu$} & 
 \multicolumn{1}{l|}{$R_g$ EOM} & \multicolumn{1}{l|}{$R_{ee}$ EOM} & \multicolumn{1}{l|}{$R_g$} & 
 \multicolumn{1}{l|}{$R_g$ ex.} & 
 \multicolumn{1}{l|}{$\nu$} & 
 \multicolumn{1}{l|}{$R_g$ EOM} & $R_{ee}$ EOM \\ \hline
\multicolumn{1}{|l|}{S(26-45)}   & \multicolumn{1}{l|}{10.66\scriptsize{$\pm$0.47}} & 
\multicolumn{1}{l|}{10.48\scriptsize{$\pm$0.07}}     
& \multicolumn{1}{l|}{0.509\scriptsize{$\pm$0.003}} & 
\multicolumn{1}{l|}{12.53}     & \multicolumn{1}{l|}{31.05}        & \multicolumn{1}{l|}{11.06\scriptsize{$\pm$0.68}} & 
\multicolumn{1}{l|}{11.04\scriptsize{$\pm$0.08}}     & 
\multicolumn{1}{l|}{0.533\scriptsize{$\pm$0.003}} & 
\multicolumn{1}{l|}{12.91}     & \multicolumn{1}{l|}{33.04}        & \multicolumn{1}{l|}{12.13\scriptsize{$\pm$0.52}} & 
\multicolumn{1}{l|}{12.66\scriptsize{$\pm$0.11}}     & 
\multicolumn{1}{l|}{0.595\scriptsize{$\pm$0.004}} & 
\multicolumn{1}{l|}{13.57}     & 36.26        \\ \hline
\multicolumn{1}{|l|}{S(45-64)}   & \multicolumn{1}{l|}{12.70\scriptsize{$\pm$0.37}} & 
\multicolumn{1}{l|}{13.18\scriptsize{$\pm$0.03}}     
& \multicolumn{1}{l|}{0.613\scriptsize{$\pm$0.001}} & 
\multicolumn{1}{l|}{14.05}     & \multicolumn{1}{l|}{36.14}        & \multicolumn{1}{l|}{12.36\scriptsize{$\pm$0.33}} & 
\multicolumn{1}{l|}{12.66\scriptsize{$\pm$0.05}}     & 
\multicolumn{1}{l|}{0.595\scriptsize{$\pm$0.002}} & 
\multicolumn{1}{l|}{13.99}     & \multicolumn{1}{l|}{35.8}         & \multicolumn{1}{l|}{12.95\scriptsize{$\pm$0.57}} & 
\multicolumn{1}{l|}{13.31\scriptsize{$\pm$0.01}}     & 
\multicolumn{1}{l|}{0.618\scriptsize{$\pm$0.000}} & 
\multicolumn{1}{l|}{14.1}      & 36.56        \\ \hline
\multicolumn{1}{|l|}{S(67-86)}   & \multicolumn{1}{l|}{13.14\scriptsize{$\pm$0.35}} & 
\multicolumn{1}{l|}{13.90\scriptsize{$\pm$0.02}}     
& \multicolumn{1}{l|}{0.637\scriptsize{$\pm$0.001}} & 
\multicolumn{1}{l|}{14.41}     & \multicolumn{1}{l|}{34.6}         & \multicolumn{1}{l|}{12.51\scriptsize{$\pm$0.39}} & 
\multicolumn{1}{l|}{12.78\scriptsize{$\pm$0.04}}     & 
\multicolumn{1}{l|}{0.599\scriptsize{$\pm$0.001}} & 
\multicolumn{1}{l|}{14.57}     & \multicolumn{1}{l|}{35.78}        & \multicolumn{1}{l|}{12.46\scriptsize{$\pm$0.28}} & 
\multicolumn{1}{l|}{13.12\scriptsize{$\pm$0.02}}     & 
\multicolumn{1}{l|}{0.611\scriptsize{$\pm$0.001}} & 
\multicolumn{1}{l|}{13.88}     & 32.33        \\ \hline
\multicolumn{1}{|l|}{S(66-81)}   & \multicolumn{1}{l|}{11.94\scriptsize{$\pm$0.53}} & 
\multicolumn{1}{l|}{12.39\scriptsize{$\pm$0.02}}     
& \multicolumn{1}{l|}{0.634\scriptsize{$\pm$0.001}} & 
\multicolumn{1}{l|}{12.96}     & \multicolumn{1}{l|}{32.99}        & \multicolumn{1}{l|}{11.30\scriptsize{$\pm$0.72}} & 
\multicolumn{1}{l|}{11.66\scriptsize{$\pm$0.03}}     & 
\multicolumn{1}{l|}{0.605\scriptsize{$\pm$0.001}} & 
\multicolumn{1}{l|}{12.87}     & \multicolumn{1}{l|}{32.91}        & \multicolumn{1}{l|}{10.84\scriptsize{$\pm$0.47}} & 
\multicolumn{1}{l|}{11.25\scriptsize{$\pm$0.03}}     & 
\multicolumn{1}{l|}{0.587\scriptsize{$\pm$0.001}} & 
\multicolumn{1}{l|}{12.48}     & 30.45        \\ \hline
\multicolumn{1}{|l|}{S(87-105)}  & \multicolumn{1}{l|}{13.01\scriptsize{$\pm$1.00}} & 
\multicolumn{1}{l|}{13.83\scriptsize{$\pm$0.09}}     
& \multicolumn{1}{l|}{0.635\scriptsize{$\pm$0.003}} & 
\multicolumn{1}{l|}{14.3}      & \multicolumn{1}{l|}{32.19}        & \multicolumn{1}{l|}{}      & 
\multicolumn{1}{l|}{}          & 
\multicolumn{1}{l|}{}      & 
\multicolumn{1}{l|}{}          & \multicolumn{1}{l|}{}             & \multicolumn{1}{l|}{13.15\scriptsize{$\pm$0.87}} & 
\multicolumn{1}{l|}{13.80\scriptsize{$\pm$0.39}}     & 
\multicolumn{1}{l|}{0.634\scriptsize{$\pm$0.012}} & 
\multicolumn{1}{l|}{14.15}     & 31.3         \\ \hline
\multicolumn{1}{|l|}{S(82-96)}   & \multicolumn{1}{l|}{11.29\scriptsize{$\pm$0.37}} & 
\multicolumn{1}{l|}{11.89\scriptsize{$\pm$0.01}}     
& \multicolumn{1}{l|}{0.614\scriptsize{$\pm$0.000}} & 
\multicolumn{1}{l|}{13.01}     & \multicolumn{1}{l|}{30.02}        & \multicolumn{1}{l|}{11.14\scriptsize{$\pm$0.47}} & 
\multicolumn{1}{l|}{11.52\scriptsize{$\pm$0.01}}     & 
\multicolumn{1}{l|}{0.599\scriptsize{$\pm$0.000}} & 
\multicolumn{1}{l|}{13.08}     & \multicolumn{1}{l|}{30.76}        & \multicolumn{1}{l|}{10.91\scriptsize{$\pm$0.38}} & 
\multicolumn{1}{l|}{11.31\scriptsize{$\pm$0.02}}     & 
\multicolumn{1}{l|}{0.590\scriptsize{$\pm$0.001}} & 
\multicolumn{1}{l|}{12.52}     & 26.4         \\ \hline
\multicolumn{1}{|l|}{S(106-128)} & \multicolumn{1}{l|}{12.57\scriptsize{$\pm$1.07}} & 
\multicolumn{1}{l|}{13.33\scriptsize{$\pm$0.08}}     
& \multicolumn{1}{l|}{0.618\scriptsize{$\pm$0.003}} & 
\multicolumn{1}{l|}{14.1}      & \multicolumn{1}{l|}{31.31}        & \multicolumn{1}{l|}{}      & 
\multicolumn{1}{l|}{}          & 
\multicolumn{1}{l|}{}      & 
\multicolumn{1}{l|}{}          & \multicolumn{1}{l|}{}             & \multicolumn{1}{l|}{12.59\scriptsize{$\pm$0.68}} & 
\multicolumn{1}{l|}{13.54\scriptsize{$\pm$0.20}}     & 
\multicolumn{1}{l|}{0.625\scriptsize{$\pm$0.006}} & 
\multicolumn{1}{l|}{14.03}     & 31.02        \\ \hline
\multicolumn{1}{|l|}{S(110-125)} & \multicolumn{1}{l|}{11.59\scriptsize{$\pm$0.14}} & 
\multicolumn{1}{l|}{12.24\scriptsize{$\pm$0.01}}     
& \multicolumn{1}{l|}{0.611\scriptsize{$\pm$0.000}} & 
\multicolumn{1}{l|}{13.25}     & \multicolumn{1}{l|}{29.28}        & \multicolumn{1}{l|}{11.30\scriptsize{$\pm$0.32}} & 
\multicolumn{1}{l|}{11.74\scriptsize{$\pm$0.02}}     & 
\multicolumn{1}{l|}{0.591\scriptsize{$\pm$0.001}} & 
\multicolumn{1}{l|}{13.2}      & \multicolumn{1}{l|}{29.45}        & \multicolumn{1}{l|}{11.54\scriptsize{$\pm$0.20}} & 
\multicolumn{1}{l|}{12.01\scriptsize{$\pm$0.01}}     & 
\multicolumn{1}{l|}{0.601\scriptsize{$\pm$0.000}} & 
\multicolumn{1}{l|}{13.05}     & 28.31        \\ \hline
\multicolumn{1}{|l|}{S(129-146)} & \multicolumn{1}{l|}{13.02\scriptsize{$\pm$0.89}} & 
\multicolumn{1}{l|}{13.53\scriptsize{$\pm$0.16}}     
& \multicolumn{1}{l|}{0.611\scriptsize{$\pm$0.005}} & 
\multicolumn{1}{l|}{14.64}     & \multicolumn{1}{l|}{34.03}        & \multicolumn{1}{l|}{}      & 
\multicolumn{1}{l|}{}          & 
\multicolumn{1}{l|}{}      & 
\multicolumn{1}{l|}{}          & \multicolumn{1}{l|}{}             & \multicolumn{1}{l|}{13.28\scriptsize{$\pm$1.04}} & 
\multicolumn{1}{l|}{14.18\scriptsize{$\pm$0.23}}     & 
\multicolumn{1}{l|}{0.631\scriptsize{$\pm$0.007}} & 
\multicolumn{1}{l|}{14.5}      & 33.4         \\ \hline
\multicolumn{1}{|l|}{S(130-143)} & \multicolumn{1}{l|}{11.13\scriptsize{$\pm$0.15}} & 
\multicolumn{1}{l|}{11.86\scriptsize{$\pm$0.01}}     
& \multicolumn{1}{l|}{0.633\scriptsize{$\pm$0.001}} & 
\multicolumn{1}{l|}{12.78}     & \multicolumn{1}{l|}{29.12}        & \multicolumn{1}{l|}{}      & 
\multicolumn{1}{l|}{}          & 
\multicolumn{1}{l|}{}      & 
\multicolumn{1}{l|}{}          & \multicolumn{1}{l|}{}             & \multicolumn{1}{l|}{11.78\scriptsize{$\pm$1.05}} & 
\multicolumn{1}{l|}{12.55\scriptsize{$\pm$0.27}}     & 
\multicolumn{1}{l|}{0.661\scriptsize{$\pm$0.010}} & 
\multicolumn{1}{l|}{12.81}     & 29.51        \\ \hline
\end{tabular}
}
\caption {\textbf {SAXS results.} Segments' dimensions in {\AA} (excluding $\nu$) at various salinity conditions. $R_g$ is extracted from Guinier approximation, $R_g$ ex. is the radius of gyration obtained from extended Guinier fit with its $\nu$ value. $R_g$ EOM and $R_{ee}$ EOM are the ensemble averages obtained from EOM.} 
\label{tab:SAXS1}
\end{table}

\begin{table}[h]
\centering
\begin{tabular}{|c|c|c|c|c|c|}
\hline
                            &                         & 20 mM Tris         & 150 mM NaCl        & 1 M NaCl           & 3 M GdnHCl            \\ \hline
\multirow{2}{*}{S segments} & SAXS                    & 0.189 $\pm$ 0.045 & 0.166 $\pm$ 0.048 & 0.111 $\pm$ 0.055 & 0.028 $\pm$ 0.050    \\ \cline{2-6} 
                            & \multirow{2}{*}{trFRET} & 0.228 $\pm$ 0.096 & 0.184 $\pm$ 0.080 & 0.144 $\pm$ 0.082 & 0.06 $\pm$ 0.11      \\ \cline{1-1} \cline{3-6} 
P segments                  &                         & 0.162 $\pm$ 0.053 & 0.155 $\pm$ 0.039 & 0.149 $\pm$ 0.046 & $-$0.013 $\pm$ 0.073 \\ \hline
\end{tabular}
\caption {The slopes' values that were obtained with linear fits for the semgent's $\nu$ values versus their NCPR (Figs. \ref{fig:S_segment_Nu} and 5). All units are one over electron charge.} 
\label{tab:slops}
\end{table}

\begin{table}[h]
\resizebox{\textwidth}{!}{%
\begin{tabular}{|c|cccc|cccc|cccc|cccc|}
\hline
           & \multicolumn{4}{c|}{0 M NaCl}& \multicolumn{4}{c|}{0.15 M NaCl}& \multicolumn{4}{c|}{1 M NaCl}& \multicolumn{4}{c|}{3 M GdnHcl}
           
           \\ 
\hline
Name       & \multicolumn{1}{c|}{a} & \multicolumn{1}{c|}{b} & \multicolumn{1}{c|}{$<R_{ee}>$}  & {FWHM}  & \multicolumn{1}{c|}{a} & \multicolumn{1}{c|}{b} & \multicolumn{1}{c|}{$<R_{ee}>$}  & {FWHM}  & \multicolumn{1}{c|}{a} & \multicolumn{1}{c|}{b} & \multicolumn{1}{c|}{$<R_{ee}>$}  & {FWHM}  & \multicolumn{1}{c|}{a} & \multicolumn{1}{c|}{b} & \multicolumn{1}{c|}{$<R_{ee}>$}  & {FWHM}   
           
           \\

\hline
S(6-25)    & 
\multicolumn{1}{c|}{0.001}        & \multicolumn{1}{c|}{0.004}         & \multicolumn{1}{c|}{18.6{\scriptsize$\pm$1.5}}   & \multicolumn{1}{c|}{19.0{\scriptsize$\pm$2.0}}  &
\multicolumn{1}{c|}{0.000}   &  \multicolumn{1}{c|}{0.004}  &
\multicolumn{1}{c|}{18.4{\scriptsize$\pm$1.5}}        & \multicolumn{1}{c|}{18.9{\scriptsize$\pm$2.0}}         & 
\multicolumn{1}{c|}{0.000}   &  \multicolumn{1}{c|}{0.005}  &
\multicolumn{1}{c|}{15.9{\scriptsize$\pm$1.5}}        & \multicolumn{1}{c|}{16.3{\scriptsize$\pm$2.0}}         & 
\multicolumn{1}{c|}{0.000}   &  \multicolumn{1}{c|}{0.003}  & 
\multicolumn{1}{c|}{20.9{\scriptsize$\pm$0.5}}        & \multicolumn{1}{c|}{21.4{\scriptsize$\pm$0.5}}         

\\ \hline
S(26-46)   & 
\multicolumn{1}{c|}{17.42} & \multicolumn{1}{c|}{0.011} &
\multicolumn{1}{c|}{21.8{\scriptsize$\pm$0.8}}        & \multicolumn{1}{c|}{14.3{\scriptsize$\pm$2.2}}         &
\multicolumn{1}{c|}{16.79} & \multicolumn{1}{c|}{0.010} &
\multicolumn{1}{c|}{21.9{\scriptsize$\pm$0.9}}        & \multicolumn{1}{c|}{15.0{\scriptsize$\pm$2.4}}         & 
\multicolumn{1}{c|}{0.000}   &  \multicolumn{1}{c|}{0.003}  &
\multicolumn{1}{c|}{19.8{\scriptsize$\pm$0.9}}        & \multicolumn{1}{c|}{20.2{\scriptsize$\pm$1.1}}         & 
\multicolumn{1}{c|}{20.82} &  \multicolumn{1}{c|}{0.021} &
\multicolumn{1}{c|}{23.0{\scriptsize$\pm$0.3}}        & \multicolumn{1}{c|}{11.0{\scriptsize$\pm$2.1}}           \\ \hline
S(67-86)   & 
\multicolumn{1}{c|}{24.60} & \multicolumn{1}{c|}{0.052} &
\multicolumn{1}{c|}{25.4{\scriptsize$\pm$0.2}}        & \multicolumn{1}{c|}{7.2{\scriptsize$\pm$1.1}}          & 
\multicolumn{1}{c|}{24.55} & \multicolumn{1}{c|}{0.056} &
\multicolumn{1}{c|}{25.3{\scriptsize$\pm$0.2}}        & \multicolumn{1}{c|}{7.0{\scriptsize$\pm$1.1}}          & 
\multicolumn{1}{c|}{24.26} & \multicolumn{1}{c|}{0.057} &
\multicolumn{1}{c|}{25.0{\scriptsize$\pm$0.2}}        & \multicolumn{1}{c|}{6.8{\scriptsize$\pm$1.0}}          &
\multicolumn{1}{c|}{24.55} &  \multicolumn{1}{c|}{0.066} &
\multicolumn{1}{c|}{25.1{\scriptsize$\pm$0.2}}        & \multicolumn{1}{c|}{6.4{\scriptsize$\pm$1.5}}          
\\ \hline
S(66-81)   & 
\multicolumn{1}{c|}{21.24} &  \multicolumn{1}{c|}{0.059}&
\multicolumn{1}{c|}{22.0{\scriptsize$\pm$0.2}}        & \multicolumn{1}{c|}{6.8{\scriptsize$\pm$0.7}}          & 
\multicolumn{1}{c|}{21.37} & \multicolumn{1}{c|}{0.062} &
\multicolumn{1}{c|}{22.1{\scriptsize$\pm$0.2}}        & \multicolumn{1}{c|}{6.6{\scriptsize$\pm$0.8}}          & 
\multicolumn{1}{c|}{20.86} & \multicolumn{1}{c|}{0.060} &
\multicolumn{1}{c|}{21.7{\scriptsize$\pm$0.2}}        & \multicolumn{1}{c|}{6.8{\scriptsize$\pm$0.8}}          &
\multicolumn{1}{c|}{22.00} &  \multicolumn{1}{c|}{0.084} &
\multicolumn{1}{c|}{22.5{\scriptsize$\pm$0.2}}        & \multicolumn{1}{c|}{5.7{\scriptsize$\pm$1.0}}  \\ \hline
S(87-105)  & 

\multicolumn{1}{c|}{25.73} &  \multicolumn{1}{c|}{0.011}&
\multicolumn{1}{c|}{29.1{\scriptsize$\pm$0.2}}        & \multicolumn{1}{c|}{15.2{\scriptsize$\pm$3.5}}         & 
\multicolumn{1}{c|}{24.99} & \multicolumn{1}{c|}{0.014} &
\multicolumn{1}{c|}{27.6{\scriptsize$\pm$0.2}}        & \multicolumn{1}{c|}{13.3{\scriptsize$\pm$2.6}}         & 
\multicolumn{1}{c|}{24.10} & \multicolumn{1}{c|}{0.018} &
\multicolumn{1}{c|}{26.3{\scriptsize$\pm$0.2}}        & \multicolumn{1}{c|}{12.0{\scriptsize$\pm$2.1}}         &
\multicolumn{1}{c|}{22.51} & \multicolumn{1}{c|}{0.013}&
\multicolumn{1}{c|}{25.6{\scriptsize$\pm$0.6}}        & \multicolumn{1}{c|}{13.7{\scriptsize$\pm$3.3}}          

\\ \hline
S(82-96)   & 
\multicolumn{1}{c|}{22.61} & \multicolumn{1}{c|}{0.013} &
\multicolumn{1}{c|}{25.8{\scriptsize$\pm$0.25}}        & \multicolumn{1}{c|}{13.9{\scriptsize$\pm$1.8}}         &
\multicolumn{1}{c|}{21.83} & \multicolumn{1}{c|}{0.013} &
\multicolumn{1}{c|}{25.1{\scriptsize$\pm$0.3}}        & \multicolumn{1}{c|}{13.8{\scriptsize$\pm$1.8}}         &
\multicolumn{1}{c|}{21.02} & \multicolumn{1}{c|}{0.013} &
\multicolumn{1}{c|}{24.3{\scriptsize$\pm$0.3}}        & \multicolumn{1}{c|}{13.6{\scriptsize$\pm$1.6}}         & 
\multicolumn{1}{c|}{22.56} &  \multicolumn{1}{c|}{0.019} &
\multicolumn{1}{c|}{24.8{\scriptsize$\pm$0.3}}        & \multicolumn{1}{c|}{11.5{\scriptsize$\pm$2.1}}          

\\ \hline
S(106-128) & 
\multicolumn{1}{c|}{26.02} & \multicolumn{1}{c|}{0.040} &
\multicolumn{1}{c|}{26.9{\scriptsize$\pm$0.2}}        & \multicolumn{1}{c|}{8.2{\scriptsize$\pm$2.2}}          & 
\multicolumn{1}{c|}{25.01} & \multicolumn{1}{c|}{0.048} &
\multicolumn{1}{c|}{25.8{\scriptsize$\pm$0.2}}        & \multicolumn{1}{c|}{7.5{\scriptsize$\pm$1.9}}          &
\multicolumn{1}{c|}{24.14} & \multicolumn{1}{c|}{0.050} &
\multicolumn{1}{c|}{24.9{\scriptsize$\pm$0.2}}        & \multicolumn{1}{c|}{7.4{\scriptsize$\pm$1.6}}          &
\multicolumn{1}{c|}{23.84} &  \multicolumn{1}{c|}{0.044} &
\multicolumn{1}{c|}{24.7{\scriptsize$\pm$0.2}}        & \multicolumn{1}{c|}{7.8{\scriptsize$\pm$1.7}}           

\\ \hline
S(110-125) & 
\multicolumn{1}{c|}{21.41} & \multicolumn{1}{c|}{0.050} &
\multicolumn{1}{c|}{22.3{\scriptsize$\pm$0.25}}        & \multicolumn{1}{c|}{7.3{\scriptsize$\pm$1.0}}          & 
\multicolumn{1}{c|}{21.47} & \multicolumn{1}{c|}{0.049} &
\multicolumn{1}{c|}{22.4{\scriptsize$\pm$0.2}}        & \multicolumn{1}{c|}{7.3{\scriptsize$\pm$1.0}}          & 
\multicolumn{1}{c|}{21.63} & \multicolumn{1}{c|}{0.052} &
\multicolumn{1}{c|}{22.5{\scriptsize$\pm$0.2}}        & \multicolumn{1}{c|}{7.1{\scriptsize$\pm$1.0}}          & 
\multicolumn{1}{c|}{21.34} &  \multicolumn{1}{c|}{0.047} &
\multicolumn{1}{c|}{22.4{\scriptsize$\pm$0.3}}        & \multicolumn{1}{c|}{7.6{\scriptsize$\pm$1.1}}           
\\ \hline
S(129-146) & 
\multicolumn{1}{c|}{23.42} &  \multicolumn{1}{c|}{0.080}&
\multicolumn{1}{c|}{23.9{\scriptsize$\pm$0.2}}        & \multicolumn{1}{c|}{5.8{\scriptsize$\pm$1.0}}          & 
\multicolumn{1}{c|}{23.72} & \multicolumn{1}{c|}{0.079} &
\multicolumn{1}{c|}{24.2{\scriptsize$\pm$0.2}}        &
\multicolumn{1}{c|}{5.8{\scriptsize$\pm$1.1}}          &
\multicolumn{1}{c|}{23.91} &  \multicolumn{1}{c|}{0.076}&
\multicolumn{1}{c|}{24.4{\scriptsize$\pm$0.2}}        &
\multicolumn{1}{c|}{6.0{\scriptsize$\pm$1.2}}          &
\multicolumn{1}{c|}{24.57} & \multicolumn{1}{c|}{0.053} & 
\multicolumn{1}{c|}{25.4{\scriptsize$\pm$0.2}}        & \multicolumn{1}{c|}{7.1{\scriptsize$\pm$2.2}}           

\\ \hline
S(130-143) & 
\multicolumn{1}{c|}{21.44} & \multicolumn{1}{c|}{0.011} &
\multicolumn{1}{c|}{25.3{\scriptsize$\pm$0.4}}        & \multicolumn{1}{c|}{14.8{\scriptsize$\pm$2.1}}         &
\multicolumn{1}{c|}{20.82} &  \multicolumn{1}{c|}{0.011}&
\multicolumn{1}{c|}{24.8{\scriptsize$\pm$0.5}}        & \multicolumn{1}{c|}{14.8{\scriptsize$\pm$2.1}}         &
\multicolumn{1}{c|}{20.09} &  \multicolumn{1}{c|}{0.011}&
\multicolumn{1}{c|}{24.2{\scriptsize$\pm$0.6}}        & \multicolumn{1}{c|}{14.8{\scriptsize$\pm$2.0}}         & 
\multicolumn{1}{c|}{21.97} &  \multicolumn{1}{c|}{0.016} &
\multicolumn{1}{c|}{24.6{\scriptsize$\pm$0.4}}        & \multicolumn{1}{c|}{12.4{\scriptsize$\pm$2.1}}          \\ \hline
\end{tabular}}
\caption {\textbf {S segments end-to-end distance determine by fitting trFRET measurements with radial Gauss model.} a and b are the fitting parameters of the radial Gauss model (Eq.~4) which are used for calculating $<R_{ee}>$  and full width half max (FWHM). Values presented in [\AA]. Errors were calculated by rigorous analysis with 2SD} 
\label{tab:S_FRET_Ree}
\end{table}

\begin{table}[h]
\resizebox{\textwidth}{!}{%
\begin{tabular}{|c|cc|cc|cc|cc||c|}
\hline
           & \multicolumn{2}{c|}{0 M NaCl}& \multicolumn{2}{c|}{0.15 M NaCl}& \multicolumn{2}{c|}{1 M NaCl}& \multicolumn{2}{c||}{3 M GdnHcl}&
           
           \\ 
\hline
Name       &  
\multicolumn{1}{c|}{$L_p$}  & $L_c$  &   \multicolumn{1}{c|}{$L_p$}  & $L_c$  &   \multicolumn{1}{c|}{$L_p$}  & $L_c$  &   \multicolumn{1}{c|}{$L_p$}  & $L_c$ &  $L_c^*$
           \\

\hline
S(6-25)    & 
\multicolumn{1}{c|}{X}   & X   &  
\multicolumn{1}{c|}{X}   & X   &         \multicolumn{1}{c|}{X}   & X   &          \multicolumn{1}{c|}{X}   & X   & 78.2 \\ \hline

S(26-46)   &  
\multicolumn{1}{c|}{6.2{\scriptsize$\pm$0.2}} & 44.2{\scriptsize$\pm$0.9} &  
\multicolumn{1}{c|}{5.8{\scriptsize$\pm$0.2}} & 46.0{\scriptsize$\pm$0.9} &  
\multicolumn{1}{c|}{X}   & X   &  
\multicolumn{1}{c|}{9.7{\scriptsize$\pm$0.2}} & 35.5{\scriptsize$\pm$1.8} & 78.2 \\ \hline

S(67-86)   &  
\multicolumn{1}{c|}{21.0{\scriptsize$\pm$0.4}} & 31.2{\scriptsize$\pm$0.2} &  \multicolumn{1}{c|}{20.9{\scriptsize$\pm$0.5}} & 31.1{\scriptsize$\pm$0.2} &  \multicolumn{1}{c|}{21.3{\scriptsize$\pm$0.5}} & 30.5{\scriptsize$\pm$0.2} &  \multicolumn{1}{c|}{21.6{\scriptsize$\pm$0.5}} & 30.7{\scriptsize$\pm$0.2} & 78.2 \\ \hline

S(66-81)   &  
\multicolumn{1}{c|}{13.0{\scriptsize$\pm$0.1}} & 29.4{\scriptsize$\pm$0.1} & 
\multicolumn{1}{c|}{13.1{\scriptsize$\pm$0.2}} & 29.5{\scriptsize$\pm$0.1} &  \multicolumn{1}{c|}{12.1{\scriptsize$\pm$0.1}} & 29.3{\scriptsize$\pm$0.1} &  \multicolumn{1}{c|}{15.5{\scriptsize$\pm$0.3}} & 28.7{\scriptsize$\pm$0.2} & 57.8 \\ \hline

S(87-105)  & 
\multicolumn{1}{c|}{11.5{\scriptsize$\pm$0.3}} & 49.2 {\scriptsize$\pm$0.9}&  \multicolumn{1}{c|}{12.1{\scriptsize$\pm$0.3}} & 43.3 {\scriptsize$\pm$0.6}&  \multicolumn{1}{c|}{12.1{\scriptsize$\pm$0.2}} & 40.0 {\scriptsize$\pm$0.4}& 
\multicolumn{1}{c|}{4.3{\scriptsize$\pm$0.6}} & 87.5 {\scriptsize$\pm$13.2}& 78.2\\ \hline

S(82-96)   & 
\multicolumn{1}{c|}{13.4{\scriptsize$\pm$0.2}} & 36.7 {\scriptsize$\pm$0.2}&  \multicolumn{1}{c|}{12.8{\scriptsize$\pm$0.2}} & 36.1 {\scriptsize$\pm$0.2}&  \multicolumn{1}{c|}{12.0{\scriptsize$\pm$0.2}} & 35.5 {\scriptsize$\pm$0.2}&  \multicolumn{1}{c|}{12.9{\scriptsize$\pm$0.2}} & 35.0 {\scriptsize$\pm$0.2}& 57.8 \\ \hline

S(106-128) & 
\multicolumn{1}{c|}{26.0{\scriptsize$\pm$0.7}} & 31.9 {\scriptsize$\pm$0.2}&  \multicolumn{1}{c|}{24.3{\scriptsize$\pm$0.6}} & 30.8 {\scriptsize$\pm$0.2}&  \multicolumn{1}{c|}{20.9{\scriptsize$\pm$0.4}} & 30.5 {\scriptsize$\pm$0.2}&  \multicolumn{1}{c|}{16.5{\scriptsize$\pm$0.4}} & 32.3 {\scriptsize$\pm$0.3}& 78.2 \\ \hline

S(110-125) & 
\multicolumn{1}{c|}{17.9{\scriptsize$\pm$0.2}} & 27.5 {\scriptsize$\pm$0.1}&  \multicolumn{1}{c|}{18.5{\scriptsize$\pm$0.3}} & 27.6 {\scriptsize$\pm$0.1}&  \multicolumn{1}{c|}{18.4{\scriptsize$\pm$0.2}} & 27.6 {\scriptsize$\pm$0.1}&  \multicolumn{1}{c|}{15.0{\scriptsize$\pm$0.2}} & 28.8 {\scriptsize$\pm$0.1}& 61.2 \\ \hline

S(129-146) & 
\multicolumn{1}{c|}{20.3{\scriptsize$\pm$0.3}} & 29.2 {\scriptsize$\pm$0.1}&  \multicolumn{1}{c|}{19.6{\scriptsize$\pm$0.3}} & 29.9 {\scriptsize$\pm$0.1}&  \multicolumn{1}{c|}{21.9{\scriptsize$\pm$0.4}} & 29.5 {\scriptsize$\pm$0.1}&  \multicolumn{1}{c|}{16.4{\scriptsize$\pm$0.4}} & 33.5 {\scriptsize$\pm$0.3}& 81.6 \\ \hline

S(130-143) & 
\multicolumn{1}{c|}{13.1{\scriptsize$\pm$0.2}} & 35.9{\scriptsize$\pm$0.2} &  \multicolumn{1}{c|}{12.6{\scriptsize$\pm$0.2}} & 35.6{\scriptsize$\pm$0.2} &  \multicolumn{1}{c|}{12.3{\scriptsize$\pm$0.2}} & 34.8{\scriptsize$\pm$0.3} &  \multicolumn{1}{c|}{13.6{\scriptsize$\pm$0.2}} & 33.9{\scriptsize$\pm$0.2} & 54.4 \\ \hline

\end{tabular}}
\caption {\textbf {S segments persistence length determine by fitting trFRET measurements with worm like chain (WLC) model.} Persistence length $L_p$ and reduced contour length $L_c$ are the result of the WLC model. $L_c^*$ is the theoretical contour length calculate as $3.4 \cdot N$. All values presented in [\AA]. Errors were calculated by rigorous analysis with 2SD}.
\label{tab:S_FRET_WLC}
\end{table}

\begin{figure}[h]
\centering
\includegraphics[width=0.8\linewidth]{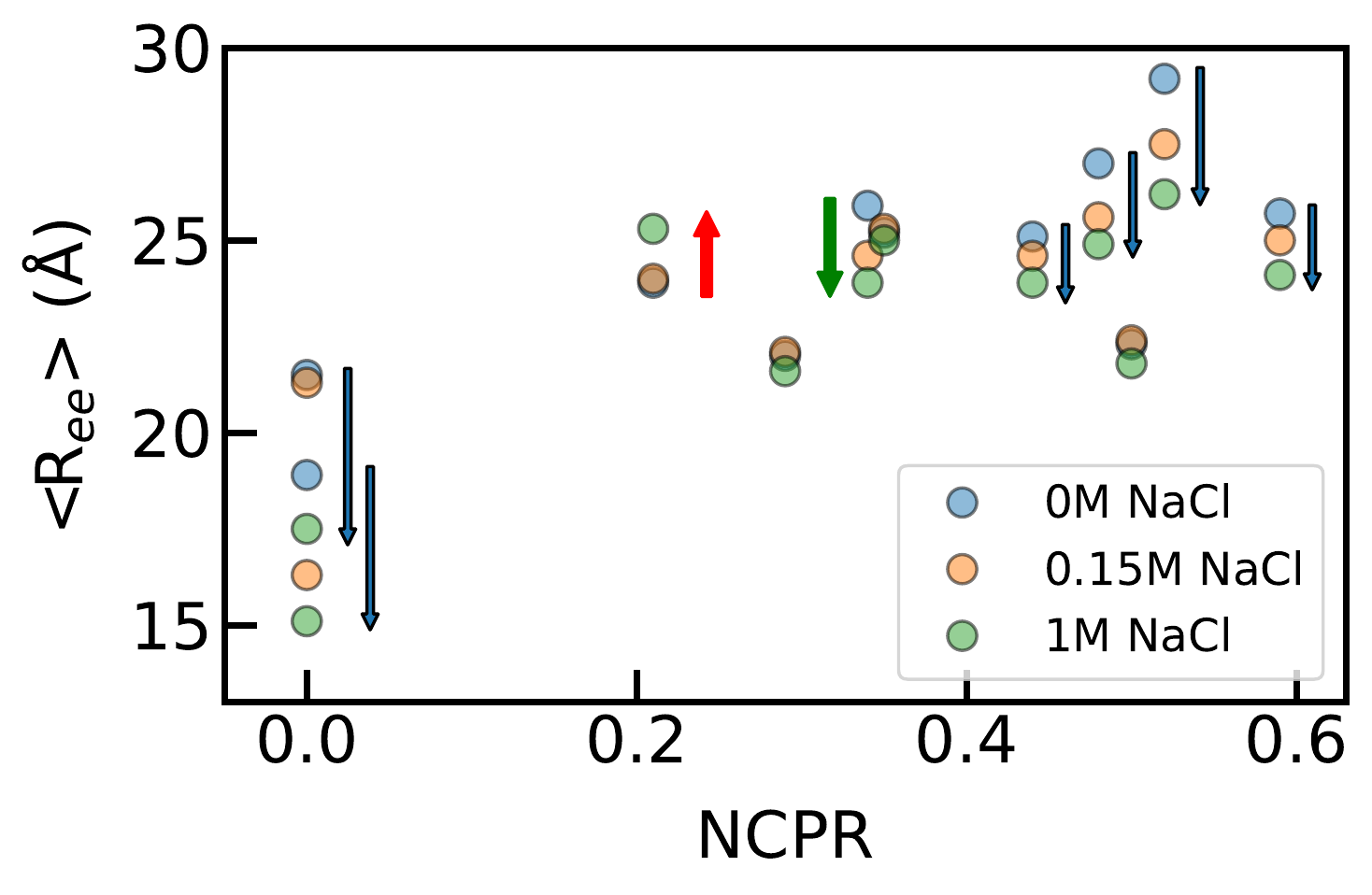}
\caption{S segment end-to-end mean distance as extracted from the radial Gauss model for the trFRET measurements. Arrows indicate the direction of increasing salinity. \deleted{PP indicate segments containing di-Proline residues.} \added{Red arrow indicate the expansion of S(129-146). Green arrow indicate the contraction of S(129-146) after changing three Lys residues at positions 143-145 to Gly. } }
\label{fig:GaussMeanSalt}
\end{figure}

\begin{figure}[h]
\centering
\includegraphics[width=0.6\linewidth]{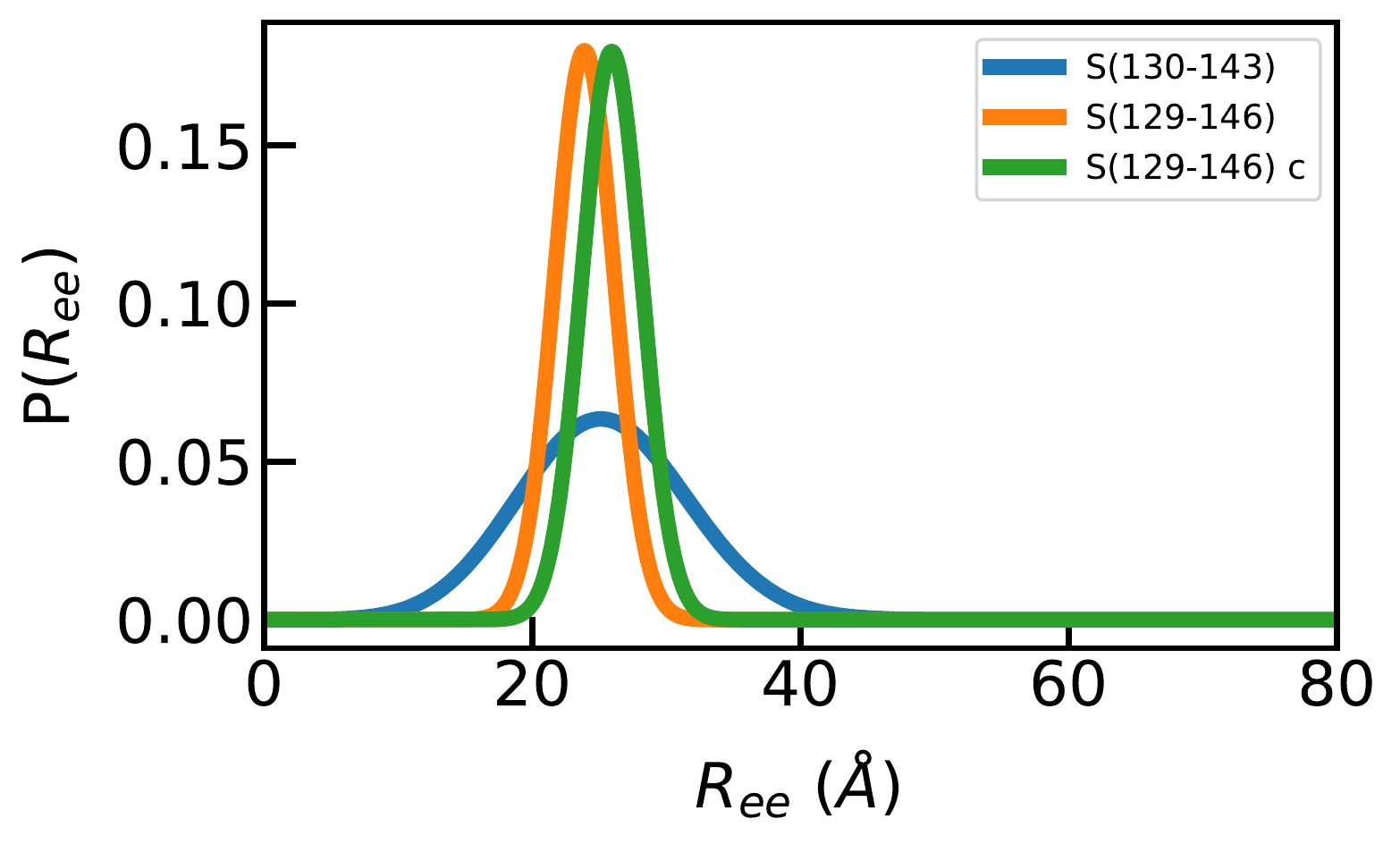}
\caption{ Radial Gauss probability function of trFRET measurements. Segment S(130-143) is shorter then S(129-146) but showing equal $<R_{ee}>$ and larger width. This we relate to loop formation due to three Lys at positions 144-146. \added{Segments S(129-146)c shows increasing $R_{ee}$ after changing three Lys to Gly}}
\label{fig:7_7p}
\end{figure}

\begin{figure}[h]
\centering
\includegraphics[width=0.5\linewidth]{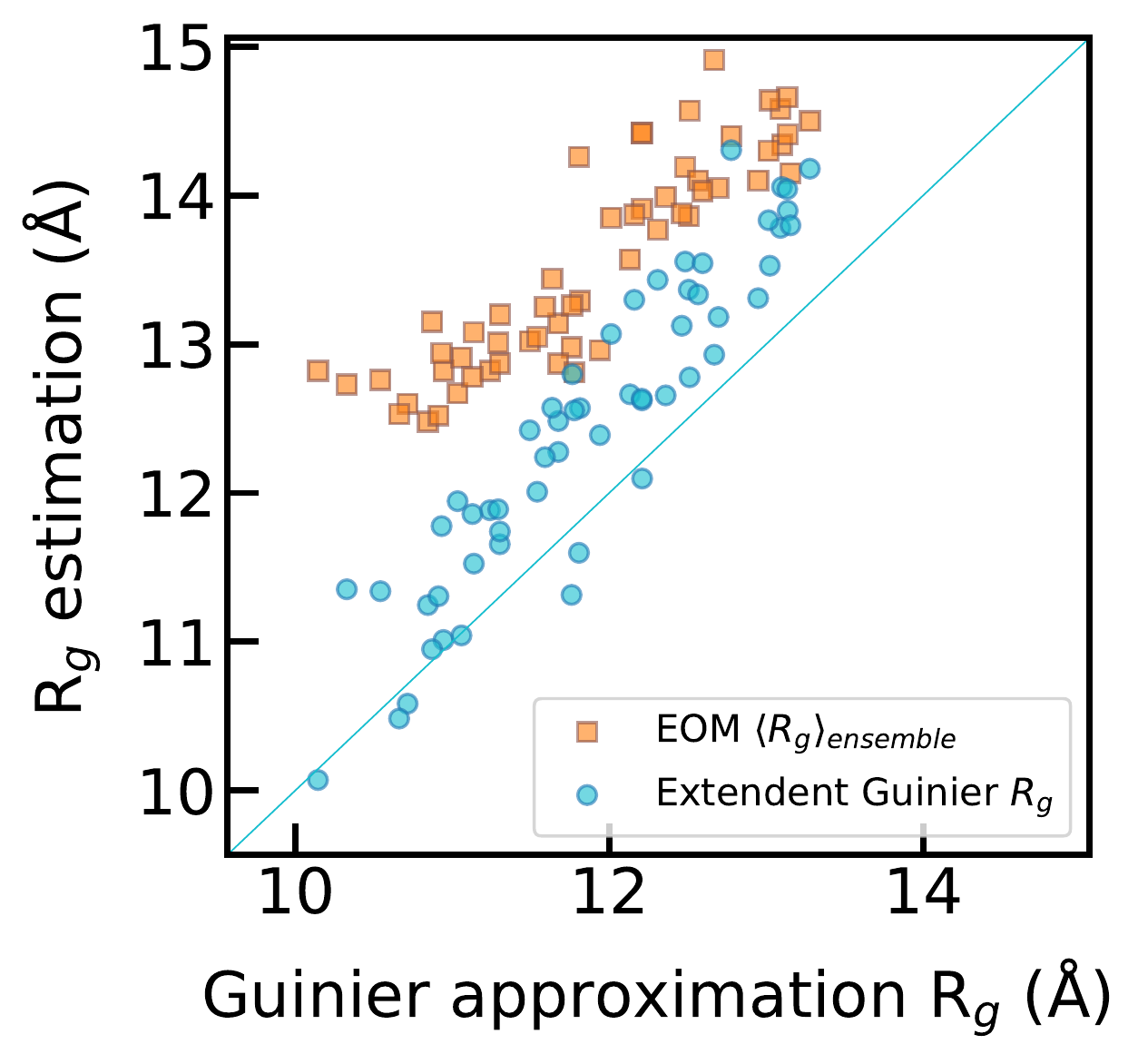}
\caption{Estimation of the Radius of gyration obtained from EOM (orange squares) or from extended Guinier fit (cyan circles) vs. the $R_g$ obtained from the regular Guinier approximation fit. The solid cyan line is the plot of $y=x$. It can be seen that EOM overestimates the $R_g$ more that the Extended Guinier fit. However, there is a satisfying correlation and the Extended Guinier $R_g$ values are close to the EOM ones.}
\label{fig:SI-Rg-estimation}
\end{figure}

\begin{figure}[h]
\centering
\includegraphics[width=1.0\linewidth]{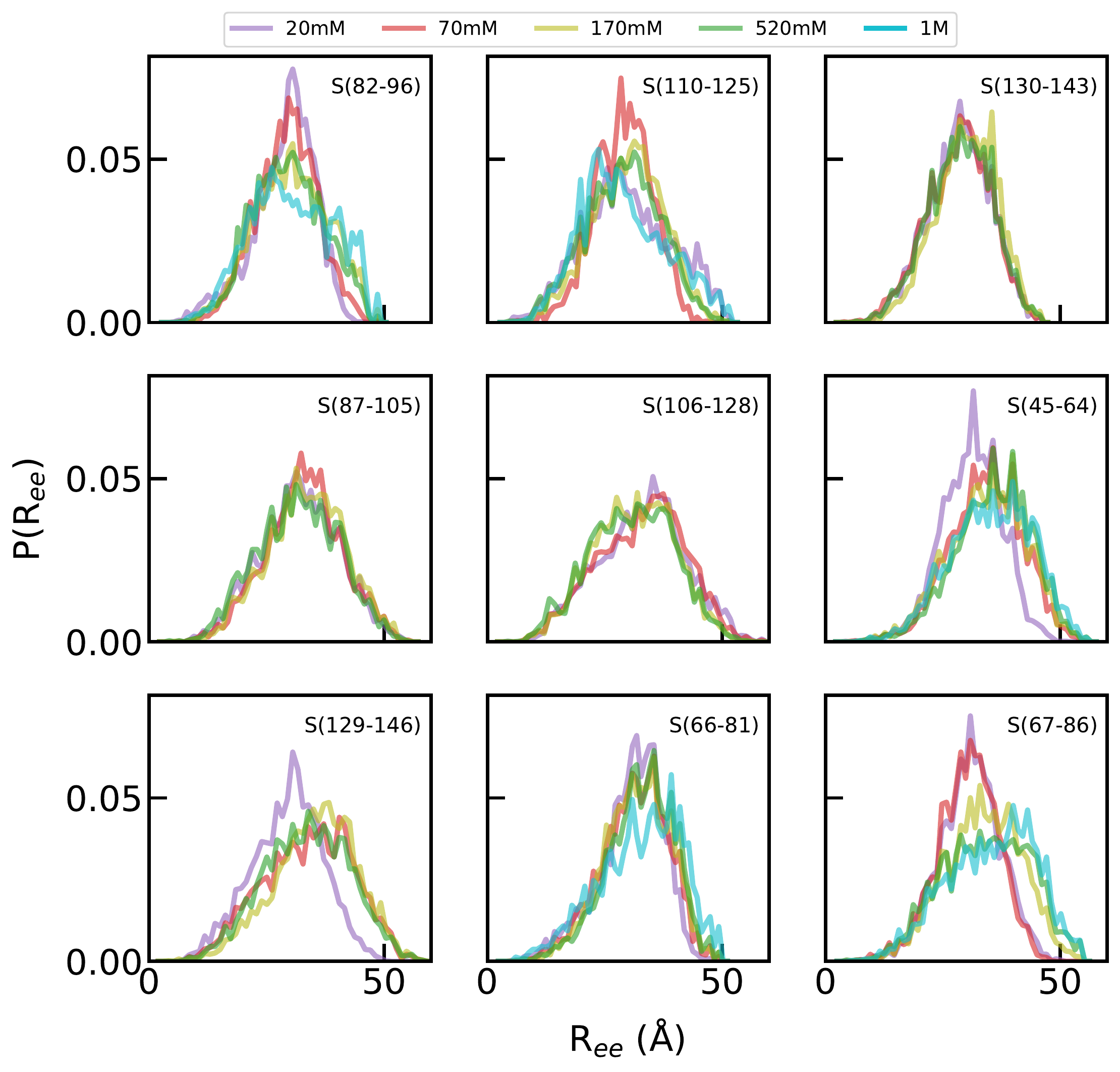}
\caption{Qualitative comparison between $R_{ee}$ probability distributions obtained from EOM on SAXS data at different salinity. The relevant segment is on the top right side of each panel. Segments S(45-64), S(67-86) and s(129-146) vary the most upon salt addition. Then, segments  S(66-81), S(82-96) and S(110-125). Less sensitive segments are S(87-105), S(106-128) and S(130-143).}
\label{fig:SI-EOM-salt}
\end{figure} 

\begin{figure}[h]
\centering
\includegraphics[width=1.0\linewidth]{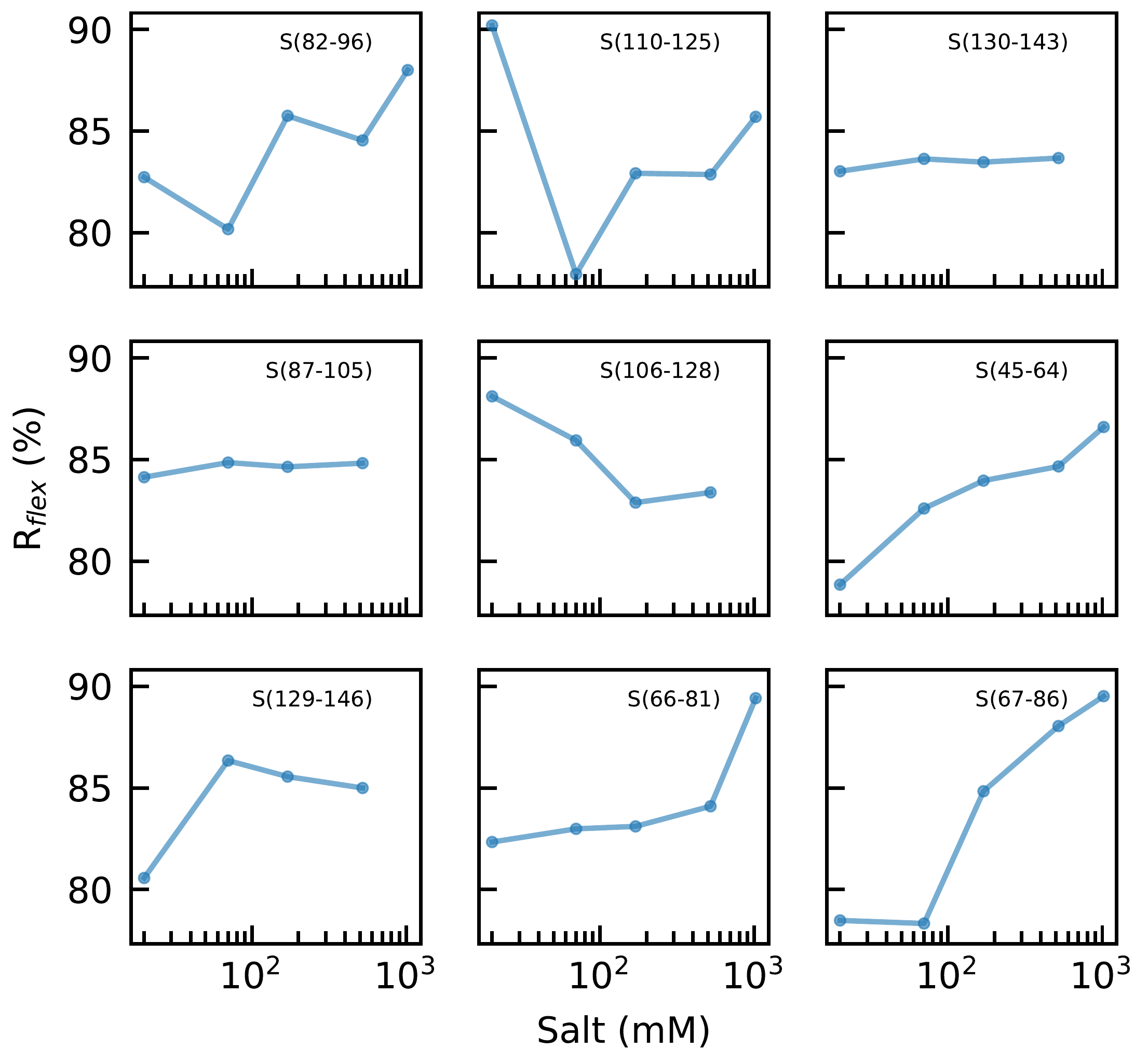}
\caption{Comparison between $R_{ee}$ probability distributions' flexibility obtained from EOM on SAXS data at different salinity. The relevant segment is on the top side of each panel. Segments S(45-64), S(67-86) and s(129-146) vary the most upon salt addition. Then, segments  S(66-81), S(82-96) and S(110-125). Less sensitive segments are S(87-105), S(106-128) and S(130-143).}
\label{fig:SI-EOM-R_flex}
\end{figure}

\begin{figure}[h]
\centering
\includegraphics[width=1.0\linewidth]{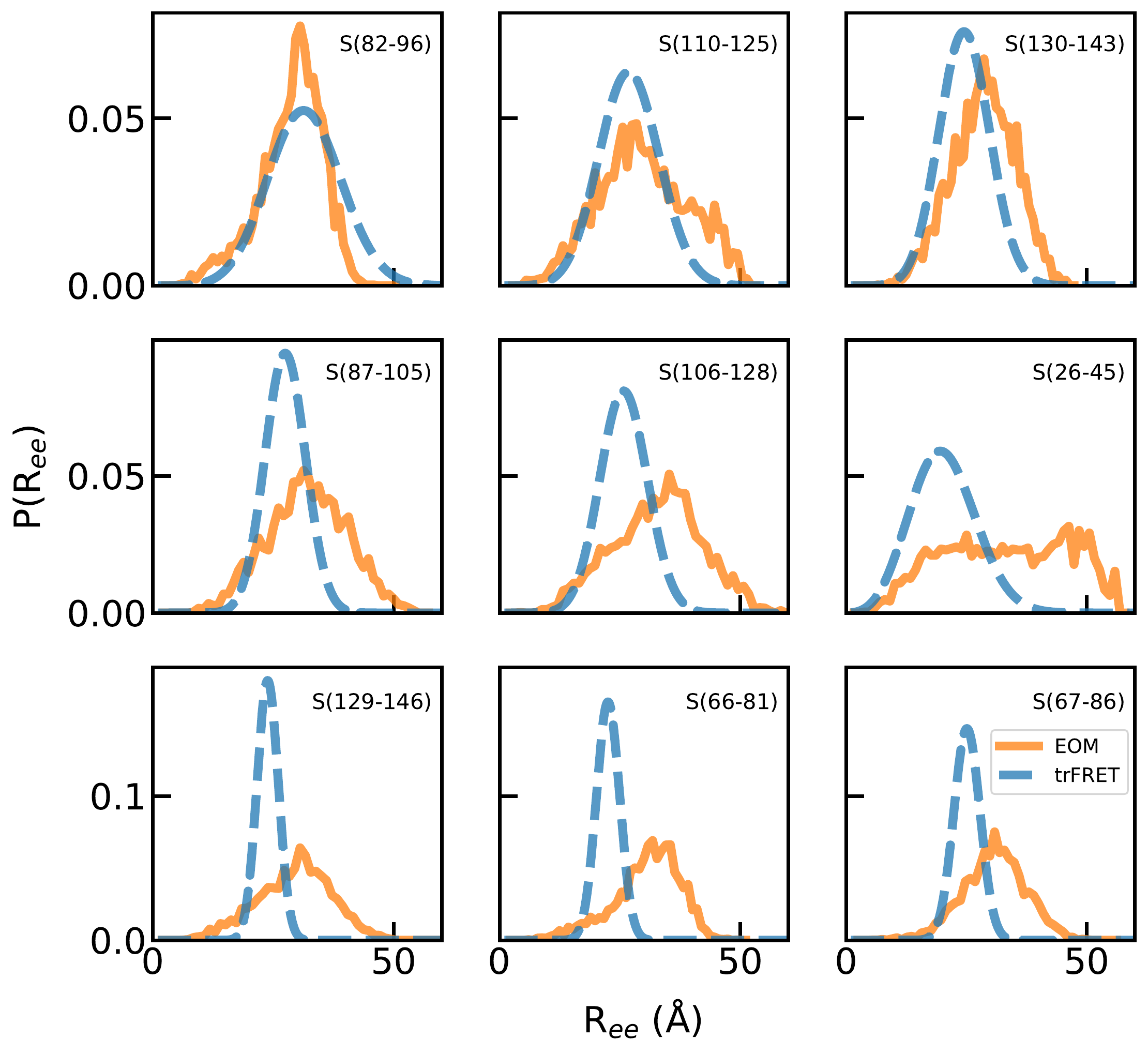}
\caption{Qualitative comparison between $R_{ee}$ probability distributions obtained from trFRET according to Eq.~4 (blue dashed lines) to $R_{ee}$ distributions from EOM on SAXS data (orange solid line). The relevant segment is on the top right side of each panel.}
\label{fig:SI-EOM-vs-FRET}
\end{figure}

\maxdeadcycles=200

\begin{figure}[h]
\centering
\includegraphics[width=0.7\linewidth]{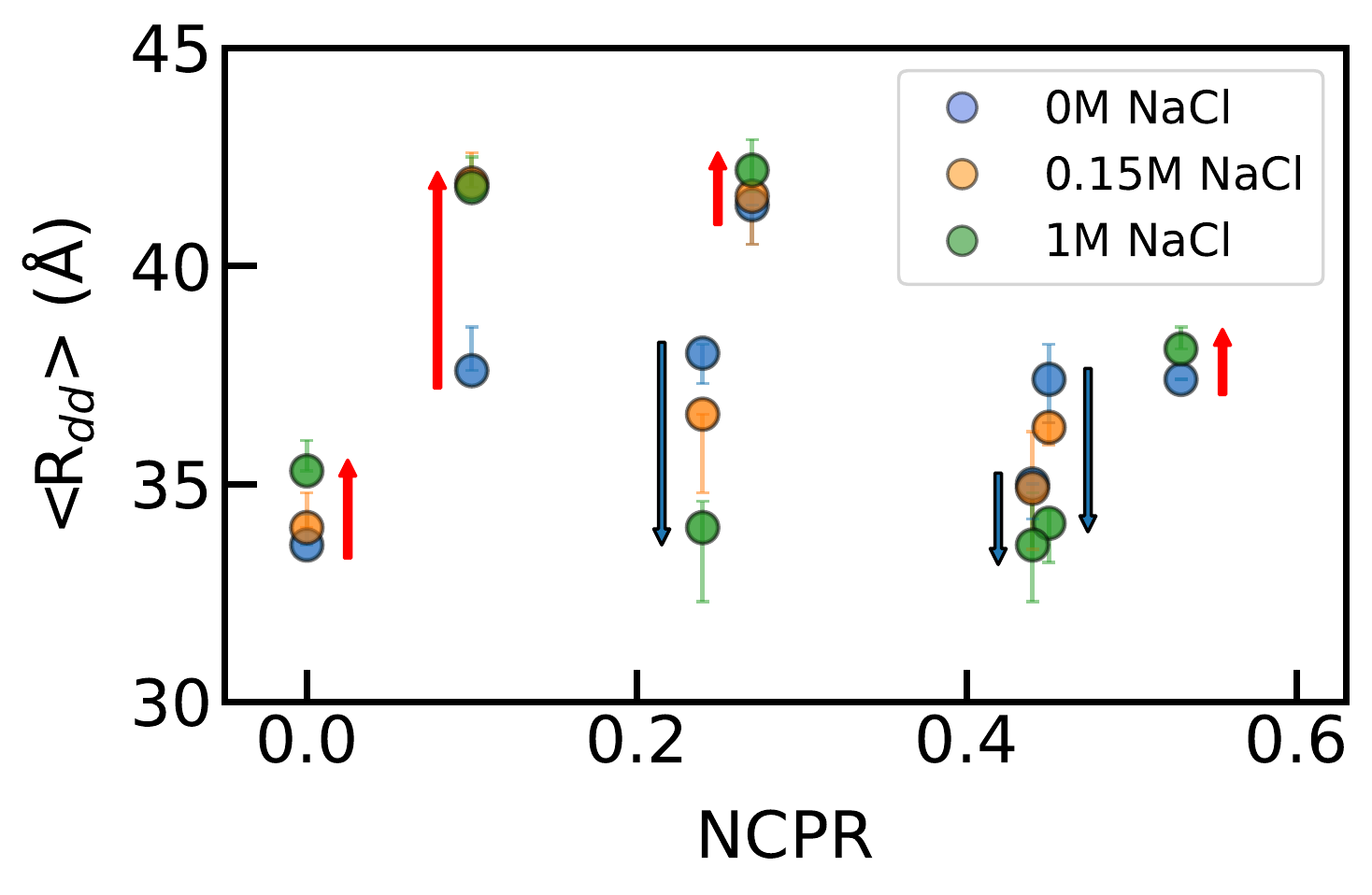}
\caption{P segment mean dye-to-dye distance as extracted with radial Gaussian model from trFRET measurements. Arrows indicate the direction of increasing salinity. Confidence intervals were determined by a rigorous error analysis procedure for 1 std.}
\label{fig:P_GaussMeanSalt}
\end{figure}

\begin{figure}[h]
\centering
\includegraphics[width=0.7\linewidth]{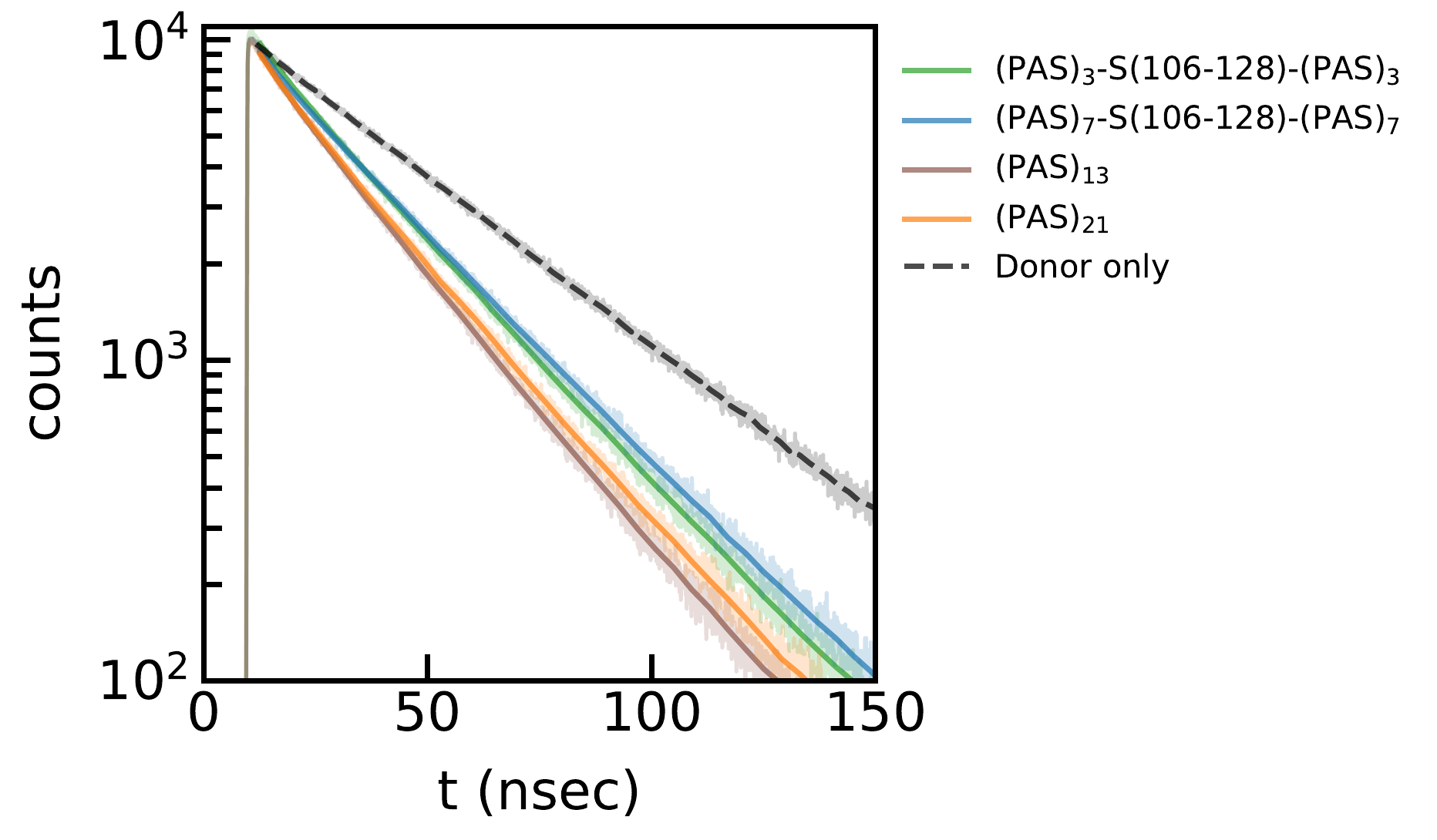}
\caption{Fluorescence decays of donor only (DO) (dashed line) and donor in the presence of an acceptor (DA) (continues line). Increasing PAS chain length results in a very small change in the fluorescence decay.}
\label{fig:SI-PAS-FRET-decay}
\end{figure}

\begin{figure}[h]
\centering
\includegraphics[width=0.7\linewidth]{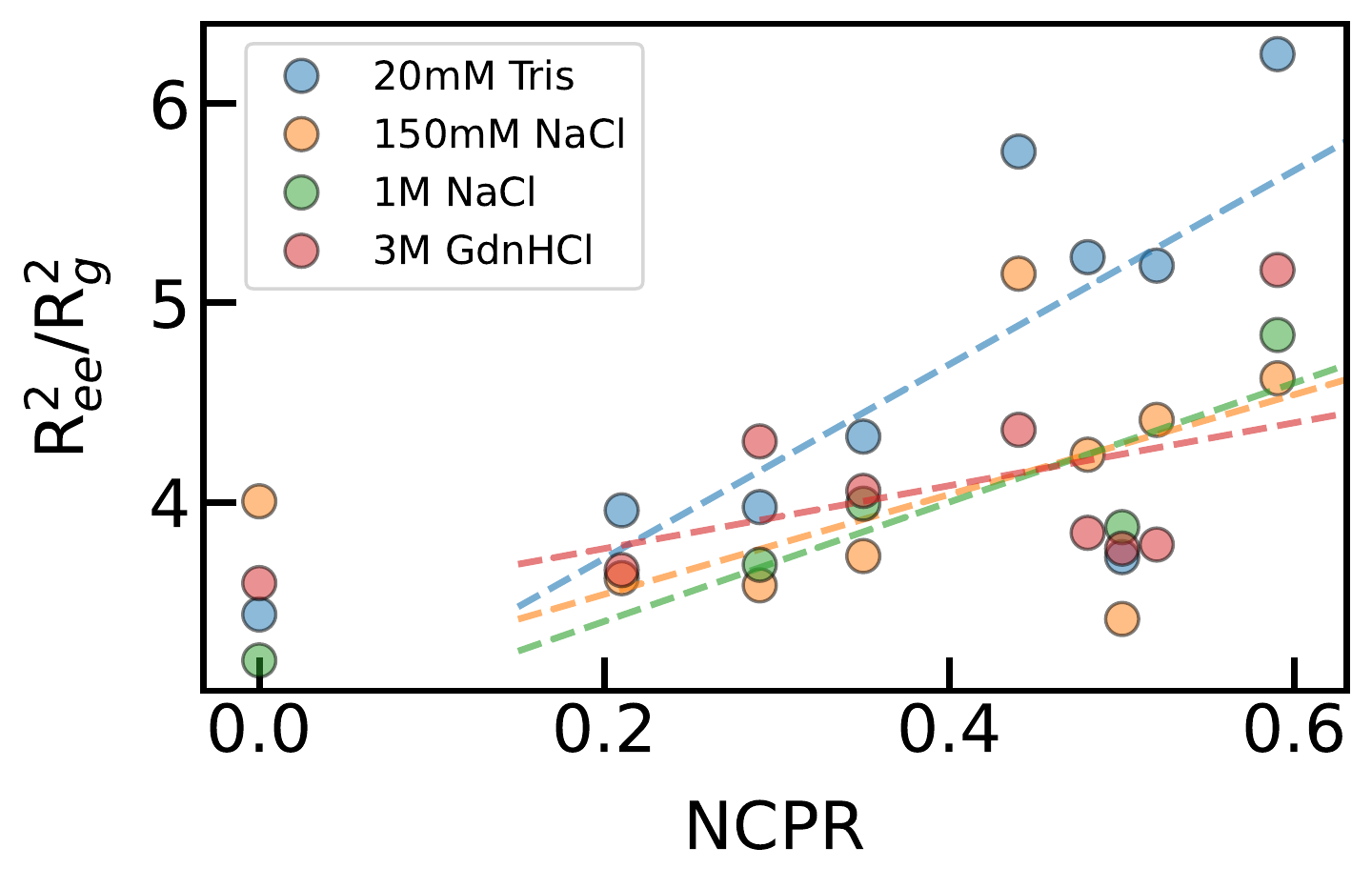}
\caption{$R_{ee}^2$ divided by $R_g^2$ versus the fractional charge at different salinity conditions.}
\label{fig:ReeRg}
\end{figure}

\begin{table}[h]
\centering
\resizebox{\textwidth}{!}{%
\begin{tabular}{|c|cccc|cccc|cccc|cccc|}
\hline
           & \multicolumn{4}{c|}{0 M NaCl}         & \multicolumn{4}{c|}{0.15 M NaCl}      & \multicolumn{4}{c|}{1 M NaCl}         & \multicolumn{4}{c|}{3 M GdnHcl}       \\ \hline
           & \multicolumn{1}{c|}a & 
           \multicolumn{1}{c|}b &
           \multicolumn{1}{c|}{$<R_{dd}>$} & FWHM & \multicolumn{1}{c|}a & 
           \multicolumn{1}{c|}b &
           \multicolumn{1}{c|}{$<R_{dd}>$} & FWHM &
           \multicolumn{1}{c|}a & 
           \multicolumn{1}{c|}b &
           \multicolumn{1}{c|}{$<R_{dd}>$} & FWHM &
           \multicolumn{1}{c|}a & 
           \multicolumn{1}{c|}b &
           \multicolumn{1}{c|}{$<R_{dd}>$} & FWHM

           \\ \hline
P(23-6)    & 

\multicolumn{1}{c|}{0.000}    & 
\multicolumn{1}{c|}{0.001}    &
\multicolumn{1}{c|}{33.6{\scriptsize$\pm$1.0}} & 
\multicolumn{1}{c|}{34.2{\scriptsize$\pm$2.6}} &
\multicolumn{1}{c|}{0.002}    & 
\multicolumn{1}{c|}{0.001}    &
\multicolumn{1}{c|}{34.0{\scriptsize$\pm$1.0}}     & 
\multicolumn{1}{c|}{34.8{\scriptsize$\pm$3.0}}    &
\multicolumn{1}{c|}{0.000}    & 
\multicolumn{1}{c|}{0.001}    &
\multicolumn{1}{c|}{35.3{\scriptsize$\pm$1.0}}     & 
\multicolumn{1}{c|}{36.2{\scriptsize$\pm$3.2}}    &
\multicolumn{1}{c|}{0.000}    &
\multicolumn{1}{c|}{0.001} &
\multicolumn{1}{c|}{38.0{\scriptsize$\pm$1.2}}     & 
\multicolumn{1}{c|}{39.0{\scriptsize$\pm$5.0}}

\\ \hline
P(23-43)   & 
\multicolumn{1}{c|}{0.000}    & 
\multicolumn{1}{c|}{0.001}    &
\multicolumn{1}{c|}{37.6{\scriptsize$\pm$1.0}} & 
\multicolumn{1}{c|}{38.6{\scriptsize$\pm$5.1}} &
\multicolumn{1}{c|}{0.002}    & 
\multicolumn{1}{c|}{0.001}    &
\multicolumn{1}{c|}{41.9{\scriptsize$\pm$0.8}}     & 
\multicolumn{1}{c|}{43.0{\scriptsize$\pm$5.2}}    &
\multicolumn{1}{c|}{0.009}    & 
\multicolumn{1}{c|}{0.001}    &
\multicolumn{1}{c|}{41.8{\scriptsize$\pm$0.8}}     & 
\multicolumn{1}{c|}{43.0{\scriptsize$\pm$5.0}}    &
\multicolumn{1}{c|}{0.000}    &
\multicolumn{1}{c|}{0.001}   &
\multicolumn{1}{c|}{37.3{\scriptsize$\pm$1.1}}     & 
\multicolumn{1}{c|}{38.1{\scriptsize$\pm$7.5}}

\\ \hline
P(64-43)   & 
\multicolumn{1}{c|}{26.67}    & 
\multicolumn{1}{c|}{0.002}    &
\multicolumn{1}{c|}{41.4{\scriptsize$\pm$1.3}} & 
\multicolumn{1}{c|}{32.9{\scriptsize$\pm$15.4}} &
\multicolumn{1}{c|}{28.29}    & 
\multicolumn{1}{c|}{0.002}    &
\multicolumn{1}{c|}{41.6{\scriptsize$\pm$2.5}}     & 
\multicolumn{1}{c|}{32.0{\scriptsize$\pm$17.7}}    &
\multicolumn{1}{c|}{9.68}    & 
\multicolumn{1}{c|}{0.001}    &
\multicolumn{1}{c|}{42.2{\scriptsize$\pm$0.7}}     & 
\multicolumn{1}{c|}{41.3{\scriptsize$\pm$12.6}}    &
\multicolumn{1}{c|}{34.14}    &
\multicolumn{1}{c|}{0.007}   &
\multicolumn{1}{c|}{38.0{\scriptsize$\pm$2.1}}     & 
\multicolumn{1}{c|}{18.7{\scriptsize$\pm$18.2}}    

\\ \hline
P(64-80)   & 
\multicolumn{1}{c|}{30.29}    & 
\multicolumn{1}{c|}{0.004}    &
\multicolumn{1}{c|}{38.0{\scriptsize$\pm$1.0}} & 
\multicolumn{1}{c|}{24.8{\scriptsize$\pm$14.7}} &
\multicolumn{1}{c|}{28.16}    & 
\multicolumn{1}{c|}{0.004}    &
\multicolumn{1}{c|}{36.6{\scriptsize$\pm$0.9}}     & 
\multicolumn{1}{c|}{25.1{\scriptsize$\pm$12.7}}    &
\multicolumn{1}{c|}{20.23}    & 
\multicolumn{1}{c|}{0.003}    &
\multicolumn{1}{c|}{34.0{\scriptsize$\pm$1.5}}     & 
\multicolumn{1}{c|}{28.2{\scriptsize$\pm$8.3}}    &
\multicolumn{1}{c|}{30.73}    &
\multicolumn{1}{c|}{0.013}  &
\multicolumn{1}{c|}{33.2{\scriptsize$\pm$0.6}}     & 
\multicolumn{1}{c|}{14.4{\scriptsize$\pm$8.5}}    
 
\\ \hline
P(80-96)   & 
\multicolumn{1}{c|}{33.12}    & 
\multicolumn{1}{c|}{0.006}    &
\multicolumn{1}{c|}{37.4{\scriptsize$\pm$2.0}} & 
\multicolumn{1}{c|}{19.8{\scriptsize$\pm$18.2}} &
\multicolumn{1}{c|}{20.25}    & 
\multicolumn{1}{c|}{0.001}    &
\multicolumn{1}{c|}{39.8{\scriptsize$\pm$1.2}}     & 
\multicolumn{1}{c|}{37.6{\scriptsize$\pm$14.5}}    &
\multicolumn{1}{c|}{35.08}    & 
\multicolumn{1}{c|}{0.009}    &
\multicolumn{1}{c|}{38.1{\scriptsize$\pm$2.7}}     & 
\multicolumn{1}{c|}{17.3{\scriptsize$\pm$18.2}}    &
\multicolumn{1}{c|}{x}    &
\multicolumn{1}{c|}{x}  &
\multicolumn{1}{c|}{x}     & 
\multicolumn{1}{c|}{x}

\\ \hline
P(126-109) & 
\multicolumn{1}{c|}{32.40}    & 
\multicolumn{1}{c|}{0.006}    &
\multicolumn{1}{c|}{37.4{\scriptsize$\pm$1.6}} & 
\multicolumn{1}{c|}{20.7{\scriptsize$\pm$16.6}} &
\multicolumn{1}{c|}{29.01}    & 
\multicolumn{1}{c|}{0.004}    &
\multicolumn{1}{c|}{36.3{\scriptsize$\pm$1.0}}     & 
\multicolumn{1}{c|}{23.8{\scriptsize$\pm$13.2}}    &
\multicolumn{1}{c|}{24.14}    & 
\multicolumn{1}{c|}{0.003}    &
\multicolumn{1}{c|}{34.1{\scriptsize$\pm$1.4}}     & 
\multicolumn{1}{c|}{25.4{\scriptsize$\pm$9.4}}    &
\multicolumn{1}{c|}{21.37}    &
\multicolumn{1}{c|}{0.003} &
\multicolumn{1}{c|}{34.4{\scriptsize$\pm$1.4}}     & 
\multicolumn{1}{c|}{27.9{\scriptsize$\pm$9.0}}

\\ \hline
P(126-141) & 
\multicolumn{1}{c|}{28.32}    & 
\multicolumn{1}{c|}{0.005}    &
\multicolumn{1}{c|}{35.2{\scriptsize$\pm$0.85}} & 
\multicolumn{1}{c|}{22.4{\scriptsize$\pm$22.5}} &
\multicolumn{1}{c|}{14.94}    & 
\multicolumn{1}{c|}{0.002}    &
\multicolumn{1}{c|}{35.0{\scriptsize$\pm$1.4}}     & 
\multicolumn{1}{c|}{31.9{\scriptsize$\pm$7.9}}    &
\multicolumn{1}{c|}{12.13}    & 
\multicolumn{1}{c|}{0.002}    &
\multicolumn{1}{c|}{33.6{\scriptsize$\pm$1.6}}     & 
\multicolumn{1}{c|}{31.5{\scriptsize$\pm$6.5}}    &
\multicolumn{1}{c|}{27.43}    &
\multicolumn{1}{c|}{0.006}   &
\multicolumn{1}{c|}{32.7{\scriptsize$\pm$1.2}}     & 
\multicolumn{1}{c|}{19.5{\scriptsize$\pm$10.9}}

\\ \hline
\end{tabular}}
\caption {\textbf {P segments end-to-end distance determine by fitting trFRET measurements with radial Gauss model.} a and b are the fitting parameters of the radial Gauss model (Eq.~4) which are used for calculating $<R_{dd}>$  and full width half max (FWHM). Values presented in [\AA]. Errors were calculated by rigorous analysis with 2SD} 
\label{tab:P_FRET_Ree}
\end{table}

\begin{table}[h]
\centering
\begin{tabular}{c|c|c|c|c|}
\cline{2-5}
                                 & 0 M NaCl & 0.15 M  NaCl & 1 M    NaCl & 3 M    GdnHCl \\ \hline
\multicolumn{1}{|c|}{P(6-23)}    & 0.63    & 0.63        & 0.63       & 0.70       \\ \hline
\multicolumn{1}{|c|}{P(23-43)}   & 0.63    & 0.64        & 0.63       & 0.69       \\ \hline
\multicolumn{1}{|c|}{P(43-64)}   & 0.68    & 0.68        & 0.67       & 0.65       \\ \hline
\multicolumn{1}{|c|}{P(64-80)}   & 0.72    & 0.71        & 0.68       & 0.67       \\ \hline
\multicolumn{1}{|c|}{P(80-96)}   & 0.70    & 0.73        & 0.73       & X          \\ \hline
\multicolumn{1}{|c|}{P(109-126)} & 0.71    & 0.69        & 0.66       & 0.67       \\ \hline
\multicolumn{1}{|c|}{P(126-141)} & 0.71    & 0.69        & 0.68       & 0.72       \\ \hline
\end{tabular}
\caption {\textbf {P segments scaling exponent determine by direct fitting fluorescence decay to the SAW model.}} 
\label{tab:P_FRET_Nu}
\end{table}

\begin{table}[h]
    \centering
    \resizebox{\textwidth}{!}{%
    \begin{tabular}{|l|l|l|l|}
        \hline Name & sequence & N & Mw  \\

        \hline (PAS)$_7$ & K(DNS)PASPASPASPASPASPASPAS(NaphA) & 23 & 2362.65 \\
        \hline (PAS)$_{13}$ & PASPASPASK(DNS)PASPASPASPASPASPASPAS(NaphA)PASPASPAS & 23 & 3894.29 \\
        \hline (PAS)$_{21}$ & PASPASPASPASPASPASPASK(DNS)PASPASPASPASPASPASPAS(NaphA)PASPASPASPASPASPASPAS & 23 & 5936.48 \\

        \hline (PAS)$_3$-S(6-25)-(PAS)$_3$ & PASPASPASK(DNS)FTSVGSITSGYSQSSQVFGR(NaphA)PASPASPAS & 22 & 4184.61 \\
        \hline (PAS)$_7$-S(6-25)-(PAS)$_7$ & PASPASPASPASPASPASPASK(DNS)FTSVGSITSGYSQSSQVFGR(NaphA)PASPASPASPASPASPASPAS & 22 & 6226.81 \\
    
        \hline (PAS)$_3$-S(106-128)-(PAS)$_3$ & PASPASPASK(DNS)
SEDTKEEEEGGEGEEEDTKE(NaphA)PASPASPAS & 22 & 4345.49 \\
        \hline (PAS)$_7$-S(106-128)-(PAS)$_7$ & PASPASPASPASPASPASPASK(DNS)SEDTKEEEEGGEGEEEDTKE(NaphA)PASPASPASPASPASPASPAS & 22 & 6387.68\\
        
        \hline
    \end{tabular}}
    \caption {\textbf {Segments with tethered PAS repeats used for trFRET and SAXS.} For SAXS measurements the peptide including only donor were used}
    \label{tab:PAS_Sequence}
\end{table}

\begin{table}[h]
\resizebox{\textwidth}{!}{%
\begin{tabular}{c|cccc|cccc|cccc|cccc|}
\cline{2-17}
                                               & \multicolumn{4}{c|}{0 M NaCl}                                                                  & \multicolumn{4}{c|}{0.15 M NaCl}                                                               & \multicolumn{4}{c|}{0.5 M NaCl}                                                                 & \multicolumn{4}{c|}{3 M GdnHcl}                                                                \\ \cline{2-17} 
                                               & \multicolumn{1}{c|}{$\tau_{DO}$} & \multicolumn{1}{c|}{$\tau_{DA}$} & \multicolumn{1}{c|}{ET}   & $\nu$    & \multicolumn{1}{c|}{$\tau_{DO}$} & \multicolumn{1}{c|}{$\tau_{DA}$} & \multicolumn{1}{c|}{ET}   & $\nu$    & \multicolumn{1}{c|}{$\tau_{DO}$} & \multicolumn{1}{c|}{$\tau_{DA}$} & \multicolumn{1}{c|}{ET}   & $\nu$     & \multicolumn{1}{c|}{$\tau_{DO}$} & \multicolumn{1}{c|}{$\tau_{DA}$} & \multicolumn{1}{c|}{ET}   & $\nu$    \\ \hline
\multicolumn{1}{|c|}{(PAS)$_7$}                   & \multicolumn{1}{c|}{39.28}  & \multicolumn{1}{c|}{17.93}  & \multicolumn{1}{c|}{54.4} & 0.499 & \multicolumn{1}{c|}{38.82}  & \multicolumn{1}{c|}{17.47}  & \multicolumn{1}{c|}{55.0} & 0.496 & \multicolumn{1}{c|}{38.03}  & \multicolumn{1}{c|}{16.91}  & \multicolumn{1}{c|}{55.5} & 0.494  & \multicolumn{1}{c|}{32.56}  & \multicolumn{1}{c|}{18.93}  & \multicolumn{1}{c|}{41.9} & 0.547 \\ \hline
\multicolumn{1}{|c|}{(PAS)$_{13}$}                  & \multicolumn{1}{c|}{40.69}  & \multicolumn{1}{c|}{23.19}  & \multicolumn{1}{c|}{43.0} & 0.543 & \multicolumn{1}{c|}{40.27}  & \multicolumn{1}{c|}{22.64}  & \multicolumn{1}{c|}{43.8} & 0.54  & \multicolumn{1}{c|}{39.56}  & \multicolumn{1}{c|}{22.08}  & \multicolumn{1}{c|}{44.2} & 0.538  & \multicolumn{1}{c|}{33.72}  & \multicolumn{1}{c|}{22.81}  & \multicolumn{1}{c|}{32.4} & 0.586 \\ \hline
\multicolumn{1}{|c|}{(PAS)$_{21}$}                  & \multicolumn{1}{c|}{41.24}  & \multicolumn{1}{c|}{23.5}   & \multicolumn{1}{c|}{43.0} & 0.543 & \multicolumn{1}{c|}{40.74}  & \multicolumn{1}{c|}{24}     & \multicolumn{1}{c|}{41.1} & 0.55  & \multicolumn{1}{c|}{40.11}  & \multicolumn{1}{c|}{23.13}  & \multicolumn{1}{c|}{42.3} & 0.545  & \multicolumn{1}{c|}{34.2}   & \multicolumn{1}{c|}{23.72}  & \multicolumn{1}{c|}{30.6} & 0.594 \\ \hline

\multicolumn{1}{|c|}{(PAS)$_{3}$-S(6-25)-(PAS)$_{3}$}    & \multicolumn{1}{c|}{30.85}  & \multicolumn{1}{c|}{11.12}  & \multicolumn{1}{c|}{64.0} & 0.467 & \multicolumn{1}{c|}{29.01}  & \multicolumn{1}{c|}{9.92}   & \multicolumn{1}{c|}{65.8} & 0.46  & \multicolumn{1}{c|}{28.3}   & \multicolumn{1}{c|}{8.61}   & \multicolumn{1}{c|}{69.6} & 0.443  & \multicolumn{1}{c|}{28.63}  & \multicolumn{1}{c|}{20.55}  & \multicolumn{1}{c|}{28.2} & 0.613 \\ \hline
\multicolumn{1}{|c|}{(PAS)$_{7}$-S(6-25)-(PAS)$_{7}$}    & \multicolumn{1}{c|}{30.85}  & \multicolumn{1}{c|}{6.71}   & \multicolumn{1}{c|}{78.2} & 0.401 & \multicolumn{1}{c|}{29.01}  & \multicolumn{1}{c|}{5.15}   & \multicolumn{1}{c|}{82.2} & 0.378 & \multicolumn{1}{c|}{28.3}   & \multicolumn{1}{c|}{4.74}   & \multicolumn{1}{c|}{83.3} & 0.371  & \multicolumn{1}{c|}{28.63}  & \multicolumn{1}{c|}{16.04}  & \multicolumn{1}{c|}{44.0} & 0.546 \\ \hline

\multicolumn{1}{|c|}{(PAS)$_{3}$-S(106-128)-(PAS)$_{3}$} & \multicolumn{1}{c|}{41.81}  & \multicolumn{1}{c|}{26.93}  & \multicolumn{1}{c|}{35.6} & 0.58  & \multicolumn{1}{c|}{40.9}   & \multicolumn{1}{c|}{25.96}  & \multicolumn{1}{c|}{36.5} & 0.577 & \multicolumn{1}{c|}{40.28}  & \multicolumn{1}{c|}{20.36}  & \multicolumn{1}{c|}{49.5} & 0.525  & \multicolumn{1}{c|}{35.03}  & \multicolumn{1}{c|}{20.55}  & \multicolumn{1}{c|}{41.3} & 0.557 \\ \hline
\multicolumn{1}{|c|}{(PAS)$_{7}$-S(106-128)-(PAS)$_{7}$} & \multicolumn{1}{c|}{41.87}  & \multicolumn{1}{c|}{27.35}  & \multicolumn{1}{c|}{34.7} & 0.584 & \multicolumn{1}{c|}{41.18}  & \multicolumn{1}{c|}{27.59}  & \multicolumn{1}{c|}{33.0} & 0.591 & \multicolumn{1}{c|}{40.59}  & \multicolumn{1}{c|}{26.02}  & \multicolumn{1}{c|}{35.9} & 0.579  & \multicolumn{1}{c|}{33.94}  & \multicolumn{1}{c|}{23.78}  & \multicolumn{1}{c|}{29.9} & 0.605 \\ \hline
\end{tabular}}
\caption {\textbf { PAS tethered segments scaling exponent determine by fitting trFRET measurements with SAW model.} $<ET>$ is calculate by $1-\tau_{DA}/\tau_{DO}$ where $\tau_{DO}$ and $\tau_{DA}$ are the fluorescence life time of segments containing donor only or donor and acceptor , respectively. Lifetime values presented in [nsec]} 
\label{tab:PAS_FRET_Nu}
\end{table}

\begin{table}[h]
\resizebox{\textwidth}{!}{%
\begin{tabular}{|c|cc|cc|cc|cc|}
\hline
                     & \multicolumn{2}{c|}{0M NaCl}                    & \multicolumn{2}{c|}{0.15M NaCl}                 & \multicolumn{2}{c|}{0.5M NaCl}                  & \multicolumn{2}{c|}{3M GdnHcl}                  \\ \hline
                     & \multicolumn{1}{c|}{$<R_{dd}>$} &FWHM& \multicolumn{1}{c|}{$<R_{dd}>$} &FWHM& \multicolumn{1}{c|}{$<R_{dd}>$} & FWHM& \multicolumn{1}{c|}{$<R_{dd}>$} & FWHM\\ \hline
7PAS                 & \multicolumn{1}{c|}{22.0}        & 5.8          & \multicolumn{1}{c|}{21.9}        & 5.7          & \multicolumn{1}{c|}{21.8}        & 5.7          & \multicolumn{1}{c|}{23.4}        & 7.9          \\ \hline
3PAS-7PAS-3PAS       & \multicolumn{1}{c|}{24.1}        & 6.8          & \multicolumn{1}{c|}{23.9}        & 6.8          & \multicolumn{1}{c|}{23.8}        & 7            & \multicolumn{1}{c|}{25.5}        & 7.9          \\ \hline
7PAS-7PAS-7PAS       & \multicolumn{1}{c|}{24.6}        & 8.1          & \multicolumn{1}{c|}{24.5}        & 8            & \multicolumn{1}{c|}{24.4}        & 8.3          & \multicolumn{1}{c|}{26.1}        & 8.6          \\ \hline
3PAS-S(6-25)-3PAS    & \multicolumn{1}{c|}{21.4}        & 22.2         & \multicolumn{1}{c|}{22.9}        & 23.5         & \multicolumn{1}{c|}{22.8}        & 23.2         & \multicolumn{1}{c|}{23.9}        & 8.2          \\ \hline
7PAS-S(6-25)-7PAS    & \multicolumn{1}{c|}{21.8}        & 22.3         & \multicolumn{1}{c|}{X}        & X         & \multicolumn{1}{c|}{X}        & X         & \multicolumn{1}{c|}{23.8}        & 8.9          \\ \hline
3PAS-S(106-128)-3PAS & \multicolumn{1}{c|}{26.1}        & 8.6          & \multicolumn{1}{c|}{25.5}        & 8.3          & \multicolumn{1}{c|}{22.4}        & 20.8         & \multicolumn{1}{c|}{24.9}        & 12.8         \\ \hline
7PAS-S(106-128)-7PAS & \multicolumn{1}{c|}{27.3}        & 12.9         & \multicolumn{1}{c|}{26.5}        & 9.9          & \multicolumn{1}{c|}{26.1}        & 11.6         & \multicolumn{1}{c|}{26.1}        & 8.2          \\ \hline
\end{tabular}}
\caption {\textbf {Tethered PAS segments end-to-end distance by fitting trFRET measurements with radial Gauss model.} Values  presented in [\AA]}
\label{tab:PAS_FRET_Ree}
\end{table}

\begin{table}[h]
\resizebox{\textwidth}{!}{%
\begin{tabular}{|l|lll|lll|lll|lll|lll|lll|}
\hline
                               & \multicolumn{3}{c|}{Tris}                                               & \multicolumn{3}{c|}{50mM NaCl}                                          & \multicolumn{3}{c|}{150mM NaCl}                                         & \multicolumn{3}{c|}{500mM NaCl}                                         & \multicolumn{3}{c|}{1M NaCl}                                            & \multicolumn{3}{c|}{3M GdnHCl}                                          \\ \hline
Segment                        & \multicolumn{1}{l|}{$R_g$}      & \multicolumn{1}{l|}{$R_g$ ex.}   & $\nu$        & \multicolumn{1}{l|}{$R_g$}      & \multicolumn{1}{l|}{$R_g$ ex.}   & $\nu$        & \multicolumn{1}{l|}{$R_g$}      & \multicolumn{1}{l|}{$R_g$ ex.}   & $\nu$        & \multicolumn{1}{l|}{$R_g$}      & \multicolumn{1}{l|}{$R_g$ ex.}   & $\nu$        & \multicolumn{1}{l|}{$R_g$}      & \multicolumn{1}{l|}{$R_g$ ex.}   & $\nu$        & \multicolumn{1}{l|}{$R_g$}      & \multicolumn{1}{l|}{$R_g$ ex.}   & $\nu$        \\ \hline
(PAS)$_3$S(106-128)(PAS)$_3$ & \multicolumn{1}{l|}{}        & \multicolumn{1}{l|}{}         &          & \multicolumn{1}{l|}{1.632} & \multicolumn{1}{l|}{1.734} & 0.556 & \multicolumn{1}{l|}{1.729} & \multicolumn{1}{l|}{1.813} & 0.572 & \multicolumn{1}{l|}{1.725} & \multicolumn{1}{l|}{1.789} & 0.567 & \multicolumn{1}{l|}{1.658}  & \multicolumn{1}{l|}{1.674} & 0.544 & \multicolumn{1}{l|}{1.685} & \multicolumn{1}{l|}{1.771} & 0.564 \\ \hline
(PAS)$_7$S(106-128)(PAS)$_7$ & \multicolumn{1}{l|}{2.089} & \multicolumn{1}{l|}{2.255} & 0.555 & \multicolumn{1}{l|}{2.267} & \multicolumn{1}{l|}{2.366} & 0.569 & \multicolumn{1}{l|}{2.352} & \multicolumn{1}{l|}{2.440}  & 0.578   & \multicolumn{1}{l|}{2.335} & \multicolumn{1}{l|}{2.403} & 0.574 & \multicolumn{1}{l|}{}        & \multicolumn{1}{l|}{}         &          & \multicolumn{1}{l|}{2.255} & \multicolumn{1}{l|}{2.365} & 0.569 \\ \hline
(PAS)$_7$                         & \multicolumn{1}{l|}{1.278} & \multicolumn{1}{l|}{1.270} & 0.559 & \multicolumn{1}{l|}{1.352} & \multicolumn{1}{l|}{1.344} & 0.584 & \multicolumn{1}{l|}{1.339} & \multicolumn{1}{l|}{1.349} & 0.585 & \multicolumn{1}{l|}{1.347} & \multicolumn{1}{l|}{1.340} & 0.582 & \multicolumn{1}{l|}{1.329} & \multicolumn{1}{l|}{1.213} & 0.540 & \multicolumn{1}{l|}{}        & \multicolumn{1}{l|}{}         &          \\ \hline
(PAS)$_{13}$                        & \multicolumn{1}{l|}{1.766}   & \multicolumn{1}{l|}{1.829} & 0.570 & \multicolumn{1}{l|}{1.897} & \multicolumn{1}{l|}{1.951} & 0.592 & \multicolumn{1}{l|}{1.878} & \multicolumn{1}{l|}{1.941}  & 0.590 & \multicolumn{1}{l|}{1.877} & \multicolumn{1}{l|}{1.910} & 0.584 & \multicolumn{1}{l|}{1.832} & \multicolumn{1}{l|}{1.849} & 0.573 & \multicolumn{1}{l|}{1.852}  & \multicolumn{1}{l|}{1.884}  & 0.580 \\ \hline
(PAS)$_{21}$                        & \multicolumn{1}{l|}{2.461} & \multicolumn{1}{l|}{2.568} & 0.591 & \multicolumn{1}{l|}{2.521} & \multicolumn{1}{l|}{2.584} & 0.592 & \multicolumn{1}{l|}{2.627} & \multicolumn{1}{l|}{2.629} & 0.597 & \multicolumn{1}{l|}{2.551} & \multicolumn{1}{l|}{2.592} & 0.593  & \multicolumn{1}{l|}{2.505} & \multicolumn{1}{l|}{2.510} & 0.584 & \multicolumn{1}{l|}{2.497} & \multicolumn{1}{l|}{2.547} & 0.588 \\ \hline
(PAS)$_7$S(6-25)(PAS)$_7$    & \multicolumn{1}{l|}{}        & \multicolumn{1}{l|}{}         &          & \multicolumn{1}{l|}{}        & \multicolumn{1}{l|}{}         &          & \multicolumn{1}{l|}{}        & \multicolumn{1}{l|}{}         &          & \multicolumn{1}{l|}{}        & \multicolumn{1}{l|}{}         &          & \multicolumn{1}{l|}{}        & \multicolumn{1}{l|}{}         &          & \multicolumn{1}{l|}{2.366} & \multicolumn{1}{l|}{2.420}  & 0.576 \\ \hline
\end{tabular}}
\caption {\textbf {SAXS results of PAS tethered segments.} The $R_g$ is in nm and $R_g$ ex. is the radius of gyration
obtained from extended Guinier fit with its $\nu$ value.}
\label{tab:PAS_SAXS}
\end{table}

\begin{figure}[h]
\centering
\includegraphics[width=0.7\linewidth]{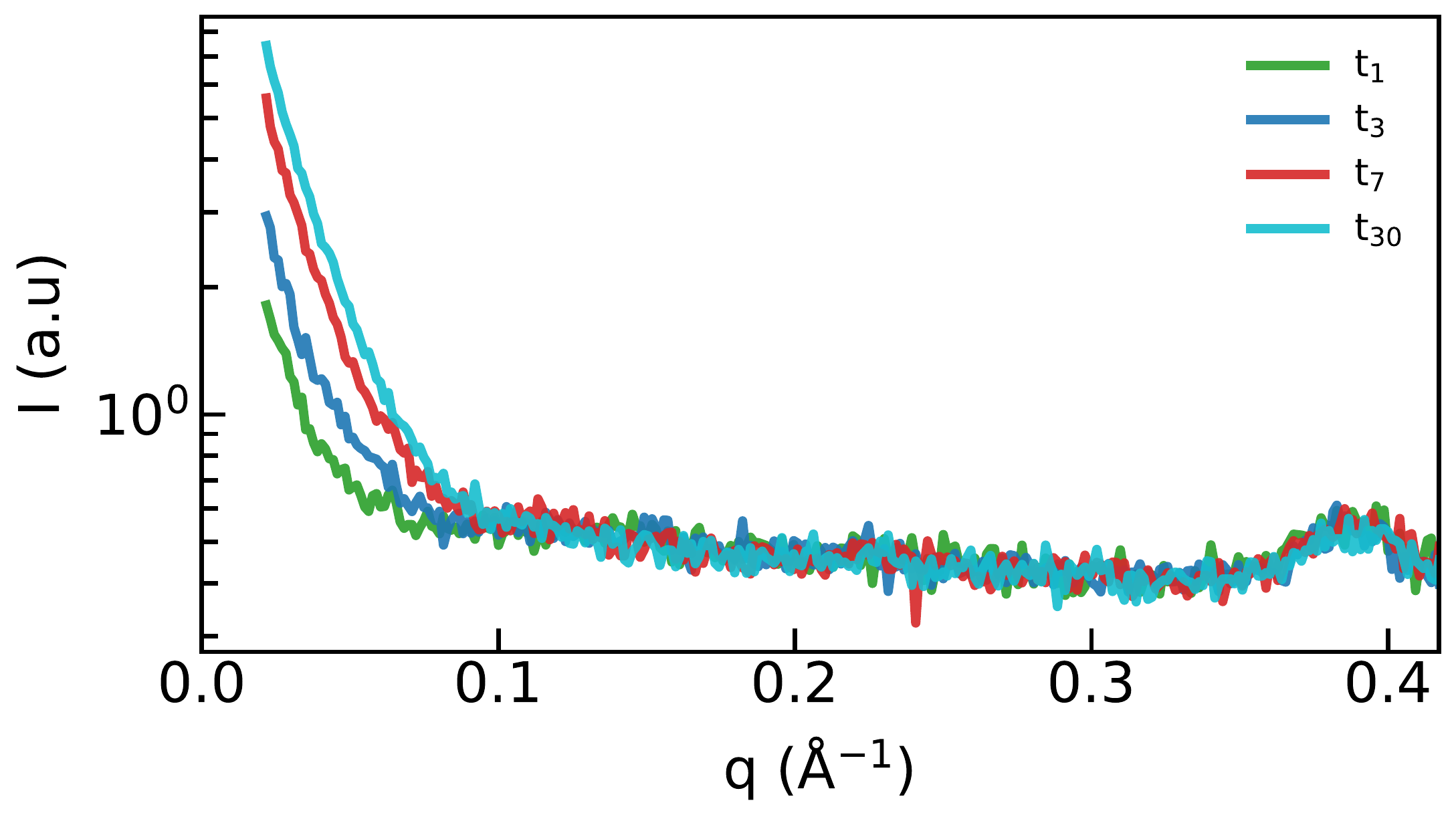}
\caption{Aggregation of segment S(6-25). In-house SAXS unsubtracted measurements at Tel-Aviv University. Each line is a 30 minutes measurement of the same peptide solution at different measurement order, according to $t_n$. For instance, the red curve ($t_7$) was measured for 30 min, after 3.5 hours.} 
\label{fig:SI-Pep1}
\end{figure}

\begin{figure}[h]
\centering
\includegraphics[width=0.6\linewidth]{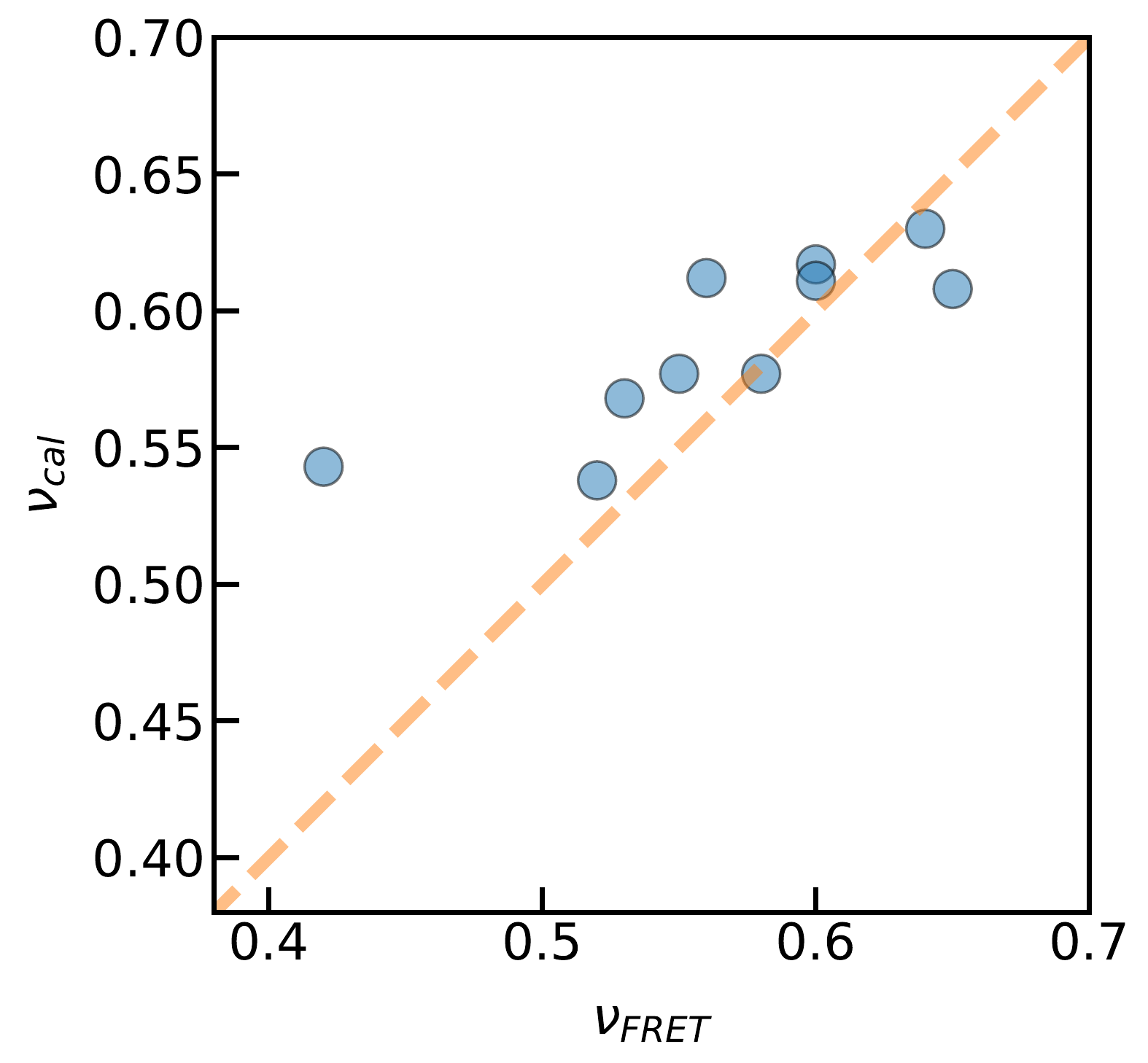}
\caption{Comparison between the scaling exponent obtain by trFRET and by using $\nu_{cal}=-0.0423\cdot SHD+0.0074\cdot SCD+0.701$ from ref \cite{zheng2020hydropathy}.
} 
\label{fig:SI-NuFRET_VS_Nucal}
\end{figure}

\begin{figure}[h]
\centering
\includegraphics[width=0.7\linewidth]{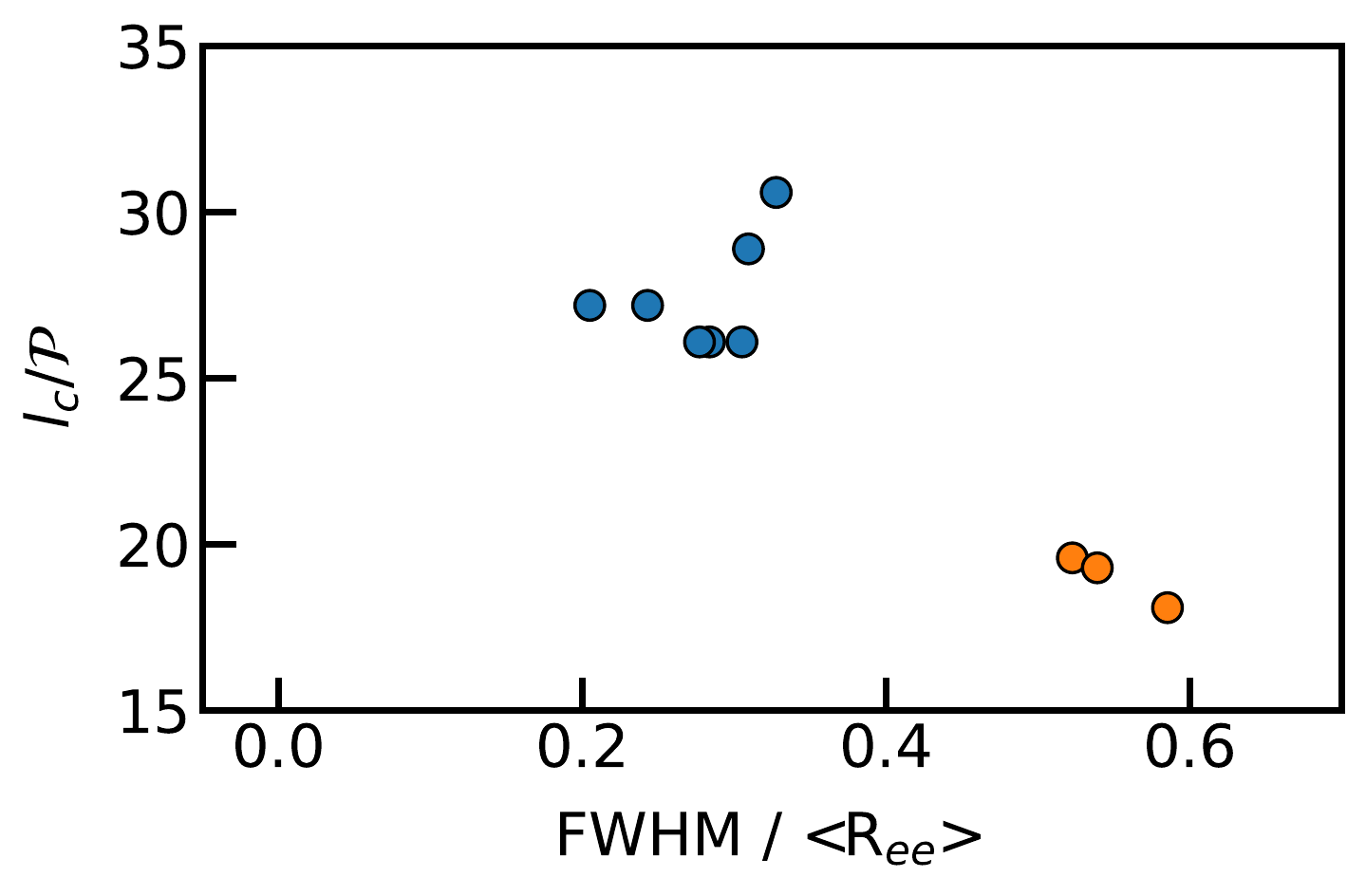}
\caption{\added{Correlation between the relative distribution width and $l_c/\mathcal{P}$. Here, $l_c$ is the segments' contour length and $\mathcal{P}$ counts the number of negatively charged clusters, with at least two neighbouring amino-acids, in the segment.     
}}
\label{fig:SI-ClusterCorrelation}
\end{figure}



\end{document}